\documentclass[reprint, superscriptaddress, amsmath, amssymb, aps, nofootinbib]{revtex4-2}
\usepackage[utf8]{inputenc}
\usepackage{amsmath}
\usepackage{amsthm}
\usepackage[normalem]{ulem}
\usepackage{graphicx}
\usepackage{svg}
\usepackage{ulem}
\usepackage[T1]{fontenc}
\usepackage{mathtools}
\graphicspath{{}}
\newtheorem{theorem}{Theorem}
\begin{document}
\title{Synthetic dimension-induced pseudo Jahn-Teller effect \\
in one-dimensional confined fermions}% Force line breaks with \\
\date{\today}% It is always \today, today,
\author{A. Becker}
\affiliation{Center for Optical Quantum Technologies, Department of Physics, University of Hamburg, 
Luruper Chaussee 149, 22761 Hamburg Germany}
\affiliation{The Hamburg Centre for Ultrafast Imaging,
University of Hamburg, Luruper Chaussee 149, 22761 Hamburg, Germany}
\author{G. M. Koutentakis}
\affiliation{Institute of Science and Technology Austria (ISTA), am Campus 1, 3400 Klosterneuburg, Austria}
\author{P. Schmelcher}
\affiliation{Center for Optical Quantum Technologies, Department of Physics, University of Hamburg, 
Luruper Chaussee 149, 22761 Hamburg Germany}
\affiliation{The Hamburg Centre for Ultrafast Imaging,
University of Hamburg, Luruper Chaussee 149, 22761 Hamburg, Germany}

\begin{abstract}
We demonstrate the failure of the adiabatic Born-Oppenheimer approximation to describe the ground state of a quantum impurity within an ultracold Fermi gas despite substantial mass differences between the bath and impurity species. Increasing repulsion leads to the appearance of non-adiabatic couplings between the fast bath and slow impurity degrees of freedom which reduce the parity symmetry of the latter according to the pseudo Jahn-Teller effect. The presence of this mechanism is associated to a conical intersection involving the impurity position and the inverse of the interaction strength which acts as a synthetic dimension. We elucidate the presence of these effects via a detailed ground state analysis involving the comparison of {\it ab initio} fully-correlated simulations with effective models. Our study suggests ultracold atomic ensembles as potent emulators of complex molecular phenomena.
\end{abstract}
\maketitle
\section{Introduction}
Landau's postulate, which emerged from a discussion with Teller in 1934, that the symmetry causing an energetically degenerate state, is spontaneously lifted, led to the Jahn-Teller effect stating 
\textit{the configuration of any non-linear polyatomic system in a degenerate electronic state undergoes spontaneous distortions that remove the degeneracy} \cite{JahnTeller1937, Englman1972, Bersuker2006, Köppel2009, Bersuker2021}.
This effect was demonstrated theoretically and experimentally in various areas of physics, such as solid state physics, molecular physics and material science, as well as in biology and chemistry \cite{Kahn1998, Feringa2000, Rocha2005,magoni2023, AngeliTosatti2019, Freitag2013, Bacci1980}. 
An extension of the Jahn-Teller effect was found in pseudo-degenerate systems, in which strong vibronic couplings between any two electronic states with an arbitrary non-vanishing energetic gap cause an instability and distortion of the polyatomic system. 
This is known as pseudo Jahn-Teller effect \cite{OepikPryce1957, BersukerPolinger1989, Bersuker2013, Bersuker2021}. 
Furthermore, it has been shown that the (pseudo) Jahn-Teller effect is the only origin for spontaneous symmetry breaking in those systems \cite{BersukerPolinger1989, Bersuker2006, Bersuker2016}. 

%\textcolor{black}{[This next paragraph should be for how cold atoms enable the study of molecular phenomena. The logic in the previous version was a little bit unfocussed. I propose you first start generally with cold atoms and what they are good for and then you say that here we exploit them to study phenomena related with non-adiabaticity. The blue part is not focussed on the experiments we want. First you should mention impurity experiments and species selective trapping and then you mention few body experiments.]}\\

Ultracold quantum gases have proven to be a pristine platform for quantum simulation due to their high degree of versatility and controllability~\cite{PethickSmith2001,PitaevskiiStringari, Bloch_2008, Bloch2012, Gross2017, SowinskiGarciaMarch2019, Schaefer2020, mistakidis_volosniev2022}. 
Furthermore, recently a lot of theoretical~\cite{Tempere2009, Scelle2013, Lemeshko2016, Erdmann2018, ErdmannMistakidis2019,RammelmuellerHuber2023a, RammelmuellerHuber2023b} and experimental~\cite{Chevy2010, Spethmann2012,  Massignan2013,Massignan2014} effort has been devoted to the understanding of the properties of impurities immersed in Bose and Fermi gases.
The emergent properties of such ensembles are analogous to polarons.
In the condensed matter setting these correspond to dressed states of electrons by the vibrations of the surrounding material, playing an important role in understanding electron transport of their host material~\cite{Froehlich1954, Feynman1955, AlexandrovDevreese2010, Sidler2016}.
Due to the excellent tunability of interaction strength in ultracold atoms Bose and Fermi polarons have been studied extensively in the strong interaction limit~\cite{Schirotek2009,Nascimbene2009,Palzer2009,Punk2009,Massignan2011,Schmidt2011,Schmidt2012,Kohstall2012,Ngampruetikorn2012,Zhang2012,Catani2012,Fukuhara2013,Burovski2014,Ardila2016, Schmidt2016B,Grusdt2017, Volosniev2017, ScazzaValtolina2017,Gamayun2018,Guenther2018,Schmidt_2018,MistakidisKatsimigaKoutentakis2019,MistakidisKatsimiga2019,MistakidisKoutentakis2021}, beyond the regimes available within material science.
The above leads to the question whether ultracold atoms can be employed to elucidate the qualitative features of the electronic structure of molecular systems. 
Such investigations can possibly lay the groundwork for observing new phenomena or designing (artificial) molecules with desired properties.
%Generally, a distinction is made between bosons and fermions according to their statistics, which makes it possible to investigate mixtures of different species respectively particles \cite{PethickSmith2001, PitaevskiiStringari}. 

%\textcolor{green}{I think this is not what you had in mind...I guess you want other references?! Do you want more these kind of references: potential landscape ->  M. Greiner, O. Mandel, T. Esslinger, T. W. H ̈ansch, and I. Bloch, Nature 415, 39 (2002).\\
%40K / 6Li -> C.-H. Wu, J. W. Park, P. Ahmadi, S. Will, and M. W. Zwierlein, Phys. Rev. Lett. 109, 085301 (2012);\\
%6 Li -> E. Wille, F. Spiegelhalder, G. Kerner, D. Naik, A. Trenkwalder, G. Hendl, F. Schreck, R. Grimm, T. Tiecke, J. Walraven, S. J. J. M. F. Kokkelmans, E. Tiesinga, and P. S. Julienne, Phys. Rev. Lett. 100, 053201 (2008)
% A. Moerdijk, B. Verhaar, and A. Axelsson, Phys. Rev. A 51, 4852 (1995).\\
%Sr -> M. Takamoto, F.-L. Hong, R. Higashi, Y. Fujii, M. Imae, and H. Katori, J. Phys. Soc. Jpn. 75, 104302 (2006).}
As a first step towards achieving this goal here we propose a one-dimensional system characterized by large mass imbalance in order to study effects associated with non-adiabaticity and the pseudo Jahn-Teller effect. 
The experimental realizability of large mass imbalanced systems has been demonstrated in~\cite{Taglieber2008} for $^6$Li$-$$^{40}$K mixtures. Such a two species mixture provides the opportunity to tune the interaction between both components via Fano-Feshbach and confinement induced resonances~\cite{Tiecke2010, Naik_2011, Cetina2016} and apply species-selective trapping geometries~\cite{Voigt2009}.
\textcolor{black}{The control of individual atoms \cite{Nelson2007} led to the possibility to design intriguing states of matter such as the Mott insulator with just a few bosonic particles \cite{Sherson2010, Bakr2011}. Later on this was also realised with fermionic $^6$Li atoms \cite{Serwane2011}, which allowed experimentalists to study the emergence of fundamental many-body effects like Cooper pairing atom-by-atom \cite{Holten2022}. In general, the few-particle platform, which is the subject of our study, turned out to be fruitful for the understanding of the fundamental processes of many-body quantum physics \cite{ZuernWenzMurmann2013, Wenz_2013, Murmann2015, Bayha2020, Holten2021}.}
Our setup consists of a few-body bath of fermions interacting with a single massive impurity. The confinement of the two species is controlled independently by a distinct harmonic trapping confinement.
%We focus on two-numerical methods to solve the underlying Schr\"odinger equation: Once, we use the numerically exact {\it ab initio} method for one-dimensional fermionic few-particle systems ML-MCTDHF \cite{CaoBolsinger2017, Erdmann2018, ErdmannMistakidis2019, KoutentakisMistakidis2019, KoutentakisMistakidis2020}, as well as a multi-channel Born-Oppenheimer \cite{SzaboOstlund1996, Larsen2020, Baer2006} motivated by the large mass difference between impurity and bath particles.
We examine the non-adiabatic physics in our system by comparing the results of the adiabatic Born-Oppenheimer (BO) approximation with the numerically exact {\it ab initio} Multi-Layer Multi-Configuration Time-Dependent Hartree method for atomic mixtures (ML-MCTDHX) \cite{KrönkeCao2013, CaoLushuai2013, CaoBolsinger2017}. 
%Thus, we can determine the comparison of the ML-MCTDHF simulations, as well as studying the non-adiabatic physics in our system. 
%This is done in a detailed ground-state analysis, where we compare the results of the adiabatic approximation with the exact calculations. 
Even on the basic level of the ground-state energy and one-body density, we point out large deviations among the two approaches evincing significant non-adiabatic contributions which become more prominent for increasing interaction strength. 
The presence of non-adiabaticity is further indicated by the correlation properties of the two-body bath-impurity densities and the inter-species entanglement captured by the von Neumann entropy.
%Finally, we use the high degree of experimental controllability in an ultracold atom setup and vary the trapping frequency and mass of the impurity to demonstrate their influence on the non-adiabaticity for the considered ground-state properties. 

The decrease of these non-adiabatic effects is found to be more sensitive on a increase of the trapping frequency of the impurity as compared to an increase of its mass, assuming a common increment value.
%This is a counterintuitive result given that the adiabatic BO approximation becomes exact at the infinitely heavy impurity limit. 
Given that the adiabatic BO approximation becomes exact for each infinite mass and infinite trapping frequency, this might not be the expected behaviour and indeed, this approximate approach shows a diverging behaviour from the exact one.
%
%These sections point out the importance of the non-adiabatic contribution. 
The above results can be explained in terms of the pseudo Jahn-Teller effect.
In particular, we demonstrate that the bath-impurity system is effectively described by a $E \otimes b$ system known to exhibit the pseudo Jahn-Teller effect.
In addition a detailed symmetry analysis shows that a conical intersection emerges when the impurity position and the inverse of the interaction strength are employed as the slow coordinates of the system provided that the number of bath particles is odd. 
The inverse of the interaction strength in this context can be interpreted as an additional synthetic dimension. Up to now synthetic dimensions have been observed in various fields: such as Rydberg atoms \cite{Hummel2021}, optical lattices \cite{Levi2011, Kolkowitz2016}, photonics \cite{Ozawa2019}, the study of gauge fields \cite{Celi2014} and quantum simulations \cite{Boada2012}.
Analyzing the potential energy curves for fixed interaction within a multi-channel BO approach reveals that more than one conical intersection might occur.
%This is a first indication for the existence of the pseudo Jahn-Teller effect, since an energetic degeneracy is not possible in a one-dimensional system \cite{Bersuker2013}. 
%Therefore, we point out the analogy between this system and an examplary $E \otimes b$ system for the Jahn-Teller effect. Especially, we determine the associated symmetry breaking in our system from the potential energy curves, which we detect finally in the state of the impurity. 
%Finally, it is shown how the Born-Huang term, which occurs in the multi-channel Born-Oppenheimer approx, can be used as a measure of non-adiabaticity in the underlying system. 
The associated (quasi-)degeneracy points among the potential energy curves are explained in terms of resonant bath particle transport through the impurity when the impurity resides in the vicinity of specific positions. These positions can be identified by the sharp increase of the bath momentum in their vicinity and thus are captured by the Born-Huang correction of the lowest potential energy surface.
%In one-dimensional systems, energetically degenerate states are excluded~\cite{Messiahbook} and as such only the pseudo Jan. 
%Initially, the setup of interest is represented, next we perform a detailed %ground-state analysis to get a deeper understanding of the underlying %system. 
%We compare the results from the adiabatic Born-Oppenheimer (BO) %approximation with a numerically exact multi-channel BO approach %\cite{SzaboOstlund1996, Larsen2020, Baer2006} and our Multi-Layer Multi-%Configuration Time-Dependent Hartree method for fermions (ML-MCTHDF), which %exploits the permutation symmetry of identical fermions and is applicable to %investigate the static and dynamical properties of fermionic few-body %systems \cite{CaoBolsinger2017, CaoLushuai2013, KrönkeCao2013}. 
%This allows us to point out the decisive non-adiabatic contributions, which %$are a first precursor for the elusive pseudo Jahn-Teller effect.
%Finally, we prove the existence of the pseudo Jahn-Teller effect.

Our work is organized as follows. 
Section \ref{sec:setup} introduces the underlying impurity setup, where we introduce a one-dimensional Hamiltonian describing the coupling of our few-body bath and the impurity via s-wave interaction. In Sec.~\ref{sec:methods}, we present the {\it ab-initio} methods ML-MCTDHF and a multi-channel BO ansatz, which we apply to solve the Schr\"odinger equation. Especially, we point out the relation between the adiabatic BO approximation and our multi-channel BO ansatz. 
The basic ground-state analysis is performed in Sec.~\ref{sec:ground_state_properties}. Comparing the adiabatic BO approximation with the exact numerical result shows deviations for the impurity energy as well as for the one-body density. In section \ref{sec:correlation_properties} we focus on the correlation properties in terms of the von Neumann entropy and two-body density. The dependence of the observed effects on the impurity parameters is addressed in Sec.~\ref{different-parameters}. The existence and implications of the pseudo Jahn-Teller effect in our system are discussed in Sec.~\ref{sec:pjte}. We finish with our conclusions and outline further perspectives in Sec.~\ref{sec:outlook}. In Appendix~\ref{appendix:MCBO}, the detailed derivation of the non-adiabatic couplings for the multi-channel BO ansatz is provided. This is followed in Appendix~\ref{entanglementBO} by an analysis of entanglement within the adiabatic BO approximation.
Appendix~\ref{sec:convergence} contains a convergence analysis of our two {\it ab initio} methods: ML-MCTDHF and the multi-channel BO approach in combination with a configuration interaction method. Appendix~\ref{proof_theorem} contains a proof of an important theorem used in the main text. Appendix~\ref{appendix-jahn-teller} reviews the relevant for us properties of two interacting confined fermions. The final Appendix~\ref{conical_appendix} explicates our analysis of the emergence of a $E \otimes \epsilon$ conical intersection.

\section{Setup and Hamiltonian} \label{sec:setup}
\begin{figure}
    \centering
    \includegraphics[width=0.5\textwidth]{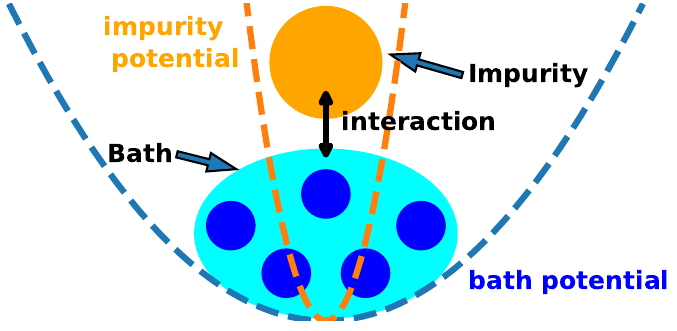}
    \caption{Schematic illustration of our many-body setup, consisting of two different fermionic species. The majority atoms, referred to as the bath, interact with a heavy and tightly-confined impurity. Both species are trapped in a species-dependent harmonic potential.}
    \label{fig:schematic representation}
\end{figure}
We consider a two-species setup of mass-imbalanced and spin-polarized fermions confined in a one-dimensional  (species-dependent) parabolic trap.
In particular, we focus on the particle imbalanced case where a lighter majority species, denoted as $B$ interacts with a single heavy impurity $I$, via $s$-wave repulsion.
%The Hamiltonian of our system  represented in units of the harmonic oscillator of the majority species ($\hbar=m=\omega=1$) reads
The Hamiltonian of our system reads $\hat{H} = \hat{H}_B + \hat{H}_I + \hat{H}_{BI}$ with the corresponding terms being
\begin{equation}
\begin{split}
 \hat{H}_{\sigma}=&\sum_{j=1}^{N_\sigma}\left[ -\frac{\hbar^2}{2m_\sigma}\left(\frac{\partial}{\partial x_j^\sigma}\right)^2+ V_{\sigma}(x_j^{\sigma})\right],\\
 \hat{H}_{BI}=&\sum_{k=1}^{N_B}\sum_{j=1}^{N_I}g\delta(x_k^B-x_j^I), 
 \label{eqn:Hamilt_Full}
 \end{split}
\end{equation}
where $\sigma \in \{ B, I \}$ and $V_{\sigma}(x)= \frac{1}{2}m_\sigma \omega_\sigma^2 x^2$ denotes the harmonic trapping potential for the associated species. 
%\textcolor{black}{[Wouldn't it be good to indicate already here what values we are using for $m_I,~,m_B,~\omega_I~\mathrm{and}~\omega_B$?}
%\textcolor{green}{We examine the system in units of the bath species and its corresponding trapping potential. The impurity values are chosen by default as $m_I=4.0m_B$ and $\omega_I=4.0\omega_B$.} 
\textcolor{black}{In addition, $N_B$ denotes the number of bath atoms (herewith we mainly focus on the $N_B = 5$ case) and $N_I = 1$ is the number of impurities.}
The effective interaction strength $g$ corresponds to the 1D s-wave contact-interaction strength between the two distinct species \cite{Bloch_2008}. Due to the Pauli principle, intra-species interaction is excluded since two fermions cannot occupy the same state. 
Let us note here that $g$ is dependent on the experimental transverse confinement length and the 3D s-wave scattering length \cite{Olshanii1998} and thus is tunable via  confinement and Fano-Feshbach resonances \cite{ChinChengGrimm2010}.
Here we consider an impurity species $I$ that is heavier and more tightly trapped compared to the bath species $B$, implying $m_I \gg m_B$ and $\omega_I \gg \omega_B$, motivating a BO-like approach \cite{BornOppenheimer1927}. In what follows, we mainly focus on the case $m_I = 4 m_B$ and $\omega_I = 4 \omega_B$ motivated by corresponding state-of-the-art experiments with $^6$Li-$^{23}$Na mixtures \cite{Schuster2012}, which are representative of the qualitative behaviour in this regime. In Sec.~\ref{different-parameters} we provide a more detailed analysis of the effects of varying $m_B$ and $\omega_B$ in the ground state of the system.

\section{Methodology and computational approach} \label{sec:methods}

In the following, we present the methods we employ for the ground state study of our system.
Therefore, we have to solve the stationary Schr\"odinger equation corresponding to the Hamiltonian of Eq.~\eqref{eqn:Hamilt_Full}. First, we will address the fully correlated numerical ML-MCTDHX approach. Second, we describe our multi-channel BO approach, which is motivated by the significant mass imbalance $m_I/m_B$ in our system. Both methods are numerically exact {\it ab-initio} approaches suitable for the solution of multi-component fermionic systems \cite{CaoBolsinger2017}. 
%Once, we have a BO-like approach matching the physical structure, on the other hand ML-MCTDHX is a well-proven method for all kind of physical few-particle quantum systems. In  the next paragraph, both methods and their underlying structure will be presented. 
\subsection{The ML-MCTDHX method}
\label{MLX_main_text}

ML-MCTDHX is a variational, {\it ab initio} and numerically exact approach for the simulation of the non-equilibrium quantum dynamics of bosonic and fermionic particles and mixtures thereof, containing a single or both types of particles \cite{KrönkeCao2013, CaoLushuai2013, CaoBolsinger2017}. 
ML-MCTDHX relies on a multi-layered ansatz that variationally optimizes the involved quantum basis at different levels of the complex structure of the total many-body wavefunction.

In particular, the total many-body wavefunction, $|\Psi(t)\rangle$ is represented as a linear combination of $j=1,2,...,D$ distinct orthonormal functions for each involved species, $|\Psi_j^\sigma(t)\rangle$, with $\sigma=B,I$
\begin{equation}
    |\Psi(t)\rangle = \sum_{j_B, j_I = 1}^D A_{j_B, j_I}(t) |\Psi_{j_B}^B(t)\rangle |\Psi_{j_I}^I(t)\rangle,
    \label{eqn:Species_decomposition_MLX}
\end{equation}
where $A_{j_B, j_I}(t)$ are the corresponding time-dependent expansion coefficients.
This expansion is formally identical to a truncated Schmidt decomposition of rank $D$, given by
\begin{equation}
    |\Psi(t)\rangle = \sum_{k=1}^D \sqrt{\lambda_k(t))} |\tilde{\Psi}_k^B(t)\rangle |\tilde{\Psi}_k^I(t)\rangle.
    \label{eqn:Schmidt_Composition_MLX}
\end{equation}
The time-dependent expansion coefficients $\lambda_k(t)$ are denoted as Schmidt weights and $|\tilde{\Psi}_k^{\sigma}(t)\rangle$ are the corresponding Schmidt modes. 
Especially, $\lambda_k$ and $|\tilde{\Psi}_k^{\sigma}(t)\rangle$ represent the eigenvalues and eigenstates of the $\sigma$-species reduced density matrix, namely
\begin{equation}
\begin{split}
\rho_{\sigma}^{(N_\sigma)}&(x_1, \dots, x_{N_\sigma}, x'_1, \dots, x'_{N_\sigma}, t)=
\int \prod_{j = 1}^{N_{\bar \sigma}} \mathrm{d} x_j^{\bar \sigma}~ \\
&\times \Psi^*(x_1^{\sigma}=x'_1, \dots, x_{N_\sigma}^{\sigma}=x'_{N_\sigma}, x^{\bar \sigma}_1, \dots, x^{\bar \sigma}_{N_{\bar \sigma}}, t) \\
&\times \Psi(x_1^{\sigma}=x_1, \dots, x_{N_\sigma}^{\sigma}=x_{N_\sigma}, x^{\bar \sigma}_1, \dots, x^{\bar \sigma}_{N_{\bar \sigma}}, t),
\end{split}
\end{equation}
where $\bar{\sigma} \neq \sigma$.
Notice that within ML-MCTDHX this density matrix can be expressed as $\rho_{\sigma}^{(N_\sigma)}(x_1, \dots, x_{N_\sigma}, x'_1, \dots, x'_{N_\sigma}, t)=\langle x_1, \dots, x_{N_\sigma}  | \hat{\rho}^{(N_{\sigma})}_{\sigma}(t)| x'_1, \dots, x'_{N_\sigma} \rangle$, where the density matrix operator reads
\begin{equation}
\hat{\rho}_{\sigma}^{(N_\sigma)}(t) = \sum_{ j_{\sigma},j'_{\sigma} = 1 \atop j_{\bar{\sigma}} = 1}^{D} 
\underbrace{A_{j_{\sigma}, j_{\bar{\sigma}}}^*(t) A_{j_{\bar{\sigma}}}, j'_{\sigma}(t)}_{\equiv \left[\hat{\rho}_{\sigma}^{(N_\sigma)}(t)\right]_{j_{\sigma}, j'_{\sigma}}} 
| \Psi_{j_{\sigma}}^{\sigma}(t) \rangle \langle \Psi_{j'_{\sigma}}^{\sigma}(t) |.
\end{equation}
Therefore, $\lambda_k$ and $| \tilde{\Psi}_k^{\sigma} (t) \rangle$ can be evaluated by diagonalizing the matrix $\left[\hat{\rho}_{\sigma}^{(N_\sigma)}(t)\right]_{j_{\sigma}, j'_{\sigma}}$ for $j_{\sigma},j'_{\sigma}=1, \dots, D$.
The truncated Schmidt decomposition of Eq.~\eqref{eqn:Schmidt_Composition_MLX} exhibits a finite bipartite entanglement of the system among the bath and impurity species, if at least two $\lambda_k(t)$'s are non-vanishing. 
In the case, $\lambda_1(t) = 1$ and $\lambda_k(t) = 0$ for $k = 2, \dots, D$ the total wavefunction $|\Psi(t)\rangle$ is a tensor product of the species states and the system is non-entangled.
%\textcolor{red}{Within Eq.~\eqref{eqn:Species_decomposition_MLX} the $\sigma$-species reduced density matrices read
%\begin{equation}
%\hat{\rho}_{\sigma}^{N_\sigma}(t) = \sum_{ j_{\sigma},j'_{\sigma} = 1 \atop j_{\bar{\sigma}} = 1}^{D} 
%A_{j_{\sigma}, j_{\bar{\sigma}}}^*(t) %A_{j_{\bar{\sigma}}, j'_{\sigma}}(t) 
%| \Psi_{j_{\sigma}}^{\sigma}(t) \rangle \langle \Psi_{j'_{\sigma}}^{\sigma}(t) |,
%\end{equation}
%\textcolor{black}{[It is correct as it was]}
%with $\bar{\sigma} \neq \sigma$, and thus $| \tilde{\Psi}^{\sigma}_k (t) \rangle$ and $| \Psi^{\sigma}_j (t) \rangle$ are related via a unitary transformation. [If you agree with my corrections delete this part.]}
The expansion of Eq.~\eqref{eqn:Schmidt_Composition_MLX} can be thought of as an expansion in terms of entanglement modes with $D$ controlling the maximum number of allowed entanglement modes of the system.

The multi-layered structure of our ansatz stems from the fact that each species function, $| \Psi_j^{\sigma}(t) \rangle$ is expanded  in terms of a time-dependent number-state basis set $|\vec{n}(t)\rangle^\sigma$ leading to
\begin{equation}
    |\Psi_j^\sigma(t)\rangle = \sum_{\vec{n}} B_{j, \vec{n}}^\sigma(t) |\vec{n}(t)\rangle^{\sigma}.
    \label{eqn:Species_Wave_Function}
\end{equation}
On this level, $|\vec{n}(t)\rangle^{\sigma}$ could be determinants or permanents for a fermionic or bosonic species $\sigma$ respectively. Further, $B_{j, \Vec{n}}^\sigma(t)$ corresponds to the time-dependent expansion coefficients with a particular number state $|\vec{n}(t)\rangle^{\sigma}$, which is built from $d^\sigma$ time-dependent variationally optimized Single-Particle Functions (SPFs) given by 
$\phi_l^\sigma(t)$, $l=1,2,..., d^\sigma$ with $\Vec{n}=(n_1,...,n_{d^{\sigma}})$ corresponding to the contribution numbers. 
On the lowest layer of the ML-MCTDHX variational ansatz, the SPFs are represented on a time-independent primitive basis. 
For the underlying case of spinless fermions, this refers to a $\mathcal{M}$ dimensional Discrete Variable Representation (DVR) represented by $\{|k\rangle\}$. 
Hence, the SPF of the $\sigma$-species are given by 
\begin{equation}
    |\phi_j^{\sigma}(t)\rangle = \sum_{k=1}^{\mathcal{M}}C_{jk}^{\sigma}(t)|k\rangle.
    \label{eqn:DVR_SPF}
\end{equation}
In our investigation we choose $\mathcal{M}=150$ grid points of a harmonic oscillator DVR. 

To determine the variationally-optimal ground state of the Hamiltonian \eqref{eqn:Hamilt_Full} corresponding to the ($N_B+N_I$)-body wavefunction $|\Psi(t)\rangle$ of Eq.~\eqref{eqn:Species_decomposition_MLX}--\eqref{eqn:DVR_SPF}, the corresponding ML-MCTDHX equations of motion are derived by employing the Dirac-Frenkel \cite{Dirac1930annihilation, frenkel1934wave} variational principle 
\begin{equation}
    \langle \delta \Psi (t)| i \hbar \frac{\partial}{\partial t}-H|\Psi(t)\rangle = 0, 
\end{equation}
%and taking into account the wavefunction expansion of Eq.~\eqref{eqn:Species_decomposition_MLX}--\eqref{eqn:DVR_SPF}, 
for details see~\cite{CaoBolsinger2017}. 
Hence, we have to solve numerically $D^2$ linear differential equations of motion for $A_{j_B, j_I}(t)$, which are coupled to $D\left(\frac{(N_B+d^B-1)!}{N_B!(d_B-1)!}+\frac{(N_I+d^I-1)!}{N_I!(d_I-1)!}\right)$ and $d^B+d^I$ non-linear integro-differential equations for the expansion coefficients of the species functions $B_{j, \vec{n}}^{\sigma}(t)$ and the SPFs $C_{j, k}^{\sigma}(t)$ respectively.
Let us note here that calculating the ground state of Eq.~\eqref{eqn:Hamilt_Full} can be achieved by performing propagation in imaginary time.
Within this approach we perform a Wick rotation of the real time $t$ which leads to the imaginary time $\tau = -i t$. 
This substitution results in the energy of the propagated state of the corresponding equations of motion to decrease monotonically in time proportionately to $\propto e^{-(E(t)-E_0)t}$, where $E_0$ is the ground state energy. Therefore the ground state is obtained in the limit of large propagation times $\tau \to \infty$ if the initial state possesses a finite overlap with the ground state. 
%\sout{The required propagation time is finite in our case due to the finite numerical accuracy of the double precision arithmetic we employ.}

The key feature of ML-MCTDHX is the expansion of the system's many-body wavefunction with respect to a time-dependent and variationally optimized basis, that can adapt to the relevant inter-particle correlations at the level of single-particle, single-species and total multi-species systems. 
The involved Hilbert-space truncation is characterised by the chosen orbital configuration space, which is characterized by $C=(D;d^B;d^I)$. Furthermore, since we consider a single impurity on the $I$ species, there are no intra-species interactions thus we can set $d^I = D$ without loss of generality. In turn, the number of Schmidt modes is taken large enough $D = 12$ to account for inter-species entanglement. 
To account for the inter-species interactions of the bath we consider a large enough number of bath species orbitals $d^B = 18$ to properly account for the different states the bath atoms can occupy as a consequence of the bath-impurity entanglement.
For more information regarding the convergence of the ML-MCTDHX method, see Appendix~\ref{sec:convergence}.

\subsection{Multi-channel Born-Oppenheimer approach}

\begin{figure*}
    \centering
    \includegraphics[width=1.0\textwidth]{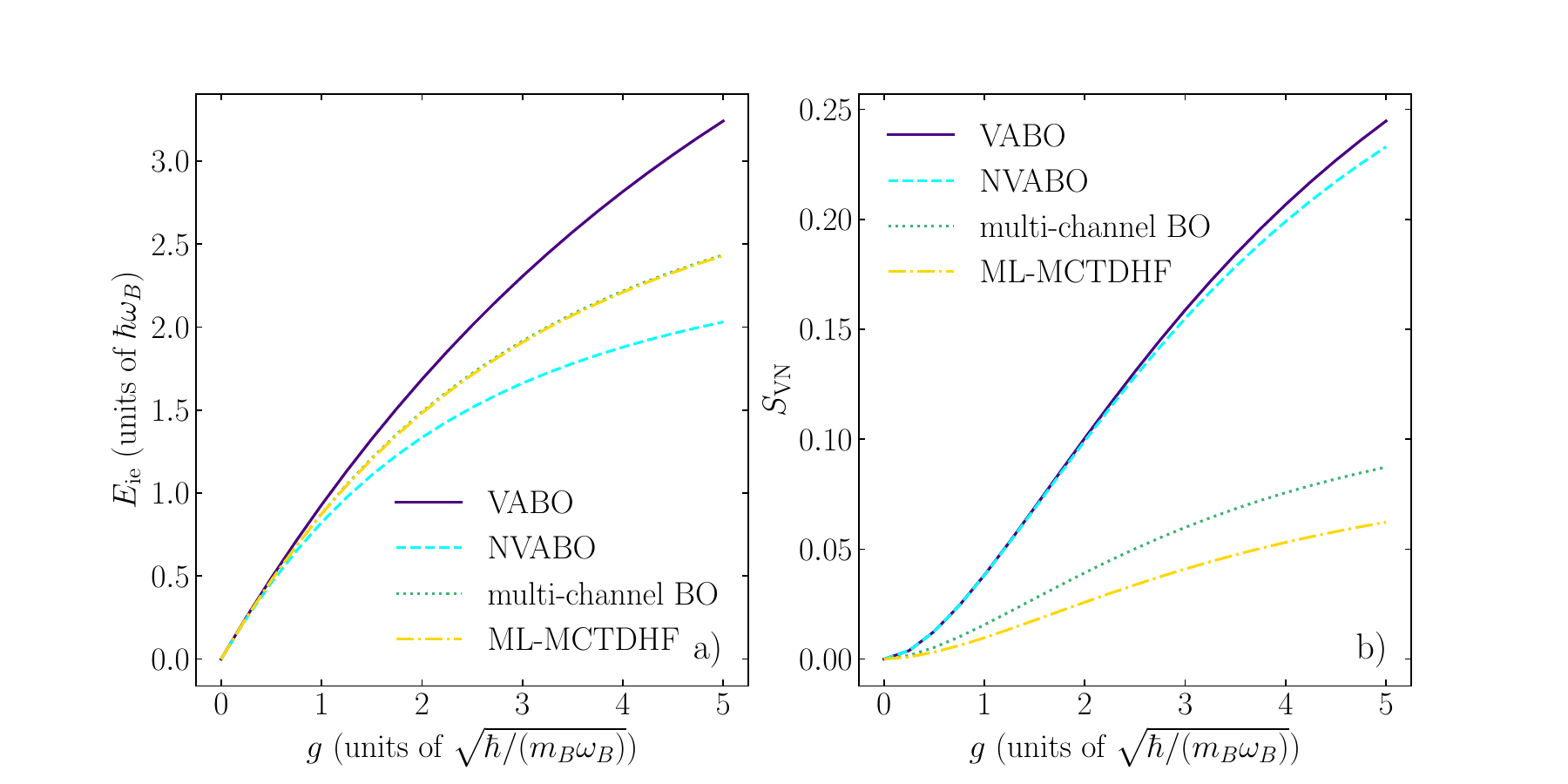}
    \caption{(a) Ground-state/Impurity-interaction energy (see Eq.~\eqref{eqn:Impurity_Energy}) and (b) von Neumann entropy, $S_{VN}$, for varying interaction strength, $g$ and within different levels of approximation (see legend). In all cases, $N_B = 5$, $m_I=4m_B$ and $\omega_I=4\omega_B$.}
    \label{fig:ground-state_energy_BO_vs_BH}
\end{figure*}

Motivated by the assumption of a heavy, less-mobile impurity the comparison of our results with a BO-like approach is justified. A general variational formulation of this approach can be established in terms of the multi-channel BO ansatz,
\begin{equation}
\begin{split}
%|\Psi \rangle = \sum_{i = 1}^{M} \int \mathrm{d}x_{I}~\Psi_{i,I} (x_I) |x_I\rangle \otimes | \Psi_{i,B} (x_I)\rangle.
\Psi(x^B_1,\dots,x^B_{N_B}, x_I) &= \\ \sum_{j = 1}^M \Psi_{j,I}(x_I) 
&\underbrace{\Psi_{j,B}(x^B_1,\dots,x^B_{N_B};x_I)}_{\equiv\langle x^B_1,\dots,x^B_{N_B} | \Psi_{j,B}(x_I) \rangle}.
\end{split}
\label{eqn:multi-channel_BornOppenheimer}
\end{equation}
Here, we have introduced an orthonormal basis for the bath species, \(| \Psi_{j,B} (x_I) \rangle\), with \(j=1,2,\dots\), exhibiting a parametric dependence on \(x_I\). 
The impurity-species wavefunctions \(\Psi_{j,I}(x_I)\) with the normalization condition \(\sum_{j = 1}^M \int \mathrm{d}x_I |\Psi_{j,I}(x_I)|^{2} = 1\) correspond to the expansion coefficients in the many-body basis of the coupled system. 
Using the multi-channel ansatz of Eq.~\eqref{eqn:multi-channel_BornOppenheimer} and employing the variational principle \( \frac{\delta\langle \Psi | \hat{H} - E | \Psi \rangle}{\delta \Psi^{*}_{k,I}(x_I)} = 0 \),  we derive the coupled set of equations
\begin{equation}
\begin{split}
E \Psi_{k,I}(x_I) &= - \frac{\hbar^2}{2 m_I} \sum_{j,l = 1}^M \left( \delta_{kj} \frac{\mathrm{d}}{\mathrm{d}x_{I}} -i A_{kj}(x_I) \right) \\
& \hspace{0.2cm} \times \left( \delta_{jl} \frac{\mathrm{d}}{\mathrm{d}x_{I}} -i A_{jl}(x_I) \right) \Psi_{l,I}(x_I) \\
& + \sum_{l = 1}^M \bigg( \langle \Psi_{k,B}(x_I) | \hat{H}_B + \hat{H}_{BI} | \Psi_{l,B} (x_I) \rangle + \\
&  \hspace{0.2cm} \delta_{kl} \frac{1}{2} m_B \omega^2_I x_I^2 + V^{\rm ren}_{kl}(x_I) \bigg) \Psi_{l,I}(x_I),
\end{split}
\label{eqn:effective_Schroendinger}
\end{equation}
for $k = 1, 2, \dots, M$ \cite{Köppel1984, Pacher1989}. 
In this step, we introduce the non-adiabatic derivative couplings \(A_{kj}(x_I) = i \langle \Psi_{k,B}(x_I) | \frac{\partial \Psi_{j,B}}{\partial x_I}(x_I) \rangle\) as an effective gauge field.
In addition, the last term in Eq.~\eqref{eqn:effective_Schroendinger} refers to the potential renormalization, which is given by
\begin{equation}
\begin{split}
    V_{kl}^{\rm ren}(x_I) =\frac{\hbar^2}{2 m_I} &\left\langle   \tfrac{\mathrm{d} \Psi_{k,B}}{\mathrm{d}x_I}(x_I) \right| 1 - \hat{\mathcal{P}}_M \left| \tfrac{\mathrm{d} \Psi_{l,B}}{\mathrm{d}x_I}(x_I) \right\rangle,
\end{split}
\label{potential_renormalization}
\end{equation}
where the projector onto the subspace spanned by \(| \Psi_{k,B} (x_I) \rangle\) is defined as
\begin{equation}
    \hat{\mathcal{P}}_{M} = \sum_{j = 1}^M | \Psi_{j,B} (x_I) \rangle \langle \Psi_{j,B}(x_I) |.
\end{equation}
In general, it is possible to define \(| \Psi_{k,B} (x_I) \rangle\) in terms of any complete wave-function basis. 
However, the convenient choice is to employ the eigenstates of \(\hat{H}_B + \hat{H}_{BI}\) for fixed $x_I$, which leads to the diagonal matrix elements 
\begin{equation}
\langle \Psi_{k,B}(x_I) | \hat{H}_B + \hat{H}_{BI} | \Psi_{l,B} (x_I) \rangle = \delta_{kl} \varepsilon_k(x_I),
\end{equation}
where the eigenvalues $\varepsilon_k(x_I)$ are the corresponding potential energy curves.

%which we have to solve numerically for a given order of truncation $M$.
In the limit of infinite channels, $M \to \infty$, the expansion of Eq.~\eqref{eqn:multi-channel_BornOppenheimer} and the equations-of-motion of Eq.~\eqref{eqn:effective_Schroendinger} are exact for any mass ratio \(m_B/m_I\).
In particular, a mass ratio \(m_B/m_I \ll 1\) is favorable, since it suppresses the off-diagonal non-adiabatic coupling terms stemming from the gauge potential $A_{kj}(x_I)$ and potential renormalization terms $V^{\rm ren}_{kl}(x_I)$.
In this case, only a few terms  contribute significantly to the exact many-body wavefunction, and consequently a relatively small value $M$ suffices for adequate convergence to the exact solution.

\subsection{Variational and non-variational adiabatic BO}
\label{sec:adiabaticBOdefinitions}

Before proceeding let us comment on the reduction of the above to the adiabatic BO approximation widely employed in molecular physics~\cite{BornOppenheimer1927}.
\textcolor{black}{If we restrict the ansatz of Eq.~\eqref{eqn:multi-channel_BornOppenheimer} to $M = 1$ term, the variational equations of motion reduce to the adiabatic BO approximation incorporating the Born-Huang correction arising from $V^{\rm ren}_{11}(x_I)$, which is therefore characterised as variational adiabatic Born-Oppenheimer (VABO) approximation. The usual adiabatic BO approach consists of dropping this additional term by considering $V_{11}^{\rm ren}(x_I) = 0$ and will be denoted as non-variational adiabatic Born-Oppenheimer (NVABO) approximation.} It can be shown that the VABO approximation accounting for $V^{\rm ren}_{11}(x_I) \neq 0$ possesses a variational character and thus yields an {\it upper} bound to the ground state energy. In contrast, the usual NVABO approximation, with $V_{11}^{\rm ren}(x_I) = 0$, yields a {\it lower} bound for the ground-state energy~\cite{Epstein1966, Baer2006} \textcolor{black}{\footnote{Since a term is dropped this leads to a lower bound, which is non-variational. A proof for this can be found in \cite{Epstein1966}:  Herein, it is shown that the standard NVABO yields a lower bound.}}. Thus, the value of the term $V_{11}^{\rm ren}(x_I)$ relative to the other parameters of the system is a good indication for the correlations among distinct potential energy curves as it provides an order of magnitude estimate for the correlation energy related to the contribution of multiple potential energy curves, $E_{\rm corr} \equiv \lim_{M \to \infty} E_M - E_{M=1}$. An important caveat here is that the lower ground-state energy of NVABO does not imply a better quality of the corresponding many-body state, but an approximation in terms of the Hamiltonian. Indeed, the variational principle guarantees that the energy of the NVABO ground state will be larger than the VABO when the exact form of the Hamiltonian is considered.
%Considering above assumptions, we use the derived representations of the non-adiabatic derivative coupling matrices \eqref{eqn:non_adiabatic_derivative_Coupling} and for the potential renormalization \eqref{eqn:potential_renormalization_terms} to solve the effective Schr\"odinger equation \eqref{eqn:effective_Schroendinger} numerically.

In the following, we will rely on the above analyzed methods to investigate the relevant ground-state properties of our system. 

\section{Basic Ground-State Properties}
\label{sec:ground_state_properties}

%After we have established the underlying Hamiltonian (see Sec. \ref{sec:setup}) and got an overview over the two different methods in Sec. \ref{sec:methods}, we focus on the analysis of the ground-state properties of Eq.~\eqref{eqn:Hamilt_Full}.
Since, we consider the case of an impurity somewhat larger than the mass of the bath particles, $m_I > m_B$, at first glance it might seem sufficient to take into account the adiabatic BO approximation.
%, which corresponds to the zeroth order of our multi-channel BO. 
Hence, in the following we will compare the ground state properties of Eq.~\eqref{eqn:Hamilt_Full} within the fully correlated ML-MCTDHX approach and the adiabatic BO approaches (VABO and NVABO), which neglects the non-adiabatic contributions of the underlying Hamiltonian.
This will give us an overview of the importance of the non-adiabaticity in our system in its most elementary ground state properties.

\begin{figure*}
    \centering
    \includegraphics[width=1.0\textwidth]{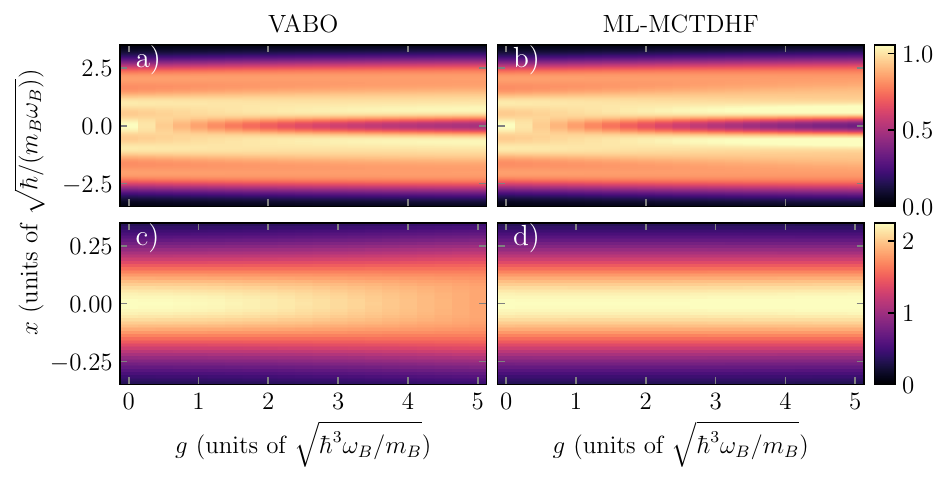}
    \caption{Interaction dependence of (a), (b) the bath, $\rho_B^{(1)}(x;t)$, and (c), (d) the impurity species one-body density matrix, $\rho_I^{(1)}(x;t)$. Panels (a) and (c) refer to the VABO approximation, whereas (b) and (d) show the ML-MCTDHX results. In all cases, $N_B = 5$, $m_I=4m_B$ and $\omega_I=4\omega_B$.}
    \label{fig:one_body_density}
\end{figure*}

%Both ab-initio methods allow us to calculate numerically exact results. 
%However, we will work out the limitations of both methods and discuss the appearing deviations.  
%To justify the corresponding choice of parameters for the numerical methods, we will include an appropriate convergence analysis for both approached in Appendix~\ref{sec:convergence} to showcase their degree of convergence.  
%\textcolor{black}{Note that numerically exact results, can be calculated with either ML-MCTDHX or the multi-channel BO approach provided that a large enough basis set is used for either method.} 
%\textcolor{black}{In the appendix part x, we show the convergence of both methods to demonstrate the equivalence of these two approaches.}

\subsection{Impurity-interaction energy}
We start our ground-state analysis by considering the impurity interaction energy, which is given by
\begin{equation}
    E_{\rm{ie}}=E_{\rm{tot}}(g)-E_{\rm{tot}}(g=0).
    \label{eqn:Impurity_Energy}
\end{equation}
Figure~\ref{fig:ground-state_energy_BO_vs_BH} (a) reveals that all approaches show that stronger repulsions lead to higher impurity-interaction energies, as the interspecies interaction energy increases. In addition, all approaches demonstrate that $E_{\rm{ie}}$ shows a linear behaviour for weak interaction and a non-linear one, characterized by a decreasing slope with increasing $g$. This is due to the change of the many-body state of the system in order to reduce the associated interaction energy penalty stemming from the spatial overlap among the impurity $I$ and bath $B$ species. The VABO approximation follows this linear trend for a larger regime of $g$ values leading to an overall stronger increase of $E_{\rm ie}$, when compared to the other approaches.
The comparison of the variational with the NVABO approximation reveals that the reason for this divergence is the inclusion of the $V_{11}^{\rm ren}(x)$ term which becomes important for $g>1$. This leads to a larger discrepancy between the lower energy bound given by the NVABO approximation and the upper (variational) bound, which is represented by NVABO incorporating the Born-Huang correction.
The multi-channel BO and ML-MCTDHX approaches are able to correct the energy accounting for non-adiabatic effects, which is a first indication for their importance in our system.
%This behaviour occurs also for the multi-channel BO approximation. Slightly discrepancies between the ML-MCTDHX and multi-channel BO approximations are caused by the stronger interactions: In this case, we hit the regime where the approximations made in the ML-MCTDHX expansion reach their limits. 
%This is evident since the corresponding energy in the ML-MCTDHX case is higher compared to the multi-channel BO result. 
%Hence, we conclude that within the adiabatic approach the 
%After, we have seen the impact of the non-adiabatic terms on the ground-state energy, we continue with the one-body density to get an overview on the spatial behaviour of each species. 
%Analysing the one-body density of both species in adiabatic BO as well as multi-channel BO, we highlight the spatial behavior and the influence of the non-adiabatic terms as a function of interaction strength. 
\begin{figure*}
    \centering
    \includegraphics{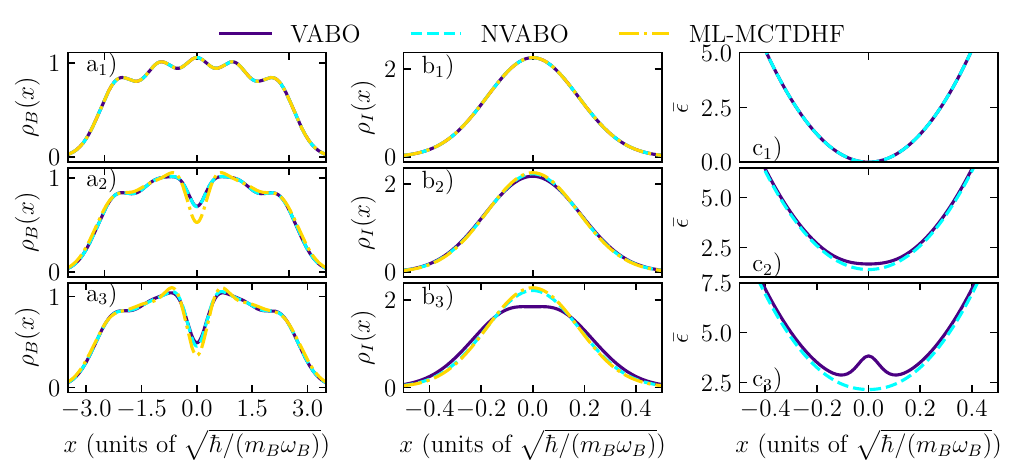}
    \caption{One-body density profiles for (a$_{i}$) the bath and (b$_{i}$) the impurity species  within different levels of approximation (see legend) and for selected values of the interaction strength, $g = g_i$, with $g_1= 0$, $g_2=1$ and $g_3 = 2$. The panels (c$_{i}$) indicate the effective potentials for the corresponding $g = g_i$, see Eq.~\eqref{ eqn:effective_potential }, within the variational and NVABO approximations. In all cases, $N_B = 5$, $m_I=4m_B$ and $\omega_I=4\omega_B$.}
    \label{fig:cuts_one_body_density}
\end{figure*}
%To get a general overview of the spatial behaviour of the bath and impurity, we compare the corresponding one-body densities in Fig. \ref{fig:one_body_density} for the adiabatic BO approximation and the multi-channel BO approach. Here, we neglect at first glance the results from our ML-MCTDHX calculation, since the deviation to the multi-channel approach is too small to point out in these two-dimensional plots, as we will see later on. 
To get further insight regarding the mechanisms for the discrepancy between the adiabatic BO result and the two numerically exact ab-initio methods, we study the one-body density profiles provided by the above-mentioned approaches.

\subsection{One-body density}
\label{sec:one-body-density}

To resolve the spatial behaviour for both the impurity and the bath we resort to the one-body density as a function of the interaction strength, see Fig.~\ref{fig:one_body_density}.
Here for the sake of brevity, we only compare the VABO approximation and ML-MCTDHX. In this section, we do not show the results of the multi-channel BO approach, since they agree very well with the results of ML-MCTDHX.
The results for the NVABO approximation will be discussed later on, since, as discussed in Sec.~\ref{sec:adiabaticBOdefinitions}, the corresponding ground state stems from the approximation of the Hamiltonian and thus there are several nuances associated with it.
For both the VABO and ML-MCTDHX, we observe an outward displacement of the majority species for increasing $g$, stemming from the gradual depletion of the density in the region around $x = 0$, see Fig.~\ref{fig:one_body_density}(a) and Fig.~\ref{fig:one_body_density}(b).
This behavior is a result of the repulsive interaction between the two species, as the heavy and tightly trapped impurity remains in the trap center, pushing the bath particles outward.
However, it is obvious that the spatial profiles of the majority species possess significant quantitative deviations among the two approaches. In particular, the VABO approach under-appreciates the density-depletion for $x = 0$ and, also, under-appreciates the development of the two density maxima at $x = \pm 0.5$ evident within ML-MCTDHX, see Fig.~\ref{fig:one_body_density}(b).
The comparison of the impurity densities also shows clear qualitative deviations. In particular, within ML-MCTDHX the profile remains essentially unchanged possessing a Gaussian shape for all $g$ values, see Fig.~\ref{fig:one_body_density}(d), while within the adiabatic BO approximation, see Fig.~\ref{fig:one_body_density}(c), the profile increasingly flattens and spreads out spatially as $g$ increases.

The above allow us to attribute the substantially larger energy of the VABO approximation when compared to ML-MCTDHX, see Fig.~\ref{fig:ground-state_energy_BO_vs_BH}(a), to the larger spatial overlap of the impurity and bath species in the former approach that increases the interaction energy. Additionally, the VABO approximation results in a larger spreading of the density of \textcolor{black}{the impurity} species, see \textcolor{black}{Fig.~\ref{fig:one_body_density}(c) and \ref{fig:cuts_one_body_density}(b${}_i$)} with $i=1,2,3$ resulting in additional contributions of potential energy when compared to ML-MCTDHX.
The above implies that the adiabatic treatment is not able to adequately describe the state of the system and thus additional contributions stemming from the non-adiabatic couplings need to be introduced in order to properly account for it.

To elucidate further the shortcomings of the adiabatic approach let us examine the behavior of the effective potential within the VABO and NVABO approximations.
This effective potential reads
\begin{equation}
    \bar{\epsilon}(x_I) = V_I(x_I) +\varepsilon_1(x_I) + V_{1,1}^{\mathrm{ren}}(x_I)-\frac{\hbar \omega_B N_B^2}{2}.
    \label{ eqn:effective_potential }
\end{equation}
The first term denotes the harmonic trapping potential for the impurity, see Eq.~\eqref{eqn:Hamilt_Full}, while the second and third terms are the potential energy curve and the Born-Huang potential renormalization, see Eq.~\eqref{potential_renormalization}.
The last term removes from $\bar{\epsilon}(x_I)$ the spatially constant energy offset stemming from the non-interacting energy of the bath species. Recall that NVABO does not contain the Born-Huang correction, i.e. $V_{1,1}^{\rm ren}(x_I) = 0$ in $\bar{\epsilon}(x_I)$.

Figure \ref{fig:cuts_one_body_density} provides the one-body densities within the VABO, NVABO and ML-MCTDHX approaches in combination with $\bar{\epsilon}(x_I)$ for three different interaction strengths.
By comparing the effective potential within the VABO approximation, see Fig.~\ref{fig:cuts_one_body_density}(c${}_i$), with $i = 1, 2, 3$, we observe that it deforms from a harmonic potential for $g = 0$, see Fig.~\ref{fig:cuts_one_body_density}(c${}_1$), to a double well structure for $g > 3$, see Fig.~\ref{fig:cuts_one_body_density}(c${}_3$).
In contrast, NVABO does not show this effect with the effective potential being parabolic for all considered interaction strengths.
This demonstrates that the Born-Huang term is responsible for the emergence of the double-well structure in Fig.~\ref{fig:cuts_one_body_density}(c${}_3$).
% This can be explained due to the fact that the ground state of the majority species for $g = 0$ exhibits an oscillatory pattern, see Fig.~\ref{fig:cuts_one_body_density}(a${}_1$), namely the Friedel oscillations stemming from fermionic character of the bath component and its harmonic confinement. This implies that when the impurity lies on the maxima of these oscillations the energy of $\epsilon_1(x_I)$ increases since the larger bath density implies larger interaction energy contribution.
% However, since the harmonic trap is tight it is the dominant energy contribution for $x \neq 0$ and thus only the peak of $\epsilon_1(x_I)$ at $x = 0$ contributes significantly to $\bar{\epsilon}(x_I)$ resulting to the observed double-well structure. Finally let us note that the effect of the Born-Huang term is typically small, but is included in Fig.~\ref{fig:cuts_one_body_density}(a${}_3$), (b${}_3$) and (c${}_3$), for completeness.}

The above analyzed behavior of the effective potential explains the density discrepancies among the ML-MCTDHX and the VABO approach. The transition from the harmonic to a double-well one gives an explanation for the displacement of the impurity from the trap center \cite{MistakidisKatsimiga2019}, see Fig.~\ref{fig:cuts_one_body_density}(b${}_i$), with $i = 2, 3$, in contrast to the ML-MCTDHX approach.
In the case, that the $V_{11}^{\rm ren}(x_I)$ is completely dropped the impurity density agrees much better to the ML-MCTDHX result.
However, notice that this term is accounted for in the exact case, which is a strong indication of the presence of sizable non-adiabatic effects that lead to the cancellation of this term within ML-MCTDHX.
Indeed, notice that dropping the Born-Huang term does not fix the sizable quantitative deviation of $\rho^{(1)}_B(x_B)$ within the variational BO approach when compared to the numerically-exact result, see Fig.~\ref{fig:cuts_one_body_density}(a${}_i$) for $i = 2, 3$.
This demonstrates the importance of non-adiabatic correlations in capturing the correct state of the system.

% In order to understand this mechanism let us note that the state of the majority species one-body density can be thought as a reduction of the interspecies two-body density, which within the adiabatic BO approximation reads 
% \begin{equation}
% \begin{split}
%  \rho_{B}^{(1)}(x_A) &= \int \mathrm{d}x_I~\rho_{BI}^{(2)}(x_B, x_I) \\
%  &= \int \mathrm{d}x_I~| \Psi_{0,I}(x_I) |^2 \\ 
%  &\underbrace{\times \langle \Psi_{0,B}(x_I) | \hat{\Psi}^{\dagger}_I(x_I) \hat{\Psi}_I(x_I) | \Psi_{0,I}(x_I) \rangle}_{\equiv C_{BI}(x_B, x_I)},
% \end{split}
% \label{one_body_density_aBO}
% \end{equation}
% where the bath-impurity correlator $C_{BI}(x_B, x_I)$, corresponds to the one-body density of the $\hat{H}_B + \hat{H}_I$ eigenstates for fixed $x_B$. Notably this quantity depends only on the ratio of the interaction strength with the majority species parameters $g/\sqrt{\hbar^3 \omega_I/m_I}$.
% Therefore, the delocalization of the impurity, see Fig.~\ref{fig:cuts_one_body_density}(b${}_2$) and \ref{fig:cuts_one_body_density}(c${}_2$), implies an averaging of this correlator over a larger $x_B$ extent resulting to the smearing of the Friedel oscillations of the bath density, see Fig.~\ref{fig:cuts_one_body_density}(b${}_1$) and \ref{fig:cuts_one_body_density}(c${}_1$).

\section{Correlation properties}
\label{sec:correlation_properties}

Having analyzed the basic ground state properties of Eq.~\eqref{eqn:Hamilt_Full}, let us elaborate on the emergent correlation patterns in terms of the two-body density and the von Neumann entropy.
\subsection{Two-body density}

Figure \ref{fig:two_body_density} addresses the bath-impurity two-body densities for the same interaction strengths as in Fig.~\ref{fig:cuts_one_body_density}.
For both the VABO and ML-MCTDHX approaches, we detect the emergence of a correlation hole in $\rho^{(2)}_{BI}(x_B, x_I)$ for $x_B \approx x_I$ as the interaction increases, see Fig.~\ref{fig:two_body_density}(a$_{i}$) and (b$_{i}$) with $i=2,3$.
Besides the emergence of this structure, the two-body densities among the two distinct approaches are significantly different.
Notice that $\rho^{(2)}_{BI}(x_B, x_I)$ within the adiabatic BO approach appears to be more pronounced oscillatory for varying $x_B$ but fixed $x_I$ when compared to the ML-MCTDHX approach, see for instance Fig.~\ref{fig:two_body_density}(a$_2$) for $x_I \approx 0.1$.
This is explained as follows. The two-body density within the VABO and NVABO approximation reads
\begin{equation}
\begin{split}
\rho^{(2)}_{BI}&(x_B, x_I) = | \Psi_{1,I}(x_I) |^2 \\
&\times \underbrace{ \int \prod_{j = 2}^{N_B} \mathrm{d}x_j~ | \Psi_{1,B}(x_B, x_2, \dots,x_{N_B};x_I)|^2}_{\equiv C_{BI}(x_B, x_I)},
\end{split}
\label{two_body_density_adiabatic}
\end{equation}
with the two-body correlator $C_{BI}(x_B, x_I)$ being the one-body density of the bath for a fixed delta-potential barrier at $x_I$. 
This implies that since the one-body density of the bath would exhibit Friedel oscillations as it corresponds to a state of spin-polarized fermions, so does $C_{BI}(x_B, x_I)$ and thus $\rho^{(2)}_{BI}(x_B, x_I)$ for fixed $x_I$.
Thus the absence of these oscillations within ML-MCTDHX, see Fig.~\ref{fig:two_body_density}(b$_2$) and \ref{fig:two_body_density}(b$_3$), provides another direct indication that more than a single bath eigenstate for fixed impurity position is involved in the exact many-body ground state.

\subsection{Von Neumann entropy}
\label{vonNeumann}

\begin{figure*}
    \centering
    \includegraphics{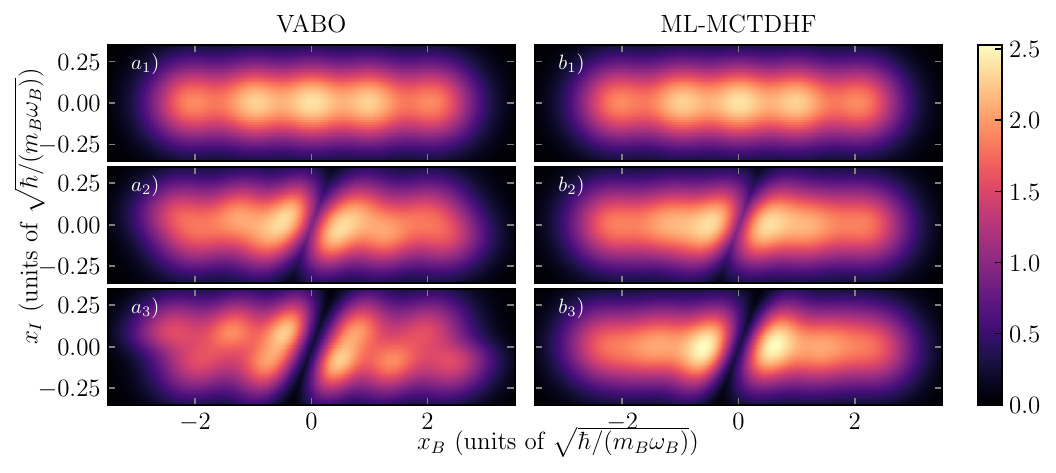}
    \caption{Bath-impurity two-body densities $\rho_{BI}^2(x_B,x_I)$ within the (a$_i$) VABO and (b$_i$) the ML-MCTDHX approach and for selected interaction strengths, $g = g_i$, with  $g_1= 0$, $g_2=1$ and $g_3 = 2$. The system refers to $N_B =5$ bath atoms interacting with an impurity with $m_I=4m_B$ and $\omega_I=4\omega_B$.}
    \label{fig:two_body_density}
\end{figure*}

The observed difference in the correlations between the adiabatic BO and ML-MCTDHX approaches (identified in the interspecies two-body densities) carries over in the entanglement among the two species.
%After elaborating the spatial correlations with the two-body density, we complete our analysis with the von Neumann entropy, which is a measurement for entanglement and depicts the correlation between the two species.  
To demonstrate this we evaluate the von Neumann entropy,
\begin{equation}
\begin{split}
     S_{\rm VN}&= - {\rm Tr}_I\left[ \hat{\rho}^{(N_B)}_B \log(\hat{\rho}^{(N_B)}_B) \right] \\
            &= - {\rm Tr}_B\left[ \hat{\rho}^{(1)}_I \log(\hat{\rho}^{(1)}_I) \right],
\end{split}
     \label{eqn:von_neumann_entropy}
\end{equation}
where $\hat{\rho}^{(N_{\sigma})}_{\sigma} = {\rm Tr}_{\bar{\sigma}}\left[ | \Psi \rangle \langle \Psi | \right]$ refer the $\sigma$-species-density matrices resulting after the other species, $\bar{\sigma} \neq \sigma$, is traced out from the many-body wavefunction $| \Psi \rangle$. 
% which can be calculated by expressing the many-body wavefunction in terms of the Schmidt decomposition
% \begin{equation}
%     |\psi \rangle = \sum_k \sqrt{\lambda_k} |\psi_k^B\rangle|\psi_k^I\rangle.
% \end{equation}
The $S_{VN}$ results are presented in Fig.~\ref{fig:ground-state_energy_BO_vs_BH}(b) for all employed approaches. 
Clearly, the adiabatic BO approximations (of both the VABO and NVABO kind) overestimate the von Neumann entropy when compared to the ML-MCTDHX case.
The underlying reason for this deviation can be traced back to the one-to-one mapping between the position of the impurity $x_I$ and the state of the bath, $| \Psi_{1,B}(x_I) \rangle$, that Eq.~\eqref{eqn:multi-channel_BornOppenheimer} implies for $M=1$.
This results in large uncertainties for the state of either species when the other is traced out giving rise to entanglement.
We show in Appendix~\ref{entanglementBO} that this entanglement depends on the impurity localization, with a more delocalized impurity resulting in larger interspecies entanglement, and on $|| \frac{\partial}{\partial x_I}| \Psi_{1,B} (x_I) \rangle \big|_{x_I = 0} ||$ parameterizing how much the bath state changes as the impurity spreads.
While the former is rather constant for increasing interaction, see Fig.~\ref{fig:cuts_one_body_density}(b$_{1}$)--\ref{fig:cuts_one_body_density}(b$_{3}$), the latter substantially increases as evidenced in Fig.~\ref{fig:two_body_density}(a$_{1}$)--\ref{fig:two_body_density}(a$_{3}$), explaining the increase of $S_{VN}$ within the adiabatic BO approaches observed in Fig.~\ref{fig:ground-state_energy_BO_vs_BH}(b).
% In particular, the density matrix of the bath within the adiabatic BO approximation reads
% \begin{equation}
% \hat{\rho}^{(N_B)}_B = \int \mathrm{d}x_I~|\Psi_{1,I}(x_I)|^2 | \Psi_{1,B} (x_I)\rangle \langle \Psi_{1,B} (x_I)|.
% \label{Adensmat_aBO}
% \end{equation}
% The above equation reveals that the impurity localization controls the degree of entanglement.
% Indeed, for fixed $g$ a more localized impurity implies a stronger weight for $| \Psi_{1,B} (0) \rangle$ and less involvement of its orthogonal state, $\frac{\partial}{\partial x_I}| \Psi_{1,B} (x_I) \rangle \big|_{x_I = 0}$, resulting in the decrease of $S_{VN}$, with the opposite being the case as the impurity density expands. \textcolor{black}{We will elaborate further about this fact in Sec.~\ref{different-parameters} where we consider variations of the confinement length scale of the impurity. For a fixed impurity localization the degree of entanglement is controlled by the normalization factor $|| \frac{\partial}{\partial x_I}| \Psi_{1,B} (x_I) \rangle \big|_{x_I = 0} || = \sum_{l = 1}^{\infty} |A_{l 1}(0)|^2$, which is an increasing function of $g$ explaining the tendency of $S_{VN}$ observed in Fig.~\ref{fig:ground-state_energy_BO_vs_BH}(b). The above arguments can be mathematically explicated, see Appendix~\ref{entanglementBO}.}
% This strong dependence of $S_{VN}$ on the position of the impurity, $x_I$, is similar to what was observed for the two-body density, see Fig.~\ref{fig:two_body_density}(a$_{i}$) and (b$_{i}$) with $i=2,3$ and Eq.~\eqref{two_body_density_adiabatic}. 
Beyond the adiabatic approximation $S_{VN}$ is shown to be substantially smaller.
%\sout{, see Fig.~\ref{fig:ground-state_energy_BO_vs_BH}(b)}
This is because the superposition of additional eigenstates of bath allows to relax the one-to-one relation among $x_I$ and bath states imposed by Eq.~\eqref{eqn:multi-channel_BornOppenheimer}.

Here, one should notice also that $S_{VN}$, within the multi-channel BO approach deviates remarkably from the ML-MCTDHX approach, see Fig.~\ref{fig:ground-state_energy_BO_vs_BH}(b), which is the first discrepancy we observe among the two-above mentioned approaches. This deviation is not physical since both approaches are asymptotically exact when the corresponding truncation parameters increase to account for all possible states. This discrepancy is rather an indication of the slow convergence of the multi-channel BO approach with increasing number of configurations $M$ (here $M = 40$ was employed), which is evident when quantities addressing the total $(N_B + 1)$-body wavefunction are concerned, see also Appendix~\ref{sec:convergence}. Nevertheless, notice that on a qualitative level the results of this approach are valid and thus it can serve as an important analysis tool, demonstrating how non-adiabatic effects modify the many-body wavefunction from the adiabatic BO approximation case to the numerically exact ML-MCTDHX case.

%Thus, we expect that the impact of the non-adiabatic terms to be decisive.
%Non-adiabaticity is a first indication for the pseudo Jahn-Teller effect, which will be investigated in greater detail in the next section. 

\section{Impact of impurity parameters}
\label{different-parameters}

\begin{figure*}
    \centering
    \includegraphics{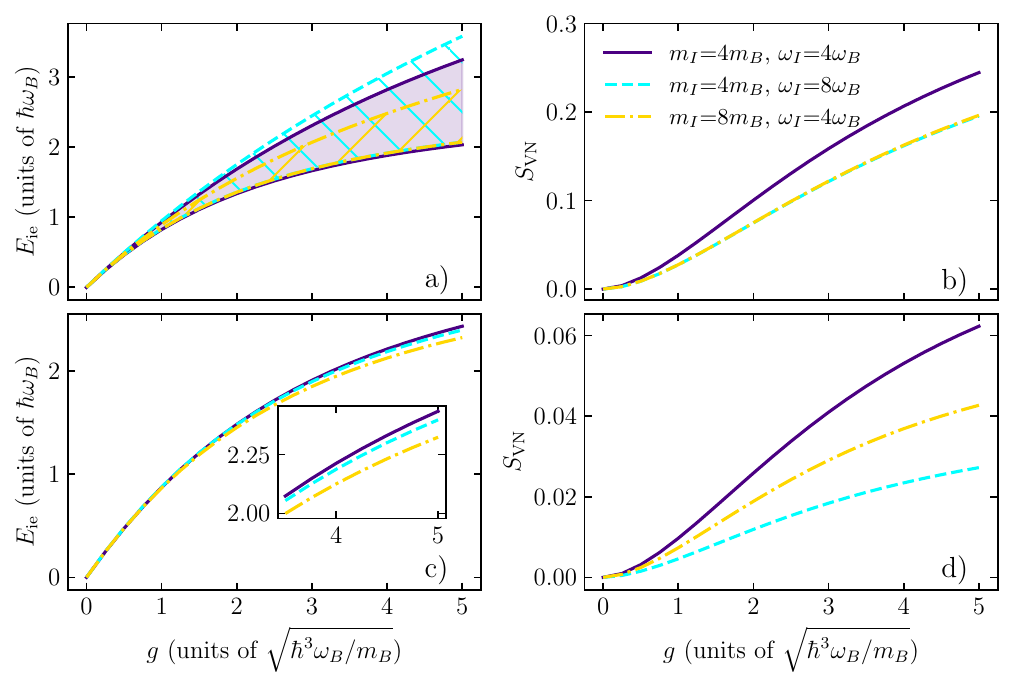}
    \caption{(a), (c) The impurity-interaction energy, \(E_{\rm{ie}}\), and (b), (d) the von Neumann entropy \(S_{\rm{VN}}\) as a function of the interaction strength, $g$, within (a), (b) the adiabatic BO approximation and (c), (d) the ML-MCTDHX approach and for different impurity parameters (see legend). The NVABO and VABO results correspond to the lower and upper bounds of $E_{\rm ie}$ respectively (see shaded regions in panel (a)), while in (b) we provide only the VABO results for visual clarity. The inset of panel (c) provides a magnification of the corresponding $E_{\rm ie}$ in the interaction interval $3.5 < g < 5$. In all cases $N_B = 5$.}
    \label{fig:variation_of_mass_trapping_frequency}
\end{figure*}

Before proceeding to the analysis of the non-adiabatic coupling terms, the importance of which is highlighted in Sec. \ref{sec:ground_state_properties} and \ref{sec:correlation_properties}, it is instructive to comment on the influence of the mass $m_I$ and trapping frequency $\omega_I$ of the impurity.
First, let us compare the impurity energy and von Neumann entropy in Fig.~\ref{fig:variation_of_mass_trapping_frequency} between the (N)VABO (a) and (b) and ML-MCTDHX (c) and (d) for different values of the impurity parameters $\omega_I$ and $m_I$. 

Figure~\ref{fig:variation_of_mass_trapping_frequency}(a) presents the energy bounds stemming from the adiabatic BO approximation to the exact energy.
It can be seen that the lower bound provided by the NVABO approximation is approximately the same for all considered parameters.
This result can be explained in terms of the effective potential, $\bar{\epsilon}(x_I)$, see Eq.~\eqref{ eqn:effective_potential }.
\textcolor{black}{As it can be seen in Fig.~\ref{fig:cuts_one_body_density}(c$_i$) for $i=1,2,3$ $\bar{\epsilon}(x_I)$, the NVABO approximation leads to a parabolic potential of frequency almost equal to $\omega^{\rm eff}_I \approx \omega_I$, with an additional energy shift stemming from the bath-impurity density-density interactions.}
This behaviour stems from the fact that the potential energy curve $\varepsilon_1(x_I)$ hardly changes for different $x_I$ since the characteristic length scale of its variation is determined by $\sqrt{\hbar/(m_B \omega_B)}$ and is thus larger than the size of the impurity wavefunction $\sim \sqrt{\hbar/(m_I \omega_I)}$.
In particular, by least-square fitting we can verify that even in the case of strong interactions, $g = 5$, the potential energy curve induced shift to the trapping frequency is $\omega^{\rm eff}_I - \omega_I \approx -0.09$ for $m_I = 4$ and $\omega_I = 4$, with this deviation further decreasing when either impurity parameter is increased (not shown here for brevity).
Therefore, within the NVABO approximation the effective Hamiltonian is
\begin{equation}
\begin{split}
    &E \Psi_{1,I}(x_I) =\\
    &\hspace{0.2cm} \left[ - \frac{\hbar^2}{2 m_I} \frac{\mathrm{d}^2}{\mathrm{d}x_I^2} + \frac{1}{2} m_I \omega_I^2 x_I^2 + \varepsilon_1(0) \right] \Psi_{1,I}(x_I),
\end{split}
\end{equation}
and thus $E_{\rm ie} \approx \varepsilon_1(0)$.

To explain the behavior of the upper bound of the impurity energy we examine the influence of the potential energy peak associated to $V^{\rm ren}_{11}(x_I)$, see Fig.~\ref{fig:cuts_one_body_density}(c${}_i$) with $i=1,2,3$.  
Notice that the spatial variation of this term does not depend on the parameters of the impurity similarly to $\varepsilon_1(x_I)$, however, its amplitude is inversely proportional to $m_I$.
Therefore, the increase of $\omega_I$ leads to the focusing of the impurity density in the spatial extent where $V^{\rm ren}_{11}(x_I)$ is large, without deteriorating the importance of this term. 
This explains the increase of the impurity-interaction energy when $\omega_I$ is twofold increased, see Fig.~\ref{fig:variation_of_mass_trapping_frequency}(a). In contrast, the twofold increase of $m_I$ causes the same focusing effect to the impurity density as its $\omega_I$ counterpart, but it suppresses the amplitude of $V^{\rm ren}_{11}(x_I)$ by a factor of two. This causes the energy increase associated to the Born-Huang term being roughly halved for $m_I = 8$, $\omega_I = 4$ when compared to $m_I = 4$, $\omega_I = 8$.
This energy decrease is evidenced by the energy difference of the VABO from the NVABO approximation in the corresponding Fig.~\ref{fig:variation_of_mass_trapping_frequency}(a).
In summary, the deviation of the energy of the VABO and NVABO approximation as expected reduces for massive impurities.
\textcolor{black}{However, as the confinement strength increases this impurity energy uncertainty, stemming from comparing the VABO and NVABO energy bounds, increases demonstrating that the adiabatic BO approximation becomes worse.} 

In the VABO approximation the von Neumann entropy decreases by the increase of either $m_I$ and $m_B$ see Fig.~\ref{fig:variation_of_mass_trapping_frequency}(b).
This effect can be attributed to the increased localization of the impurity within its parabolic trap with a characteristic length scale $\ell_I = \sqrt{\hbar/(m_I \omega_I)}$.
As the discussion in Appendix~\ref{entanglementBO} and Sec.~\ref{vonNeumann} reveals, a more localized state implies a stronger weight for $| \Psi_{1,B} (0) \rangle$ and thus a reduction of the entanglement captured by $S_{VN}$.
The fact that a twofold increase of either $m_I$ or $\omega_I$ leads to the same value of $S_{VN}$, see Fig.~\ref{fig:variation_of_mass_trapping_frequency}(b), stems from the independence of $| \Psi_{1,B} (x_I) \rangle$ on the impurity parameters. Therefore, all correlation measures depend only on $\ell_I$ which is equal in both considered cases.

In the numerically exact case of ML-MCTDHX the von Neumann entropy decreases similarly to the adiabatic BO approximation when either $m_I$ or $\omega_I$ increases, see Fig.~\ref{fig:variation_of_mass_trapping_frequency}(d).
However, the resulting increase of $S_{VN}$ is not equivalent to the adiabatic BO case.
This can be explained by the fact that increasing $\omega_I$ increases the energy gap among distinct impurity states thus the impurity is forced to occupy excited states more weakly. This means that according to the Schmidt decomposition the Schmidt weights, $\lambda_k$, with $k > 1$ should reduce and as a consequence also $S_{VN}$ reduces. In contrast a reduction of $m_I$ affects only the involved length scales and not the energy ones and thus we expect that $S_{VN}$ is more sensitive to an increase of $\omega_I$.

Less entanglement also means less opportunities to reduce the bath-impurity interaction energy below its density-overlap contribution.
Notice that the latter can be argued to be similar in both a two-fold decrease of $m_I$ or $\omega_I$ since the corresponding length scale $\ell_I' = \ell_I/\sqrt{2}$ is equal in both cases.
Indeed, the impact of a twofold increase of the impurity mass $m_I$ on the impurity energy is larger compared to a twofold increase in the $\omega_I$, see Fig.~\ref{fig:variation_of_mass_trapping_frequency}(c) and its inset.
Notably though the energy difference for different parameters is not as prominent as in the VABO approximation case, compare Fig.~\ref{fig:variation_of_mass_trapping_frequency}(c) to Fig.~\ref{fig:variation_of_mass_trapping_frequency}(a), demonstrating the important role of the non-adiabatic derivative couplings in cancelling the energy increase due to the Born-Huang term.
%\textcolor{red}{Obviously, the frequency seems not to have a decisive impact anymore after the correction.}
%\textcolor{green}{The resolution is very rough. a) We could take a more detailed look on this problem. Maybe with a better resolution we can see a reduction due to frequency. b) We expect an unchanged behaviour? This would be a contradiction to the previos statement.}

%\textcolor{black}{[Change this. It is important to comment on how the above results can be interpretted regarding the exactness of adiabatic BO approximation in the $m_I \to \infty$ limit.]}
%\textcolor{green}{Changed it, does it match now?}
%From this consideration, the decisive impact of the mass ratio for the validity of the adiabatic Born Oppenheimer approximation becomes obvious: For the limiting case $m_I 
%\to \infty$ the potential renormalization \eqref{eqn:potential_renormalization_terms} will vanish, similar to the non-adiabatic derivative couplings. Therefore, non-adiatic effects are not present anymore. 
%Fig.~\ref{fig:variation_of_mass_trapping_frequency} shows in particular the different effects of impurity mass and trapping frequency. Once more, the decisive influence of the mass is emphasized. For our further procedure, it shows for the chosen impurity mass $m_I=4.0m_A$ in our previous ground-state analysis the presence of non-adiabatic effects, which we will now work out in the following in the form of the pseudo Jahn-Teller effect. 

\section{Pseudo Jahn-Teller Effect}
\label{sec:pjte}
The Jahn-Teller effect is known for giving rise to spontaneous symmetry breaking in molecular and condensed matter physics~\cite{JahnTeller1937, Bersuker2006, Bersuker2016, BersukerPolinger1989, Bersuker2021}. 
In the case, of degenerate states, non-adiabatic couplings lift the degeneracy, which results in a ground-state possessing a lower level of symmetry. 
\textcolor{black}{In the pseudo Jahn-Teller effect, also in a non-degenerate system, the symmetry is reduced compared to the adiabatic approximation due to the non-adiabatic couplings among the fast (bath) and slow (impurity) degrees of freedom~\cite{Bersuker2013, Bersuker2021}.} 
In this section, we work out the symmetry breaking processes and the impact of non-adiabatic effects, which we have pointed out in the above ground-state analysis.

\subsection{Origin of Pseudo Jahn-Teller effect in Fermi impurity systems}
\label{pjte-general}

\begin{figure*}
    \centering
    \includegraphics[width=1.0\textwidth]{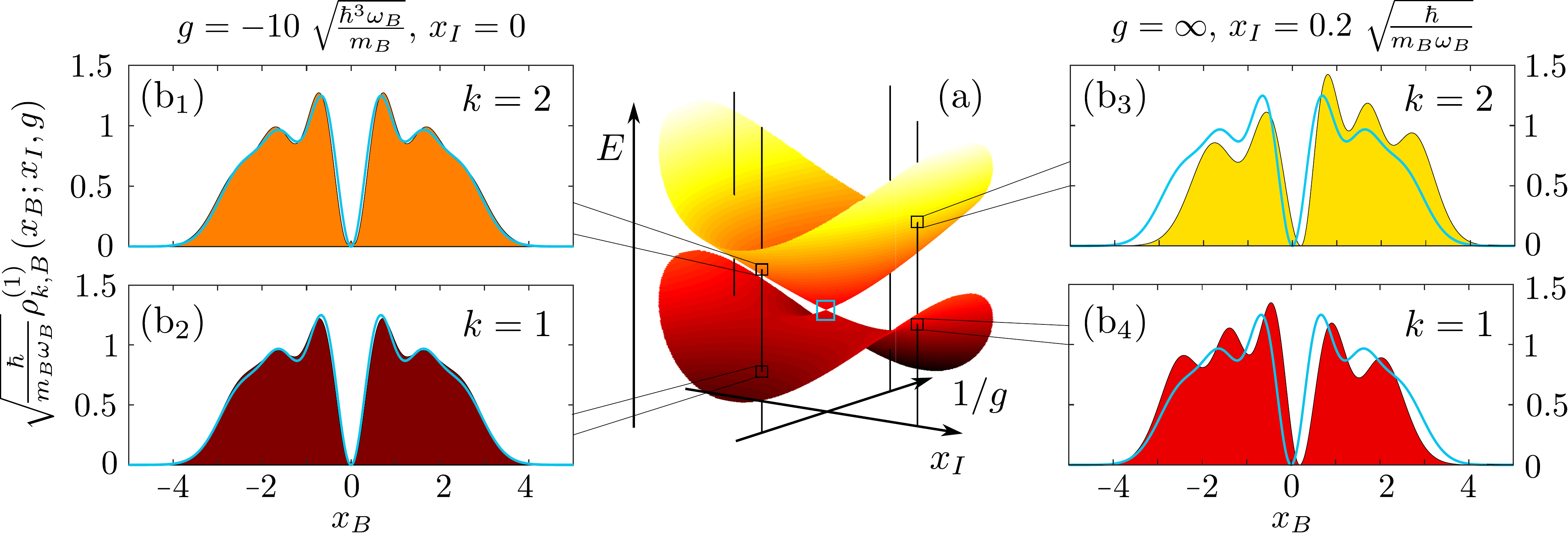}
    \caption{(a) Schematic of the conical intersection at the origin of the parametric plane ($x_I$, $1/g$), guaranteed by Theorem~\ref{theorem}. The surrounding panels (b$_i$), with $i = 1, \dots, 4$, indicate the one-body densities of the bath, corresponding to the two participating potential energy surfaces within the multi-channel BO approach for a finite displacement from the point of the conical intersection (filled areas) when compared to the point of the conical intersection (light blue curves). In all cases $N_B = 5$.}
    \label{fig:schematic_ci}
\end{figure*}

The first step in identifying the (pseudo) Jahn-Teller effect in our setup is to identify the symmetries of our setup and how these are reduced by the interaction \cite{Bersuker2006}.
In the case $g = 0$ our system possesses a parity symmetry for each individual component $\hat{\mathcal{P}}_B x_i^B = -x_i^B$ and $\hat{\mathcal{P}}_I x_I = - x_I$. However, for $g \neq 0$ this symmetry ceases to hold since the interaction term couples the two species and consequently the application of either $\hat{\mathcal{P}}_I$ or $\hat{\mathcal{P}}_B$ alters the state of the system. 
Therefore, it is interesting to examine how this reduction of symmetry affects the impurity state and especially its correlation to the state of its environment.
%\textcolor{red}{Notice that $\hat{\mathcal{P}}_I \hat{\mathcal{P}}_B$ defines a symmetry of the system and the associated conserved quantity is $\langle \Psi | \hat{X} | \Psi \rangle = 0$, with $\hat{X} = \frac{m_B \sum_{j = 1}^{N_B} x_j^B + m_I x_I}{N_B m_B + m_I}$ being the center of mass coordinate. This conservation law implies that $N_B m_B \langle \Psi | \hat{x}_B | \Psi \rangle =  - m_I \langle \Psi | \hat{x}_I | \Psi \rangle$, but neither is guaranteed to be zero. Accordingly, $\langle \Psi | \hat{x}_I | \Psi \rangle$ corresponds to the parameter identifying the degree via which the symmetries of the system components, $\hat{\mathcal{P}}_B$ and $\hat{\mathcal{P}}_I$ are broken.}
In the spirit of the original Jahn-Teller derivation~\cite{JahnTeller1937, Englman1972}, it is instructive to calculate the state of the fast degree of freedom, being the bath state, at the high-symmetry point $| \Psi_{j,B}(x_I = 0) \rangle$ and then identify the leading order coupling to the slow coordinate $x_I$.
The description of our system is simplified by recasting the Hamiltonian in terms of the shifted fast coordinates as $r_i = x_i - x_I$. This coordinate change is equivalent to the so-called Lee-Low-Pines transformation~\cite{Gurari1953, LeePines1952, LeeLow1953}. The transformed Hamiltonian reads $\hat{H}' = \hat{H}_{0r} + \hat{H}_{P_{\rm CM}} + \hat{H}_{I} + \hat{H}_{\rm coup}$. The first term refers to the bath Hamiltonian
\begin{equation}
    \hat{H}_{0r} =\sum_{j = 1}^{N_B} \left(-\frac{\hbar^2}{2 m_B} \frac{\partial ^2}{\partial  r_j^2}  + \frac{1}{2} m_B \omega^2_B r_j^2 + g  \delta(r_j) \right).
\label{H0r}
\end{equation}
Notice that $\hat{H}_{0r}$ is independent of the $x_I$ coordinate at the cost of introducing a derivative interaction term, $\hat{H}_{P_{\rm CM}}$, proportional to $1/m_I$, \textcolor{black}{corresponding to the kinetic energy of the bath particles in the transformed frame}. Namely this term reads
\begin{equation}
    \hat{H}_{P_{\rm CM}} =-\frac{\hbar^2}{2 m_I} \left( \sum_{j = 1}^{N_B} \frac{\partial}{\partial r_j} \right)^2.
\label{H_PCM}
\end{equation}
This is a typical property of a system following the Lee-Low-Pines transformation~\cite{AlexandrovDevreese2010}.\textcolor{black}{\footnote{In contrast to the bosonic case, for fermions it is more convenient not to absorb the $\propto \sum_{j = 1}^{N_B} \frac{\partial^2}{\partial r_j^2}$ appearing in Eq.~\eqref{H_PCM} as a reparametrization of the bath mass $m_B \to m_B m_I/(m_B + m_I)$ in Eq.~\eqref{H0r} as such a choice avoids difficulties in calculations. This is because this term scales $\propto N_B^2$ owing to the Pauli exclusion principle, in contrast to the overall $\propto N_B$ scaling of the center-of-mass momentum in $\hat{H}_{P_{\rm CM}}$.}}
%where $$ is the reduced mass and the effective trapping frequency is $\omega_r = \sqrt{m_B/m_r} \omega_B$.
The impurity Hamiltonian corresponds to a harmonic oscillator with modified frequency
\begin{equation}
    \hat{H}_{I} = -\frac{\hbar^2}{2 m_I} \frac{\partial ^2}{\partial  x_I^2}  + \frac{1}{2}m_I \omega^2_{I {\rm eff}} x_I^2,
    \label{impurity_hamilt}
\end{equation}
where $\omega_{I {\rm eff}} = \omega_I \sqrt{1 + \frac{m_B \omega^2_B}{m_I \omega^2_I}}$.
Finally, the coupling Hamiltonian contains a derivative and a linear coupling term
\begin{equation}
    \hat{H}_{\rm coup} = \sum_{j = 1}^{N_B} \left( \frac{\hbar^2}{m_I} \frac{\partial}{\partial r_j} \frac{\partial}{\partial x_I} + m_B \omega^2_B r_j x_I \right).
    \label{coupling_hamiltonian}
\end{equation}
The single-particle behavior \textcolor{black}{(for $N_B = 1$)} of $\hat{H}_{0 r}$ is well-known as this system admits an analytic solution~\cite{BuschEnglert1998, FarrellVanZyl2010, BudewigMistakidis2019}, see Appendix~\ref{appendix-jahn-teller}. Importantly, we know that its eigenspectrum for $g \to \infty$ features an equidistant in energy ladder of pairs of degenerate states. However, in the many-body bath case, $N_B > 1$, especially for finite $m_I/m_r$ the structure of the eigenspectrum is more involved. Nevertheless, regarding the ground state of the $\hat{H}_{0r} + \hat{H}_{P_{\rm CM}}$ system we can prove Theorem~\ref{theorem}.

\begin{theorem}
The two lowest energy eigenstates of $\hat{H}_{0 r}$ are degenerate for odd $N_B$ provided that $g \to \infty$ irrespectively of the values of the remaining system parameters, i.e. $m_I/m_B$ and $\omega_I/\omega_B$. 
\label{theorem}
\end{theorem}
\textcolor{black}{This theorem does not carry over to the case of even $N_B$ where this degeneracy can appear or not depending on the values of the system parameters.}
We will show the proof of the theorem \ref{theorem} in Appendix~\ref{proof_theorem}.

Let us now discuss the effect of the impurity on the bath state in the limit $g \to \infty$ in view of Theorem~\ref{theorem}. Notice that according to Eq.~\eqref{impurity_hamilt} the impurity lies in a harmonic oscillator potential and thus $x_I$ is delocalized within a length scale $\ell = \sqrt{\hbar / (m_I \omega_{I{\rm eff}})}$. The spreading of the impurity within its confinement potential results in a back-reaction to the bath state owing to $\hat{H}_{\rm coup}$, see Eq.~\eqref{coupling_hamiltonian}.
%For the same reasons that were outlined previously we can show that 
\textcolor{black}{According to the line of arguments in Appendix \ref{proof_theorem} we can show that $\langle \tilde{\Psi}_k | \sum_{j = 1}^{N_B} \hat{r}_j | \tilde{\Psi}_{k'} \rangle$ and $\langle \tilde{\Psi}_k | \sum_{j = 1}^{N_B} \frac{\partial}{\partial r_j} | \tilde{\Psi}_{k'} \rangle$ are non-zero only in the case that the eigenstates $| \tilde{\Psi}_{k} \rangle$ and $| \tilde{\Psi}_{k'} \rangle$ of $\hat{H}_{0r} + \hat{H}_{P_{\rm CM}}$ refer to the same $\Delta N$. $\Delta N$ corresponds to a good quantum number of the underlying system referring to the particle difference of bath atoms in the $r<0$ and $r>0$ spatial regions.} 
\textcolor{black}{Furthermore, the bath-impurity interaction Hamiltonian $\hat{H}_{\rm coup}$ does not involve a coupling among the two degenerate ground states referring to $\Delta N$ and $-\delta N$, see also Appendix~\ref{proof_theorem}, in the $g \to \infty$ limit independently of the position of the impurity, which takes non-zero values for odd $N_B$.
%\textcolor{black}{However, in this case $\langle \tilde{\Psi}_k | \sum_{j = 1}^{N_B} \hat{r}_j | \tilde{\Psi}_{k} \rangle = -\langle \tilde{\Psi}_k | \hat{\mathcal{P}_r} \sum_{j = 1}^{N_B} \hat{r}_j \hat{\mathcal{P}_r} | \tilde{\Psi}_{k} \rangle \neq 0$ since }. 
% Thus, we conclude that the ground states exhibit an exact crossing in the $g \to \infty$ limit as $x_I$ is varied.
Moving off from $x_I=0$ the coupling among the impurity and bath degrees-of-freedom lifts the degeneracy of the $x_I = 0$ ground-states owing to the different particle at $r=x_B - x_I > 0$ and $r=x_B - x_I <0$ associated with $\Delta N \neq 0$.}
Since this coupling is linear, see Eq.~\eqref{coupling_hamiltonian}, one of the potential energy curves decreases when the position of the impurity shifts from $x_I = 0$ to either positive or negative values. Thus the total many-body ground state obeys $\langle \Psi | \hat{x}_I | \Psi \rangle \neq 0$ for either value of $\Delta N$, which reduces the symmetry and corresponds to a manifestation of the Jahn-Teller effect.

In the finite but large interaction range, $g \gg 1$, the physical situation changes since the analytical continuations of the $\psi_{jL}(r)$ and $\psi_{jR}(r)$ states for finite $g$ (see appendix \ref{appendix-jahn-teller}) are not completely localized in their respective domains $L$, $R$ but show a non-vanishing amplitude in the corresponding other domain. Therefore a weak transport across the barrier at $r = 0$ is allowed. This implies a finite ``tunneling'' integral $t_i = \langle \psi_{iL} | \hat{H}_{0 r} | \psi_{iR} \rangle$ lifting the degeneracy of these two states. Notice also that this tunneling is amplified by the derivative interaction terms of Eq.~\eqref{H_PCM} and that $\hat{H}_{\rm coup}$ can also lead to coupling of these states. Precise derivations of $t_i$ can be found in the Appendix~\ref{appendix-jahn-teller} and~\ref{conical_appendix}. Therefore, the exact crossing for $g \to \infty$ outlined above becomes avoided for finite $g$. \
Provided that $t_i$ is small enough, or equivalently $g$ is large enough, the lifting of the $x_I = 0$ degeneracy does not, however, change the behavior of the system away from this high symmetry point, where the energies of the states predominantly shift due to the non-zero $\langle \tilde{\Psi}_k | \hat{x}_I | \tilde{\Psi}_{k} \rangle$ terms. The lowest-lying potential energy curve possesses a double-well structure and consequently the impurity lies in both wells in its ground state. Thus strictly speaking $\langle \Psi | \hat{x}_I | \Psi \rangle = 0$ but we can still claim that the symmetry is broken since a small symmetry-breaking perturbation would lift the degeneracy among the wells leading to $\langle \Psi | \hat{x}_I | \Psi \rangle \neq 0$. The above implies that for finite but strong enough $g$ we can identify signatures of the presence of the Jahn-Teller effect (which occurs in a strict sense only in the $g \to \infty$ limit) and identify the reduction of the symmetry of the system, despite of the absence of degeneracy at the high symmetry point. This consists a manifestation of the so-called pseudo Jahn-Teller effect.

The above can be interpreted as the emergence of a conical intersection~\cite{Domcke2004} at the $(0, 0)$ point of the parametric plane $(x_I,1/g)$, its emergence has been explicated in Appendix~\ref{conical_appendix} within the perturbative regime of both coordinates. Due to the synthetic character of the $1/g$ effective coordinate, we denote this as synthetic conical intersection \cite{Hummel2021}.
\textcolor{black}{As a summary of the results of this section we provide a schematic of the potential energy landscape in the vicinity of this synthetic conical intersection in Fig.~\ref{fig:schematic_ci}(a). A shift of the impurity position from $x_I = 0$ in the $g \to \infty$ limit lifts the degeneracy of the states possessing different $\Delta N$. This is evident in the corresponding one-body density of the bath, $\rho^{(1)}_{k, B}(x_B; x_I, g) = \langle \Psi_{k, B}(x_I, g) | \hat{\rho}^{(1)}_B | \Psi_{k, B}(x_I, g) \rangle$, see Fig.~\ref{fig:schematic_ci}(b$_3$) and Fig.~\ref{fig:schematic_ci}(b$_4$), where the number of particles on the right and the left of the impurity is directly observable by the number of $\rho^{(1)}_{k, B}(x_B; x_I, g)$ humps appearing in either region. In this case, the state with more particles in the wider region is energetically preferable, in Fig.~\ref{fig:schematic_ci}(b$_4$) this region is $x_B < x_I$ since $x_B > 0$. For $x_I = 0$ but $g$ finite the states are energetically separated in terms of their different total many-body parity, owing to the finite tunneling term $t_i \propto -1/g$, see Appendix~\ref{conical_appendix} for more details. In particular, the odd parity state (consisting of $N_B-1$ even-odd pairs of single-particle states and an additional even state) is preferred for $g>0$, since $t_i<0$, while the even state (consisting of $N_B-1$ even-odd pairs and an odd state) is preferred for $g<0$, since $t_i>0$. The parity of the states is not directly observable in $\rho^{(1)}_{k, B}(x_B; x_I, g)$, since both even and odd states will exhibit a profile symmetric around $x_B = 0$ as it can be verified in Fig.~\ref{fig:schematic_ci}(b$_1$) and Fig.~\ref{fig:schematic_ci}(b$_2$). The difference in the content of even parity single-particle states can be inferred by the slight difference of the densities with $k = 1$ and $k = 2$, stemming from the interaction dependence of these states in contrast to the odd parity ones, see also Appendix~\ref{conical_appendix}. Since in the previous sections we have working with finite $g$ this tunneling effect is the reason why the exact degeneracy at $x_I = 0$ was not observable in our results.}

The detailed analysis of the geometric properties of this conical intersection (beyond the perturbative regime), e.g. its gauge structure and Berry phase, is left as an interesting future perspective. Below we will analyze the manifestation of the pseudo Jahn-Teller effect and its relation to the non-adiabatic processes emerging in our few-body system.

\begin{figure*}
\centering
\includegraphics[width=1.\textwidth]{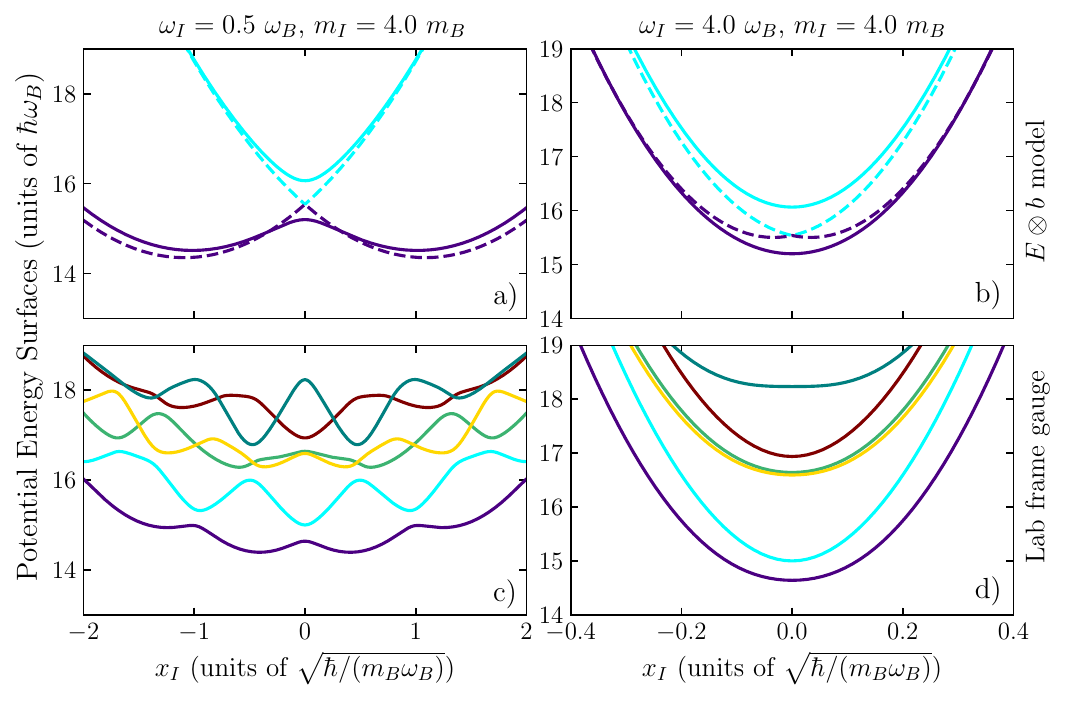}
\caption{Comparison of the lowest potential energy curves within (a), (b) the $E\otimes b$ model (crude Born-Oppenheimer approximation), see Eq.~\eqref{Etimesb_model} and (c), (d) the multi-channel BO approach for $g=5.0~\sqrt{\hbar^3\omega_B/m_B}$ and varying impurity parameters (see column label). The dashed lines in (a) and (b) show the limiting case \textcolor{black}{of the synthetic coordinate $1/g \to 0$} within the $E \otimes b$ model where the exact crossing between the two lowest adjacent potential energy curves occurs for both trapping frequencies. Notice that the $g \to \infty$ lines are offset downwards by $\delta=1$ energy units to demonstrate better the degree of pseudo-degeneracy of the involved potential energy curves for finite $g = 5$.}
\label{fig:potential_energy_surfaces}
\end{figure*}

\subsection{Pseudo Jahn-Teller effect and potential energy curves}
\label{pjte-potential-energy-surfaces}

\begin{figure*}
\centering
\includegraphics[width=1.\textwidth]{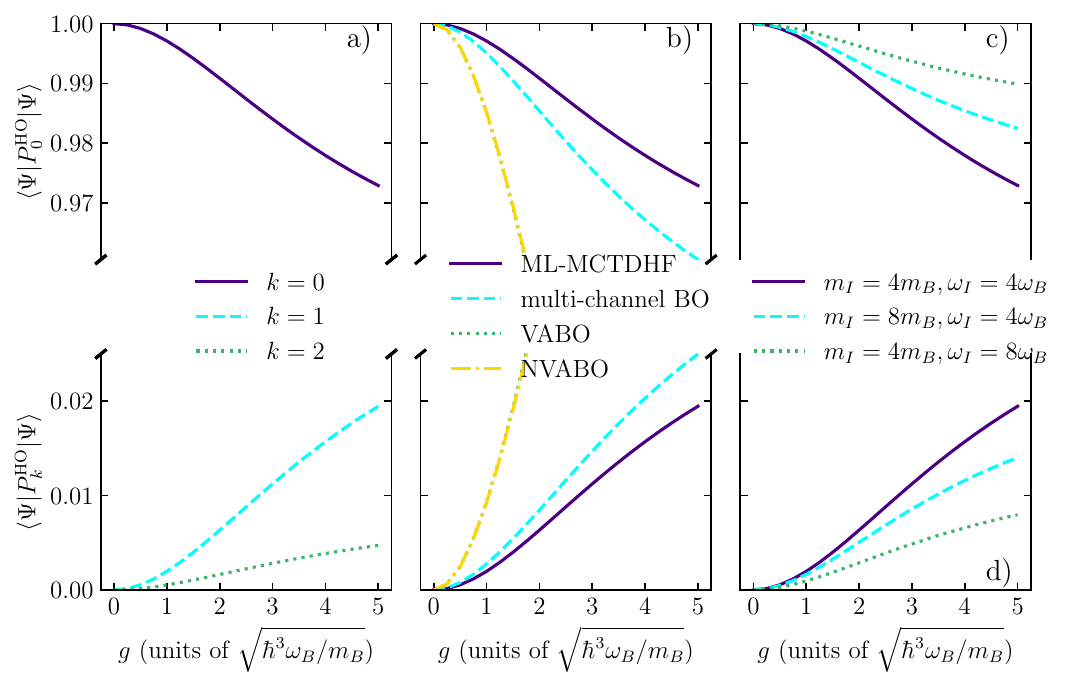}
\caption{Occupation of the $k$-th harmonic oscillator level by the impurity as captured by $\langle \Psi | \hat{P}^{\rm HO}_k | \Psi \rangle$. (a) $\langle \Psi | \hat{P}^{\rm HO}_k | \Psi \rangle$ for $k = 0$--$3$ (see legend) for $m_I=4m_B$ and $\omega_I=4\omega_B$ within ML-MCTDHX. (b) Comparison of $\langle \Psi | \hat{P}^{\rm HO}_k | \Psi \rangle$ for $k = 0$ (upper part) $k = 1$ (lower part) within different levels of approximation (see legend) and the same impurity parameters as (a). (c) Comparison of $\langle \Psi | \hat{P}^{\rm HO}_k | \Psi \rangle$ for $k = 0$ (upper part) $k = 1$ (lower part) for different impurity parameters (see legend) within ML-MCTDHX. In all cases $N_B = 5$.}
\label{fig:occupation_comparison_m4_w4}
\end{figure*}

A first step in identifying the pseudo Jahn-Teller mechanism analyzed above is to identify its effect in the potential energy curves of the system. To achieve this, we expand the transformed Hamiltonian $\hat{H}'$ (\ref{H0r}~--~\ref{coupling_hamiltonian}) in terms of the eigenstates of $\hat{H}_{0 r}+\hat{H}_{P_{\rm CM}}$ obtaining the effective Hamiltonian
\begin{equation}
\begin{split}
    \langle \tilde{\Psi}_n | &\hat{H}' | \tilde{\Psi}_m  \rangle =\\
    &-\frac{\hbar^2}{2 m_I} \delta_{n,m} \frac{\mathrm{d}^2}{\mathrm{d} x_I^2}
    - i \frac{\hbar}{m_I} P_{n,m} \frac{\mathrm{d}}{\mathrm{d} x_I} \\ 
    & + \frac{1}{2} m_I \omega^2_{I {\rm eff}}x_I^2 \delta_{n,m} + m_B \omega^2_B X_{n,m} x_I \\ 
    &+ E_n \delta_{n,m},
\end{split}
\label{Etimesb_model}
\end{equation}
where $X_{n,m} = \langle \tilde{\Psi}_n | \sum_{j=1}^{N_B} \hat{r}_j | \tilde{\Psi}_m \rangle$ and $P_{n,m} = - i \hbar \langle \tilde{\Psi}_n | \sum_{j=1}^{N_B} \frac{\mathrm{d}}{\mathrm{d} r_j} | \tilde{\Psi}_m
\rangle$. It can be shown that $X_{00} = X_{11} = 0$ and $X_{01} \neq 0$, since the eigenstates $| \tilde{\Psi}_m \rangle$ are parity symmetric for finite $g$, see also Appendix~\ref{appendix-jahn-teller}. \textcolor{black}{This effective Hamiltonian within the manifold of the two energetically lowest states $| \tilde{\Psi}_0  \rangle$ and $| \tilde{\Psi}_1 \rangle$ realizes the so-called $E \otimes b$ model, which is known to exhibit the pseudo Jahn-Teller effect. This model refers to the so-called crude Born-Oppenheimer approximation \cite{Kemper1977} being also the main tool employed for the proof of Theorem~\ref{theorem} and our arguments of Sec.~\ref{pjte-general}.} \textcolor{black}{Despite the fact that this approach is not accurate enough for quantitative comparison with the multi-channel BO approach for reasons that will be explained later on, a qualitative comparison among the two will illustrate how our above-mentioned theoretical predictions materialize within accurate numerical descriptions of our system.}

The potential energy curves stemming from this model are the eigenvalues of the effective potential 
\begin{equation}
V(x_I) = 
\left(
\begin{array}{cc}
    E_{0} + \frac{1}{2} m_I \omega^2_{I {\rm eff}}x_I^2 & m_B \omega^2_B X_{01} x_I \\
    m_B \omega^2_B X_{01} x_I & E_{1} + \frac{1}{2} m_I \omega^2_{I {\rm eff}}x_I^2 
\end{array}
\right),
\label{etimesbeffectivepotential}
\end{equation}
for varying $x_I$. The resulting potential energy curves are presented in Fig.~\ref{fig:potential_energy_surfaces}(a) and~\ref{fig:potential_energy_surfaces}(b) for $\omega_I = 0.5 \omega_B$ and $\omega_I = 4 \omega_B$ respectively. In both cases $m_I = 4 m_B$ and $g = 5$, while the two lowest-energy potential energy curves for $g \to \infty$ are also indicated by the dashed lines. Focusing especially in the case of a weaker parabolic potential, $\omega_{I} = 0.5$, the development of an avoided crossing among the first two potential energy curves at $x_I = 0$ is evident, see Fig.~\ref{fig:potential_energy_surfaces}(a). This crossing becomes an exact crossing for $g \to \infty$. In contrast to this behaviour, for higher trapping frequencies the development of this avoided crossing is not as pronounced due to the strong confinement of the impurity, see Fig.~\ref{fig:potential_energy_surfaces}(b). In this case even for $g \to \infty$ the lowest potential energy curve is almost flat and thus we have a weak symmetry breaking in terms of the $\langle \Psi | \hat{x} | \Psi \rangle$ expectation values, \textcolor{black}{even when a symmetry breaking perturbation is introduced.}

Figures~\ref{fig:potential_energy_surfaces}(c) and~\ref{fig:potential_energy_surfaces}(d) provide the potential energy curves for the same physical situations as for Fig.~\ref{fig:potential_energy_surfaces}(a) and~\ref{fig:potential_energy_surfaces}(b) respectively but within the multi-channel BO approach. We observe that for weak impurity confinement, $\omega_I = 0.5 \omega_B$, Fig.~\ref{fig:potential_energy_surfaces}(c) showcases an avoided crossing among the first two potential energy surfaces for $x_I = 0$, similarly to the case of Fig.~\ref{fig:potential_energy_surfaces}(a). However, in contrast to the two state $E\otimes b$ model more structures reminiscent of avoided crossings appear for $x_I \neq 0$. We conjecture that this occurs because additional degeneracies appear in the $g \to \infty$ case giving rise to additional instances of the pseudo Jahn-Teller effect. These are associated with many-body states where the bath atoms on the left and the right side of the impurity differ by one but possess equivalent energy (see also Sec.~\ref{sec:born-huang-adiabaticity}). This change in the confinement profile of the impurities leads to the density of the impurity being more localized in the multi-channel BO case, since the wells are narrower. Therefore, the symmetry breaking, which is associated to a doubly humped density structure as shown within Sec.~\ref{pjte-general}, becomes less apparent than within the $E \otimes b$ approach. Finally, notice that due to the fact that more than two potential energy curves are considered in the multi-channel BO, additional avoided crossings emerge among the excited potential energy curves resulting to a more convoluted potential energy landscape. Similarly to the $E \otimes b$ case the increase of $\omega_I$ leads to less prominent avoided crossings among the involved potential energy curves, compare Fig.~\ref{fig:potential_energy_surfaces}(d) and~\ref{fig:potential_energy_surfaces}(b). Therefore, in this case we do not expect an apparent symmetry breaking in the densities in accordance with the singly-peaked impurity densities identified in Fig.~\ref{fig:cuts_one_body_density}(b$_{i}$). However, the influence of the pseudo Jahn-Teller effect can be identified by carefully studying the impurity state, as we will demonstrate in Sec.~\ref{sec:jahn-teller-comparison}.

\textcolor{black}{Before proceeding let us elaborate on the shortcomings of the crude BO approach that allow only a qualitative comparison among the $E \otimes b$ and the multi-channel BO approaches which already at this level show important discrepancies.}
\textcolor{black}{First notice that within the crude BO approximation the coupling among the quasi-degenerate states is linear on both $\hat{x}_I$ and $\hat{p}_I$ due to the structure of the coupling Hamiltonian \eqref{coupling_hamiltonian} and the truncation on the two-lowest lying states at $x_I = 0$ spanning the quasi-degenerate subspace. Indeed, the multi-channel BO approximation reveals that the coupling among the bath and impurity states is significantly more complicated (not shown here for brevity) stemming from the modification of the bath state for different $x_I$ encoded in $| \Psi_{j,B}(x_I) \rangle$, see Eq.~\eqref{eqn:multi-channel_BornOppenheimer}. Since the crude BO approach involves states independent on $x_I$ an increasingly larger number of such states is required for capturing the behaviour of the system as the displacement of the impurity from $x_I = 0$  increases. Therefore, a quantitatively accurate description of the system becomes numerically challenging and difficult to intuitively interpret. For this reason the $E \otimes b$ model presented above is expected to be valid only in the region of $x_I \approx 0$ and, indeed, as Fig.~\ref{fig:potential_energy_surfaces} reveals qualitative deviations to the multi-channel approach emerge beyond this regime.}
\textcolor{black}{In addition, even for $x_I \approx 0$ the potential energy curves cannot be compared directly due to the different gauge structure of the crude and multi-channel BO approaches.} The gauge is characterized by the gauge field, $A_{jk}(x_I)$, appearing in the non-adiabatic couplings i.e. the prefactors of $\frac{\mathrm{d}}{\mathrm{d} x_I}$ in Eq.~\eqref{eqn:effective_Schroendinger} and Eq.~\eqref{Etimesb_model}. It is related to the selection of $|\Psi_{i,B}(x_I)\rangle$ states in the multi-channel BO ansatz, see Eq.~\eqref{eqn:multi-channel_BornOppenheimer}. It can be easily verified that $A_{01}(x_I) \neq P_{01}$ and even after the diagonalization of the effective potential, $V(x)$ of Eq.~\eqref{etimesbeffectivepotential}, the corresponding gauge field 
\begin{equation}
    A'_{01}(x_I) = P_{01} + \sum_{j=0}^1 U^*_{0j}(x_I) \frac{\mathrm{d}}{\mathrm{d}x_I} U_{j1}(x_I), 
\end{equation}
where $U_{jk}(x_I)$ are the matrix elements of the unitary matrix stemming from the diagonalization of Eq.~\eqref{etimesbeffectivepotential}, is different than the multi-channel BO approach, i.e. $A_{01}(x_I) \neq A'_{01}(x_I)$. 

\subsection{Indications of Jahn-Teller effect in the impurity state}
\label{sec:jahn-teller-comparison}

The coupling induced by the pseudo Jahn-Teller effect can be identified by analyzing the contributions to the impurity state. To achieve this, we evaluate the expectation values of the operators $\hat{P}^{\rm HO}_k = | \psi^{\rm HO}_{k,I} \rangle \langle \psi^{\rm HO}_{k,I} | \otimes \hat{\mathbb{I}}_B$, which project the state of the impurity to the $k$-lowest eigenstate of the harmonic oscillator with $\ell_I = \sqrt{\hbar/(m_I \omega_I)}$\textcolor{black}{, while acting as an identity operator, $\hat{\mathbb{I}}_B$, for the bath species.} \textcolor{black}{The corresponding expectation values are related to the impurity one-body density via $\langle \Psi | \hat{P}^{\rm HO}_k | \Psi \rangle = \langle \psi^{\rm HO}_{k,I} | \hat{\rho}^{(1)}_I | \psi^{\rm HO}_{k,I} \rangle$ and} are summarized in Fig.~\ref{fig:occupation_comparison_m4_w4}. In all cases the largest contribution is the ground state of the harmonic oscillator $k = 0$, which is expected due to the parabolic form of the potential energy curve even for strong $g$, see Fig.~\ref{fig:potential_energy_surfaces}(d). Our {\it ab initio} results, see Fig.~\ref{fig:occupation_comparison_m4_w4}(a), further reveal that the most strongly occupied out of the remaining harmonic oscillator levels is the $k = 1$ mode, with the remainder of the states providing significantly smaller contribution. Notice that the $k = 1$ mode is parity odd and thus its simultaneous contribution with the $k=0$ implies a state that is slightly displaced from $x = 0$ (i.e. a coherent state) in accordance to our arguments for the pseudo Jahn-Teller effect. The contribution of the $k = 2$ modes can be explained in terms of an effective modification of the confinement frequency of the impurity due to its interaction with the bath which, as claimed in Sec.~\ref{pjte-potential-energy-surfaces}, is small but non-zero.

\begin{figure*}
\includegraphics[width=1.0\linewidth]{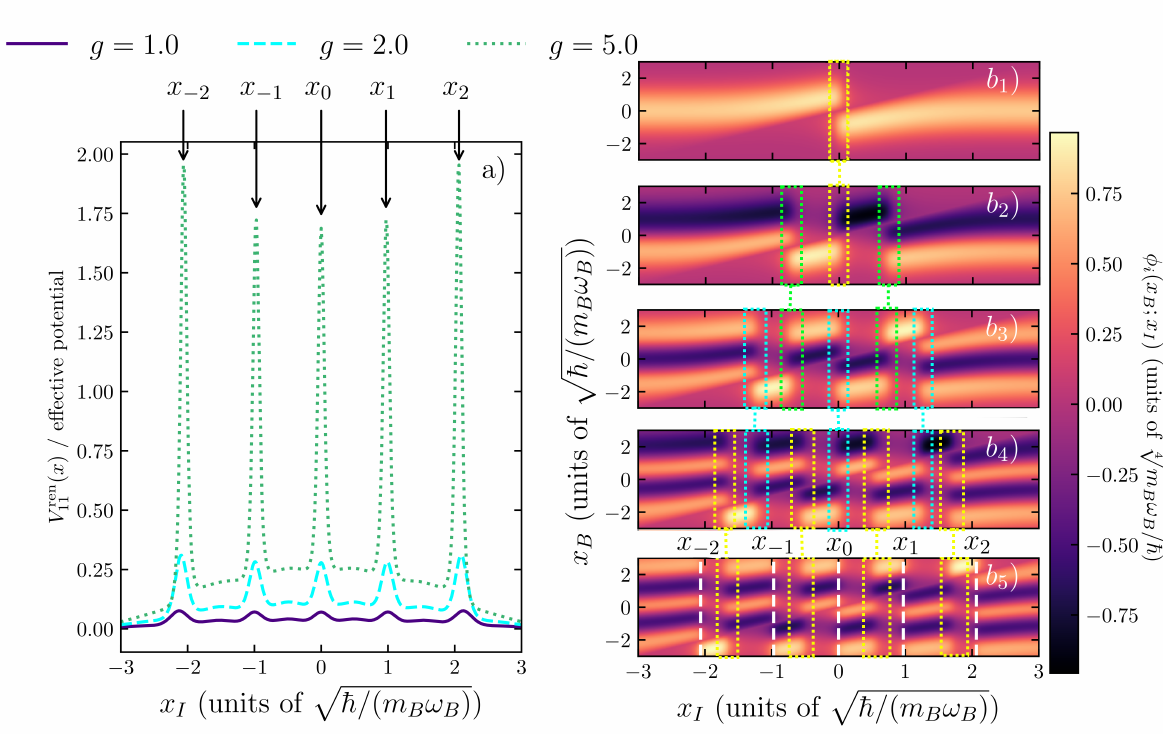}
\caption{(a) The profile of the potential renormalization (Born-Huang) term $V_{11}^{\rm ren}(x)$, see Eq.~\eqref{potential_renormalization}, for different interaction strength $g$ (see legend) within the multi-channel BO approach. (b$_i$) Parametric dependence of five lowest in energy single-particle eigenstates of $\hat{H}_B + \hat{H}_{BI}$, $\phi_i(x_B; x_I)$, with $i = 1, \dots, 5$ on the impurity position $x_I$ for $g = 5$. In all cases $N_B = 5$, $m_I = 4 m_B$ and $\omega_I = 4 \omega_B$.}
\label{fig:effective_potential}
\end{figure*}

By comparing the contribution of \textcolor{black}{the $k=0$ and $k=1$} states within different levels of approximation, see Fig.~\ref{fig:occupation_comparison_m4_w4}(b), we can see that the depletion of the $k = 0$ mode and the contribution of the $k = 1$ mode decrease as the accuracy of the approach increases. This is because the adiabatic BO approaches are affected the most by the modification of the potential energy curves. Notice also that the effect of the Born-Huang correction term is small since the VABO and NVABO approaches yield almost indistinguishable results for $g < 2$. This implies that the increase of the $k = 1$ mode is not caused by the development of the double-well structure in the effective potential identified in Fig.~\ref{fig:effective_potential}(c${}_2$) and ~\ref{fig:effective_potential}(c${}_3$), but it rather originates from the pseudo Jahn-Teller effect. As the correlations among different potential energy curves are included even partially, within the multi-channel BO approach, the population of the $k = 1$ mode decreases while it obtains a minimum but finite value within the fully correlated ML-MCTDHX approach. This is caused by the non-negligible contributions of non-adiabatic effects that increase the population of the excited state. As it can be seen already within the $E \otimes b$ approach the excited potential energy curves are more strongly confining at $x_I = 0$, see \textcolor{black}{Fig.~\ref{fig:potential_energy_surfaces}(b) (this effect is more prominent in Fig.~\ref{fig:potential_energy_surfaces}(a) albeit for different $m_I$ than the one used here)}, and thus the shift from zero of the impurity state is smaller, decreasing the population of the $k = 1$ mode.

Figure~\ref{fig:occupation_comparison_m4_w4}(c) compares the behaviour of $P^{HO}_k$ for varying $m_I$ and $\omega_I$. We observe that an increase of either $m_I$ or $\omega_I$ leads to a decrease of the depletion of $\langle \Psi | \hat{P}^{\rm HO}_0 | \Psi \rangle$ and a decrease of the contribution of $\langle \Psi | \hat{P}^{\rm HO}_1 | \Psi \rangle$. Both of these tendencies can be explained by the fact that the impurity confining potential becomes tighter since $\frac{\mathrm{d}^2V}{\mathrm{d}x^2} \propto m_I \omega_I^2$ and as a consequence the impurity gets more localized in the center of the trap competing with the pseudo Jahn-Teller effect promoting its displacement. Since this effect is quadratic on $\omega_I$ the effect of this parameter is more crucial than $m_I$.

\subsection{The Born-Huang term as a probe of non-adiabaticity}
\label{sec:born-huang-adiabaticity}

Before concluding let us address the important information gained by studying the Born-Huang term, $V_{11}^{\rm ren}(x_I)$. Its profile is provided in Fig.~\ref{fig:effective_potential}(a) for $m_I = 4 m_B$, $\omega_I = 4 \omega_B$ and varying interaction strength $g$. For all considered interactions, $V_{11}^{\rm ren}(x_I)$ possesses an inverted parabola shape with additional potential peaks at specific points denoted as $x_{k}$, with $k = 0, \pm 1, \pm 2$. The amplitude of these peaks increases strongly with increasing value of $g$, becoming the dominant feature of $V_{11}^{\rm ren}(x_I)$ at strong $g$, see Fig.~\ref{fig:effective_potential}(a) for $g= 5$. As we have claimed in Sec.~\ref{sec:one-body-density} this is the origin of the double well structure of the effective potential in the VABO approximation, see Eq.~\eqref{ eqn:effective_potential }, and it can be verified that for $g = 5$ the amplitude of the $x_0$ peak is much larger than the gap among the two lowest energy potential energy curves, see Fig.~\ref{fig:potential_energy_surfaces}(d). However, as claimed in Sec.~\ref{sec:one-body-density} the effect of these potential peaks does not appear in the impurity densities within the exact approaches and therefore it is compensated by the non-adiabatic couplings in the system.

The examination of this term within an alternative viewpoint allows us to gain a deeper understanding regarding the non-adiabatic processes present in the system. As it can be seen by Eq.~\eqref{potential_renormalization} the Born-Huang term corresponds to the change of the kinetic energy of the bath depending on the position of the impurity. This implies that the strong peaks of $V_{11}^{\rm ren}(x_I)$ at $x_k$ indicate that if the impurity resides in this region the momentum of the bath particles increases. The fast motion of bath particles can be thought as a probe of non-adiabaticity in the system.

To understand why this occurs Fig.~\ref{fig:effective_potential}(b$_i$), with $i = 1, 2, \dots, 5$, depicts the wavefunctions of the five occupied orbitals, $\phi_i(x_B; x_I)$ of $| \Psi_{1, B} (x_I) \rangle$ corresponding to the lowest energy potential energy curve \textcolor{black}{where $|\Psi_{1,B}(x_I)\rangle$ corresponds to a single Slater determinant \eqref{Slater_determinant} owing to the fact that the bath is composed of spin-polarized fermions}. By inspecting $\phi_1(x_B; x_I)$, see Fig.~\ref{fig:effective_potential}(b$_1$), we directly observe that within the region $-3 < x_I \lessapprox 0$ the wavefunction gets localized on $x_B > x_I$, while close to $x_I = 0$ the $x_B < x_I$ region starts to get occupied resulting to an equal superposition at exactly $x_I = 0$. As $x_I$ increases the region $x_B > x_I$ looses its population and at $x_I = 0.5$ only the region $x_I < x_B$ is populated. This is exactly what is expected from our discussion in Sec.~\ref{pjte-general} regarding the avoided crossing due to the pseudo Jahn-Teller effect at $x_I=0$. This is not a feature specific to $\phi_1(x_B;x_I)$ but it occurs for all of the considered single particle states, see the dotted boxes of Fig.~\ref{fig:effective_potential}(b$_i$), with $i = 1$--$5$ at $x_I \approx 0$.

Surprisingly, we can observe that this bath transport through the impurity does not occur only for $x_I = 0$ but it appears also for non-vanishing impurity displacements, see e.g. the boxes of Fig.~\ref{fig:effective_potential}(b$_2$) at $x_I \neq 0$. In this case a state with one node for $x_B > x_I$ is coupled to the state without nodes for $x_B < x_I$ as $x_I \approx 0.7$ is approached. This reveals that further exact crossings might be possible in the $g \to \infty$ limit giving rise to additional regions where the pseudo Jahn-Teller effect is exhibited for finite $g$. Thus it would be interesting to connect the avoided crossings exhibited in the potential energy curves, see Fig.~\ref{fig:potential_energy_surfaces}(c) in terms of the above mentioned regions at $x \neq 0$.

The Born-Huang term can help us in this endeavor. In Fig.~\ref{fig:effective_potential}(b$_5$) we have indicated the positions of the peaks in $V_{11}^{\rm ren}(x_I)$ on top of the profile of $\phi_5(x_B; x_I)$. It can be readily observed that these peaks correlate almost exactly to some of the points involving population transfer among the $x_B < x_I$ and $x_B > x_I$ regions. This of course makes sense since such a population transfer implies the motion of bath particles though the impurity and thus the increase of the kinetic energy of the bath component. However, not all points where transfer happens are associated with an increase of $V_{11}^{\rm ren}(x_I)$. \textcolor{black}{This can be understood as follows. Notice that each one of the cases for $\phi_5(x_B; x_I)$ where no peak in $V_{11}^{\rm ren}(x_I)$ occurs aligns with a transfer process in $\phi_4(x_B; x_I)$ (see the dashed boxes in Fig.~\ref{fig:effective_potential}(b$_4$) and the associated dashed lines connecting them to the boxes of Fig.~\ref{fig:effective_potential}(b$_5$)). Finally, notice that such regions occur also among lower-lying orbitals, see the boxes and connecting lines in Fig.~\ref{fig:effective_potential}(b$_i$) with $i = 1$--$4$.}

The above discussed kinetic energy increase is a probe for non-adiabaticity in the system since at the regions of $x_k$, with $k=0,\pm 1, \pm 2$, a strong non-adiabatic coupling among the $\phi_5(x_B; x_I)$ and $\phi_6(x_B; x_I)$ is exhibited giving rise to a large value of $A_{01}(x_I)$ (not shown here for brevity). In addition, we can confirm that the points $x_{\pm 1}$ capture well the position of the $x_I \neq 0$ avoided crossings of the first two potential energy curves in Fig.~\ref{fig:potential_energy_surfaces}(c), being consistent with the presence of the pseudo Jahn-Teller effect in this spatial region.

\section{Summary and Outlook}
\label{sec:outlook}
We have performed a comprehensive ground-state analysis of a fermionic few-particle setup consisting of five light fermionic particles interacting with a single heavy impurity. This setup demonstrates the failure of the adiabatic BO approximation. 
In particular strong deviations are identified between the numerically exact ML-MCTDHX approach and the adiabatic BO approximation in the impurity energy and one-body density, as well as the correlation properties given by the two-body density and von Neumann entropy. 
%The numerically exact calculations have been performed by ML-MCTDHF, where we confirmed the convergence due to the comparison with a multi-channel Born-Oppenheimer approach,
These results indicate the presence of strong non-adiabatic effects in our system.  
In particular, we are able to interpret these results by introducing the inverse of the interaction strength as a synthetic dimension and analyzing the emergence of the Jahn-Teller effect in the strong interaction limit of our Fermi impurity system. 
Based on this we have shown that our system approximately maps to a $E \otimes b$ system and thus exhibits the pseudo Jahn-Teller effect for finite interactions, associated with the breaking of the parity symmetry of the impurity state. 
An increasing interaction strength between the bath and impurity atoms leads to strong ``vibronic'' couplings among the slow degrees-of-freedom of the impurity with the fast motion of the bath atoms, explaining the previously identified non-adiabatic effects.
By examining the potential energy curves, we demonstrate at least one conical intersection at the trap center and for infinitely strong interactions. 
Especially, we have shown that the Born-Huang term of the lowest energy potential energy curve can be employed as a measure for the non-adiabaticity of the system indicating resonant transport of the bath particles through the impurity. 

Based on the above results several pathways of future research become evident. First, by considering the time propagation of the fermionic impurity system, the possibility arises of probing the pseudo Jahn-Teller effect during the impurity dynamics.
This can be achieved by shifting the center of the harmonic trapping potential for the impurity and tracking the induced impurity dynamics, in terms of its dipole and breathing modes.
\textcolor{black}{An intriguing perspective concerns the effect of spin-orbit coupling in systems experiencing the Jahn-Teller effect. The presence of spin-orbit coupling allows for a direct relation of our ultracold setup to molecular physics since the Jahn-Teller effect commonly appears in compounds of heavy elements \cite{Bersuker2021, Bersuker2023}. In addition, ultracold atom setups allow for addressing the particle distribution in a spin-resolved manner and thus encoding the Jahn-Teller induced symmetry-breaking process on the spin-state of the gas can be a valuable resource for experimentally addressing our findings.
Notice that the (pseudo) Jahn-Teller effect can also manifest in isotropically confined two and three-dimensional ultracold gases. In such setups the interparticle repulsion is relevant only in the case that the atoms do not have a relative angular momentum and it has been shown that the interaction energy can exceed the rotational barrier \cite{BuschEnglert1998}. This can lead to the formation of a conical intesection caused by the displacement of the impurity from the trap center breaking the rotational invariance of the confinement. }
%Next, we perform a quench of the impurity trapping confinement back to the center back into line with the center of the trap of the bath particles. 
Considering similar quasi-particle systems, the question arises about symmetry breaking effects in polarons, which can be investigated also for higher dimensional systems. One intriguing theoretical question in this context is the relation of the non-adiabaticity unveiled here with the well-known Anderson orthogonality catastrophe~\cite{Anderson1967,Anderson1967_2}. 
%Since these polarons have been proven to exist also in two and three dimensions, symmetry breaking effects can also be studied in the higher dimensional case.

\begin{acknowledgements}
This work has been funded by the Cluster of Excellence
``Advanced Imaging of Matter'' of the Deutsche Forschungsgemeinschaft
(DFG) - EXC 2056 - project ID 390715994. G. M. K. gratefully acknowledges funding from the European Union’s Horizon 2020 research and innovation programme under the Marie Skłodowska-Curie grant agreement No.~ 101034413.
\end{acknowledgements}

\appendix

\section{Details on the calculation of non-adiabatic couplings within the multi-channel BO approach}
\label{appendix:MCBO}

The assumption of a spin-polarized fermionic impurity species greatly simplifies the complexity of evaluation of the non-adiabatic couplings and the potential renormalization terms appearing in Eq.~\eqref{eqn:effective_Schroendinger}.
In particular, since the species $B$ does not possess interspecies interactions its eigenstates for fixed impurity position, $| \Psi_{k,B}(x_I) \rangle$, can be exactly represented in terms of a single Slater determinant
\begin{equation}
\begin{split}
\Psi_{k,B} (x_1,\dots& ,x_N;x_I) = 
\frac{1}{\sqrt{N!}}\sum_{j = 1}^{N!} {\rm sign}\left(P_j \right)\\ &\times\phi_{P_{j}(I^k_1)}(x_1; x_I) \dots \phi_{P_{j}(I^k_N)}(x_N; x_I),
\end{split}
\label{Slater_determinant}
\end{equation}
where \(\phi_{j}(x_B; x_I)\) for $j = 1, 2, \dots$ are the \textcolor{black}{$N_B = 1$} eigenstates of $\hat{H}_B + \hat{H}_{BI}$ for a fixed value of \(x_I\) with eigenenergy $\varepsilon^{(1)}_j(x_I)$. $I_j^k$, with $j = 1, \dots, N$ parameterize the particular set of orbitals corresponding to the $k$-th lowest in energy, $\varepsilon_k(x_I) = \sum_{j=1}^{N} \varepsilon^{(1)}_{I^k_j}(x_I)$, eigenstate of the many-body system. In order for this orbital set to be unique, we demand that it is ordered in ascending order, i.e. $I^k_j < I^k_{j'}$ for $j < j'$. Notice that the above is exactly equivalent to the standard prescription of how the many-body eigenstates of non-interacting fermions are mapped to the corresponding number states~\cite{Thouless1961}.

The Slater determinant of Eq.~\eqref{Slater_determinant}, allows us to express the non-adiabatic couplings of the many-body bath species in terms of \textcolor{black}{the $N_B=1$ orbitals $\phi_{j}(x_B;x_I)$}.
To make the presentation of this process clearer, let us introduce the notation where the many-body eigenstates \( | \Psi_{k,B}(x_I) \rangle \) are expressed in terms of the occupied orbitals as \( | \Psi_{\{I^{k}_1, \dots, I_{N}^k \}}(x_I) \rangle \).
Within this notation the non-adiabatic derivative coupling matrix reads
\begin{equation}
\begin{split}
    A_{kl}(x_I) &= A_{\{I^{k}_1, \dots, I_{N}^k \}\{I^{l}_1, \dots, I_{N}^l \}}(x_I) \\
    &=i \bigg\langle  \Psi_{\{I^{k}_1, \dots, I_{N}^k \}}(x_I) \bigg| \frac{\mathrm{d} \Psi_{\{I^{l}_1, \dots, I_{N}^l \}}}{\mathrm{d} x_I}(x_I) \bigg\rangle.
\end{split}
\end{equation}
By employing Eq.~\eqref{Slater_determinant}  we derive
\begin{equation}
\begin{split}
&A_{\{I_1,\dots,I_N\}\{J_1,\dots,J_N\}}(x_I) \\
&\hspace{0.3cm}=  
i \sum_{l = 1}^{N} \sum_{k = 1}^N (-1)^{k+l} \bigg\langle \phi_{I_k}(x_I) \bigg| \frac{\mathrm{d} \phi_{J_l}}{\mathrm{d} x_I}(x_I) \bigg\rangle\\
&\hspace{0.7cm}
\times\delta_{\{I_1, \dots, I_{k-1}, I_{k+1}, \dots, I_N \}\{J_1, \dots, J_{l-1}, J_{l+1}, \dots, J_N \}},
\end{split}
\label{nonadiabderivinter}
\end{equation}
where the Kronecker delta symbol $\delta_{S_1S_2}$ yields zero for two different sets $S_1$ and $S_2$, and one in the case they are equivalent.
For a trapped system, such as the one described by Eq.~\eqref{eqn:Hamilt_Full}, it is known that the eigenbasis consists of real valued \(\phi_j(x_B;x_I)\) states~\cite{Messiah1981}. Then by considering that the momentum operator for the impurity species is Hermitian, we obtain that the single-particle eigenfunctions fulfill
\begin{equation}
\bigg\langle \phi_{j}(x_I) \bigg| \frac{\mathrm{d} \phi_{j}}{\mathrm{d} x_I}(x_I) \bigg\rangle = 0.
\end{equation}
This fact allows us to simplify the expression of Eq.~\eqref{nonadiabderivinter} further. In particular, $A_{\{I_1,\dots,I_N\}\{J_1,\dots,J_N\}}(x_I)$ vanishes except in the case where a single orbital in the sets $\{I_1,\dots,I_N\}$ and $\{J_1,\dots,J_N\}$ is different.
The non-vanishing elements read
\begin{equation}
\begin{split}
A_{\{I_1,\dots,I_N\}\{I_1,\dots,I_{k-1},I_{k+1},\dots,I_{l-1},J_{l},I_{l+1},\dots,I_N\}}(x_I)\\ =  
i (-1)^{k + l} \bigg\langle \phi_{I_k}(x_I) \bigg| \frac{\mathrm{d} \phi_{J_l}}{\mathrm{d} x_I}(x_I) \bigg\rangle,
\label{eqn:non_adiabatic_derivative_Coupling}
\end{split}
\end{equation}
where \(J_l \neq I_k\).

According to Eq.~\eqref{eqn:effective_Schroendinger} the renormalization potential \(V_{kl}^{\rm{ren}}(x_I)\) can be written as
\begin{equation}
V_{kl}^{\rm{ren}}(x_I) = \frac{\hbar^2}{2 m_I} \bigg[B_{k l}(x_I) - \sum_{r = 1}^M A_{r k}^*(x_I) A_{r l}(x_I) \bigg],
\end{equation}
where the \(B\) matrix elements read
\begin{equation}
B_{k l}(x_I) = \bigg\langle  \frac{\mathrm{d} \Psi_{k,B}}{\mathrm{d} x_I}(x_I) \bigg| \frac{\mathrm{d} \Psi_{l,B}}{\mathrm{d} x_I}(x_I) \bigg\rangle.
\end{equation}
An analogous procedure to the one applied above for the non-adiabatic derivative couplings (see Eq.~\eqref{nonadiabderivinter}) yields that the $B$ matrix elements read
\begin{equation}
\begin{split}
&B_{\{I_1,\dots,I_N\}\{J_1,\dots,J_N\}}(x_B)\\
&=\sum_{k=1}^N \sum_{l = 1}^N (-1)^{k + l} \bigg\langle \frac{\mathrm{d} \phi_{I_k}}{\mathrm{d} x_I}(x_I) \bigg| \frac{\mathrm{d} \phi_{J_l}}{\mathrm{d} x_I}(x_I) \bigg\rangle\\
&\hspace{0.14cm}\times \delta_{\{I_1,\dots,I_{k-1},I_{k+1},\dots,I_N\}\{J_1,\dots,J_{l-1},J_{l+1},\dots,J_N\}} \\
&- i \sum_{k=1}^N \sum_{l = 1}^N (-1)^{k + l}\bigg\langle \frac{\mathrm{d} \phi_{I_k}}{\mathrm{d} x_I}(x_I) \bigg| \phi_{J_l}(x_I) \bigg\rangle\\
&\hspace{0.14cm}\times A_{\{I_1,\dots,I_{k-1},I_{k+1},\dots,I_N\}\{J_1,\dots,J_{l-1},J_{l+1},\dots,J_N\}}(x_I).
\end{split}
\end{equation}
Owing to the properties of the non-adiabatic derivative couplings, see Eq.~\eqref{eqn:non_adiabatic_derivative_Coupling}, we can distinguish three different cases where the $B$ matrix elements are non-vanishing.
First, if both sets are equivalent, we obtain
\begin{equation}
\begin{split}
&B_{\{I_1,\dots,I_N\}\{I_1,\dots,I_N\}}(x_I)\\&\hspace{0.5cm}=
\sum_{k=1}^N \bigg\langle \frac{\mathrm{d} \phi_{I_k}}{\mathrm{d} x_I}(x_I) \bigg| \frac{\mathrm{d} \phi_{I_k}}{\mathrm{d} x_I}(x_I) \bigg\rangle\\
&\hspace{0.7cm}- \sum_{k=1}^N \sum_{l = 1 \atop l \neq k}^N \bigg\langle \phi_{I_l}(x_I) \bigg| \frac{\mathrm{d} \phi_{I_k}}{\mathrm{d} x_I}(x_I) \bigg\rangle^{2}.
\end{split}
\end{equation}
Second, in the case that the sets are different by a single orbital, we have 
\begin{equation}
\begin{split}
&B_{\{I_1,\dots,I_N\}\{I_1,\dots,I_{k-1},I_{k+1},\dots,I_{l},J_{l},I_{l+1},\dots, I_N\}}(x_I) \\
&\hspace{0.5cm}= 
(-1)^{k + l} \bigg[ \bigg\langle \frac{\mathrm{d} \phi_{I_k}}{\mathrm{d} x_I}(x_I) \bigg| \frac{\mathrm{d} \phi_{J_l}}{\mathrm{d} x_I}(x_I) \bigg\rangle\\
&\hspace{2.3cm}
-\sum_{r = 1 \atop r \neq k}^{N} \bigg\langle \phi_{I_r}(x_I) \bigg| \frac{\mathrm{d} \phi_{I_k}}{\mathrm{d} x_I}(x_I) \bigg\rangle \\
&\hspace{2.9cm}\times \bigg\langle \phi_{I_r}(x_I) \bigg| \frac{\mathrm{d} \phi_{J_l}}{\mathrm{d} x_I}(x_I) \bigg\rangle \bigg],
\end{split}
\end{equation}
where we demand $J_l \neq I_k$.
Finally, the last case that $B$ is non vanishing arises when the two sets differ by two indices.
Here, the $B$ matrix elements read
\begin{equation}
\begin{split}
&B_{\{I_1,\dots, I_N\}\{I_1,\dots,I_{k-1},I_{k+1},\dots,}\\&\hspace{0.5cm}_{I_{l-1},I_{l+1},\dots,I_{r+1},J_{r},I_{r+2},\dots,I_{s+1},J_{s},I_{s+2},\dots,I_N\}}(x_I)\\
&\hspace{0.05cm} = (-1)^{k + r + l + s}\\ 
&\hspace{0.05cm}\times \bigg ( \bigg\langle \phi_{J_r}(x_I) \bigg| \frac{\mathrm{d} \phi_{I_k}}{\mathrm{d} x_I}(x_I) \bigg\rangle
\bigg\langle \phi_{I_l}(x_I) \bigg| \frac{\mathrm{d} \phi_{J_s}}{\mathrm{d} x_I}(x_I) \bigg\rangle 
\\&\hspace{0.2cm}-\bigg\langle \phi_{J_s}(x_I) \bigg| \frac{\mathrm{d} \phi_{I_k}}{\mathrm{d} x_I}(x_I) \bigg\rangle
\bigg\langle \phi_{I_l}(x_I) \bigg| \frac{\mathrm{d} \phi_{J_r}}{\mathrm{d} x_I}(x_I) \bigg\rangle \\
&\hspace{0.2cm}-\bigg\langle \phi_{J_r}(x_I) \bigg| \frac{\mathrm{d} \phi_{I_l}}{\mathrm{d} x_I}(x_I) \bigg\rangle
\bigg\langle \phi_{I_k}(x_I) \bigg| \frac{\mathrm{d} \phi_{J_s}}{\mathrm{d} x_I}(x_I) \bigg\rangle 
\\
&\hspace{0.2cm}+\bigg\langle \phi_{J_s}(x_I) \bigg| \frac{\mathrm{d} \phi_{I_l}}{\mathrm{d} x_I}(x_I) \bigg\rangle
\bigg\langle \phi_{I_k}(x_I) \bigg| \frac{\mathrm{d} \phi_{J_r}}{\mathrm{d} x_I}(x_I) \bigg\rangle \bigg),
\label{eqn:potential_renormalisation_terms}
\end{split}
\end{equation}
where each of the indices \(J_r\) and \(J_s\) has to be different to both $I_k$ and $I_l$. 

The representation of the non-adiabatic derivative coupling \eqref{eqn:non_adiabatic_derivative_Coupling} and the renormalization potential \eqref{eqn:potential_renormalisation_terms} allow for the numerical solution of the effective Schr\"odinger equation \eqref{eqn:effective_Schroendinger}.

\section{Inter-species entanglement within the adiabatic Born-Oppenheimer approximation}
\label{entanglementBO}

A perhaps not well-known fact regarding the adiabatic BO approximation is that it involves entanglement between the slow (nuclear) and fast (electronic) degrees of freedom. This aspect has also been noted in the recent quantum chemical literature~\cite{BouvrieMajtey2014, IzmaylovFranco2017}. The purpose of this section is to demonstrate that this entanglement effect also appears in our two-species one-dimensional system and provides a proper mathematical framework for our arguments in Sec.~\ref{vonNeumann}.

% To understand the reason for this behaviour let us briefly analyze $S_{VN}$ within the adiabatic BO approximation.
% Notice that the wavefunction structure of this approach, see Eq.~\eqref{eqn:multi-channel_BornOppenheimer} for $M=1$, establishes a one-to-one mapping between the position of the impurity $x_I$ and the state of the bath, $| \Psi_{1,B}(x_I) \rangle$. 
% This results to large uncertainties for the state of either species when the other is traced out giving rise to entanglement and thus substantial values of $S_{VN}$.
%Since this claim might seem counter-intuitive given the status of the adiabatic BO approximation as a {\it semi-classical} theory let us prove it.
%In particular, the density matrix of the bath within the adiabatic BO approximation reads
Our starting point is the reduced bath-density matrix within the adiabatic BO approximation where the impurity degrees-of-freedom have been traced out,
\begin{equation}
\hat{\rho}^{(N_B)}_B = \int \mathrm{d}x_I~|\Psi_{1,I}(x_I)|^2 | \Psi_{1,B} (x_I)\rangle \langle \Psi_{1,B} (x_I)|.
\label{Adensmat_aBO}
\end{equation}
The above equation reveals that the impurity localization controls the degree of entanglement.
In particular, it can be easily verified that for $\Psi_{1,I}(x_I) = \delta(x_I - x_I^0)$ the state of the bath is pure and thus there is no entanglement in the system.
%In particular, the bath density matrix results after the $x_I$ averaging of the $\hat{H}_B + \hat{H}_I$ eigenstates $| \Psi_0 (x_I) \rangle$ and 
% Indeed, for fixed $g$ a more localized impurity implies a stronger weight for $| \Psi_{1,B} (0) \rangle$ and less involvement of its orthogonal state, $\frac{\partial}{\partial x_I}| \Psi_{1,B} (x_I) \rangle \big|_{x_I = 0}$, resulting in the decrease of $S_{VN}$, with the opposite being the case as the impurity density expands. \textcolor{black}{We will elaborate further about this fact in Sec.~\ref{different-parameters} where we consider variations of the confinement length scale of the impurity.}
% This strong dependence of $S_{VN}$ on the position of the impurity, $x_I$, is similar to what was observed for the two-body density, see Fig.~\ref{fig:two_body_density}(a$_{i}$) and (b$_{i}$) with $i=2,3$ and Eq.~\eqref{two_body_density_adiabatic}. 
In the general case it is difficult to analyze the entanglement properties of the system. However, the assumption of a heavy $m_I \gg m_B$ and tightly confined $\omega_I \gg \omega_B$ impurity enable us to employ the approximation that the bath state, $| \Psi_{1,B}(x_I) \rangle$, changes at much longer length scales than $\Psi_{1,I}(x_I)$ defining the size of the impurity state. Indeed, the length scale associated to each species is $\ell_{\sigma} = \sqrt{\hbar/(m_{\sigma} \omega_{\sigma})}$ and thus $\ell_{B}$ controlling the $x_I$-dependence of $| \Psi_{1,B} (x_I)\rangle$ is much longer than the spatial region where the impurity localizes $\sim \ell_I$.
This allows us to expand the bath state in a Taylor series around the equilibrium position of the impurity $x_I = 0$ as
\begin{equation}
\begin{split}
    | \Psi_{1,B} (x_I)\rangle &= | \Psi_{1,B} (0)\rangle + x_I \frac{\partial}{\partial x_I}| \Psi_{1,B} (x_I)\rangle \bigg|_{x_I = 0} \\
     &+ \frac{1}{2} x_I^2 \frac{\partial^2}{\partial x_I^2}| \Psi_{1,B} (x_I)\rangle \bigg|_{x_I = 0} + \mathcal{O} \left( x_I^3 \right).
\end{split}
\end{equation}
This expansion allows us to evaluate the integral appearing in Eq.~\eqref{Adensmat_aBO} order by order.
Since we operate in the regime $\ell_I \ll \ell_B$, it is reasonable to consider that $\langle \Psi_{1,I} | \hat{x}_I^n | \Psi_{1,I} \rangle/\ell_B^n$ is a decreasing sequence in $n$ and thus we consider that only its first three terms for $n = 0, 1, 2$ are non-negligible.
Within this approximation the bath-density operator reads
\begin{equation}
\begin{split}
    \hat{\rho}_B^{(N_B)} &= | \Psi_{1,B} (0) \rangle \langle \Psi_{1,B} (0) |
    + \langle \Psi_{1,I} | \hat{x}_I | \Psi_{1,I} \rangle\\
    &\times \left( 
        | \Psi_{1,B} (0) \rangle \left\langle \frac{\partial \Psi_{1,B}}{\partial x_I} (0) \right| \right. \\
        &+~\left. \left| \frac{\partial \Psi_{1,B}}{\partial x_I} (0) \right\rangle \left\langle \Psi_{1,B} (0) \right| 
    \right)\\
    &+ \langle \Psi_{1,I} | \hat{x}_I^2 | \Psi_{1,I} \rangle \bigg(
    \left| \frac{\partial \Psi_{1,B}}{\partial x_I} (0) \right\rangle \left\langle \frac{\partial \Psi_{1,B}}{\partial x_I} (0) \right|\\
    &- \frac{1}{2} | \Psi_{1,B} (0) \rangle \left\langle \frac{\partial^2 \Psi_{1,B}}{\partial x_I^2} (0) \right| \\
    &- \frac{1}{2} \left| \frac{\partial^2 \Psi_{1,B}}{\partial x_I^2} (0) \right\rangle \left\langle \Psi_{1,B} (0) \right| \bigg) \\
    &+\mathcal{O}\left( \langle \Psi_{1,I} | \hat{x}_I^3 | \Psi_{1,I} \rangle\right).
\end{split}
\end{equation}
\textcolor{black}{This expression allows us to determine the matrix elements of $\hat{\rho}_B^{(N_B)}$ in the basis defined by the eigenstates of $\hat{H}_B + \hat{H}_{BI}$ (see Eq.~\eqref{eqn:Hamilt_Full}) for an impurity fixed at $x_I = 0$, namely $| \Psi_{j,B} (0) \rangle$, for $j = 1, 2, \dots$. Notice that since this basis is complete and independent of the position of the impurity state enabling us to employ the usual definitions for calculating the Schmidt modes and von Neumann entropy.
Within this prescription the above-mentioned matrix elements read}
\begin{equation}
\begin{split}
    \langle \Psi_{1,B} (0) &| \hat{\rho}_B^{(N_B)} | \Psi_{1,B} (0) \rangle =\\
    &1 - \langle \Psi_{1,I} | \hat{x}_I^2 | \Psi_{1,I} \rangle \sum_{l = 1}^{\infty} | A_{1 l}(0) |^2 \\
    &+\mathcal{O}\left( \langle \Psi_{1,I} | \hat{x}_I^4 | \Psi_{1,I} \rangle\right), \\
    \langle \Psi_{1,B} (0) &| \hat{\rho}_B^{(N_B)} | \Psi_{j,B} (0) \rangle =\\
    &-\frac{1}{2} \langle \Psi_{1,I} | \hat{x}_I^2 | \Psi_{1,I} \rangle\sum_{l = 1}^{\infty} A^*_{jl}(0) A^*_{l1}(0) \\
    &+\mathcal{O}\left( \langle \Psi_{1,I} | \hat{x}_I^4 | \Psi_{1,I} \rangle\right), \\
    \langle \Psi_{j,B} (0) &| \hat{\rho}_B^{(N_B)} | \Psi_{k,B} (0) \rangle =\\
    &\langle \Psi_{1,I} | \hat{x}_I^2 | \Psi_{1,I} \rangle
    A_{j1}(0) A_{k1}^{*}(0),\\
    &+\mathcal{O}\left( \langle \Psi_{1,I} | \hat{x}_I^4 | \Psi_{1,I} \rangle\right),
\end{split}
\label{matrixelementsBObathdensity}
\end{equation}
where $j, k=2, 3, \dots$. The non-adiabatic couplings appear in Eq.~\eqref{matrixelementsBObathdensity} since by definition they are equal to $A_{kj}(x_I) = i \langle \Psi_{k,B}(x_I) | \frac{\partial \Psi_{j,B}}{\partial x_I}(x_I) \rangle$. Moreover, we have used the completeness property of the $| \Psi_{j,B} (x_I) \rangle$ states to simplify $\langle \Psi_{k,B}(x_I) | \frac{\partial^2 \Psi_{j,B}}{\partial x_I^2}(x_I) \rangle = - \sum_{l = 1}^{\infty} A_{k l}(x_I) A_{l j}(x_I)$ and the parity symmetry for the impurity species yielding $\langle \Psi_{1,I} | \hat{x}_I^{2n+1} | \Psi_{1,I} \rangle = 0$, for all $n$.
Within first order perturbation theory for the dominant Schmidt mode we obtain
\begin{equation}
    \begin{split}
\lambda_1 =& 1 - \langle \Psi_{1,I} | \hat{x}_I^2 | \Psi_{1,I} \rangle \sum_{l = 1}^{\infty} |A_{1 l}(0)|^2 \\
 &+ \mathcal{O} (\langle \Psi_{1,I} | \hat{x}_I^4 | \Psi_{1,I} \rangle).
    \end{split}
\end{equation}
While degenerate first order perturbation theory for the remaining modes yields
\begin{equation}
    \begin{split}
\lambda_2 =& \langle \Psi_{1,I} | \hat{x}_I^2 | \Psi_{1,I} \rangle \sum_{l = 1}^{\infty} |A_{1 l}(0)|^2 \\
    &+ \mathcal{O} (\langle \Psi_{1,I} | \hat{x}_I^4 | \Psi_{1,I} \rangle), \\
\lambda_k =& \mathcal{O} (\langle \Psi_{1,I} | \hat{x}_I^4 | \Psi_{1,I} \rangle),~\text{for}~k=3, 4, \dots,
    \end{split}
\end{equation}
It can be easily verified that higher order perturbative corrections yield higher order terms in $\langle \Psi_{1,I} | \hat{x}_I^n | \Psi_{1,I} \rangle$ and are consequently negligible according to our arguments.
Therefore, we can verify that the entanglement even in the case of an impurity density with very small non-zero width is finite, as the von Neumann entropy reads
\begin{equation}
\begin{split}
S_{VN} =& \langle \Psi_{1,I} | \hat{x}_I^2 | \Psi_{1,I} \rangle \sum_{l = 1}^{\infty} |A_{1 l}(0)|^2 \\
& \times \left[ 1 - \log \left( \langle \Psi_{1,I} | \hat{x}_I^2 | \Psi_{1,I} \rangle \sum_{l = 1}^{\infty} |A_{1 l}(0)|^2 \right) \right] \\
& + \mathcal{O} (\langle \Psi_{1,I} | \hat{x}_I^4 | \Psi_{1,I} \rangle).
\end{split}
\end{equation}

Let us now comment on the above results. We see see that there are only two parameters that enter the von Neumann entropy for a narrow impurity density distribution, these are its width in terms of $\langle \Psi_{1,I} | \hat{x}_I^2 | \Psi_{1,I} \rangle$ and the sum of non-adiabatic couplings $\sum_{l = 1}^{\infty} |A_{1 l}(0)|^2$. Notice that the latter quantity is related to the norm of the derivative of $| \Psi_{1, B} (x_I) \rangle$, namely $|| \frac{\partial}{\partial x_I} \left| \Psi_{1,B} (x_I) \right\rangle|| = \sqrt{\sum_{l = 1}^{\infty} |A_{1 l}(x_I)|^2}$. Therefore, we can conclude that $S_{VN}$ increases when the impurity width increases, see also Fig.~\ref{fig:variation_of_mass_trapping_frequency}(b), and when the bath state becomes more strongly dependent on $x_I$, i.e. for higher $g$, see Fig.~\ref{fig:ground-state_energy_BO_vs_BH}(b). The latter can be independently verified by considering the Feynman-Hellmann theorem
\begin{equation}
    A_{jk}(x_I) =\frac{\langle\Psi_{j, B} (x_I) | \frac{\partial \hat{H}_{BI}}{\partial x_I} | \Psi_{k, B} (x_I)\rangle}{\varepsilon_j(x_I) - \varepsilon_k(x_I)}.
\end{equation}
This shows that the non-adiabatic couplings should increase with interaction since, first, the nominator is $\propto g$ and, second, the energy differences in the denominator decrease due to the closing of the gaps among the even and odd single particle bath eigenstates as the $g \to \infty$ limit is approached, see also Appendix~\ref{appendix-jahn-teller}.

\section{Convergence behaviour of ML-MCTDHX versus TBH}
\label{sec:convergence}

\begin{table*}
\begin{tabular}{|c|c|c|c|c|c|c|c|c|}
\hline
&\multicolumn{3}{|c|}{ML-MCTDHX} & \multicolumn{2}{|c|}{Energy-pruned CI} & \multicolumn{3}{|c|}{multi-channel BO} \\
\hline
&\shortstack{$D=d^I$=10 \\and $d^B$=10}& \shortstack{$D=d^I$=10 \\and $d^B$=15}& \shortstack{$D=d^I$=12 \\and $d^B$=18} & $E_{\rm cut} =$ 50 & $E_{\rm cut} =$ 70 & 10 pec & 24 pec & 40 pec 
\\
\hline
$E_{\rm imp}(g=1.0)$ & 15.37 & 15.37 & 15.37 & 15.41 & 15.40 & 15.38 & 15.37 & 15.37  \\
\hline
$S_{\rm VN}(g=1.0)\times 10^{-3}$ & 9.74 & 9.70 & 9.17 & 8.90 & 9.42 & 18.99 & 15.61 & 14.31\\
\hline
$E_{\rm imp}(g=2.0)$ & 16.01 & 15.98 & 15.98 & 16.10 & 16.06 & 16.00 & 15.99 & 15.98  \\
\hline
$S_{\rm VN}(g=2.0)\times 10^{-3}$ & 26.00 & 25.87 & 24.79 & 25.39 & 26.27 & 46.93 & 39.26 & 36.35\\
\hline
$E_{\rm imp}(g=5.0)$ & 17.00 & 16.93 & 16.92 & 17.20 & 17.13 & 16.98 & 16.92 & 16.92  \\
\hline
$S_{\rm VN}(g=5.0)\times 10^{-3}$ & 62.52 & 62.28 & 61.71 & 64.93 & 65.67 & 101.07 & 87.29 & 82.04\\
\hline
\end{tabular}
\caption{Impurity energy, $E_{\rm imp}$, and von Neumann entropy, $S_{VN}$, for different interaction strength, $g$, within three different numerical approaches. The dependence of these quantities on the parameters controlling the accuracy of the employed methods (see text) is also provided for estimating the degree of their convergence to the exact limit.}
\label{tab:convergence_check}
\end{table*}

To elucidate the degree of convergence of the employed ab-initio variational approaches namely the ML-MCTDHX and the multi-channel BO methods, we present here a comparative analysis of the dependence of their results in the corresponding parameters determining their numerical accuracy.

As discussed in Sec.~\ref{MLX_main_text} the parameters that define the multi-layered truncation of the many-body wavefunction are given by the orbital configuration $C = (D; d^B; d^I)$ and the choice of the primitive basis and its size $\mathcal{M}$. First, since the primitive basis size hardly affects the CPU time of the ML-MCTDHX calculations we have selected a large enough basis of $\mathcal{M} = 150$ grid points of the harmonic oscillator DVR with $\omega_{\rm DVR} = 0.72~\omega_B$, which is enough for convergence for the employed interaction scales. Second, due to the fact that we only consider a single impurity we can simplify the ansatz to $C = (D; d^B; D)$ since at least $D$ basis states for the impurity species are required to give rise to $D$ distinct Schmidt modes, see Eq.~\eqref{eqn:Schmidt_Composition_MLX}. Thus below we analyze the convergence of the ML-MCTDHX approach only in terms of the truncation in terms of entanglement modes, controlled by $D$ and the truncation in terms of bath orbitals $d^B$. 

\textcolor{black}{For our chosen configuration $C=(12, 18, 12)$, we employ $D = 12$ basis states for each species in the top layer (yielding $D^2 = 144$ coefficients). In the middle layer, the many-body basis of the bath species is constructed in terms of $d^B = 18$ single-particle states ($D \times \binom{d^B}{N_B} = 102816$ coefficients for the many-body basis, $d^B \times \mathcal{M}= 2700$ for the single-particle states). Since we consider a single impurity, the expansion in terms of $d^I = 12$ single-particle states yields $d^I \times \mathcal{M} = 1800$ coefficients. Therefore, it is evident that dynamically updating the operators acting on this large amount of coefficients, especially for the middle layer of the bath species, is computationally challenging. More specifically the most demanding of our calculations referring to an interaction strength $g = 5$, employing the above mentioned basis size, took approximately three months of computational time of an Intel Xeon X5650 CPU.}

Regarding the multi-channel BO method, the main parameter that controls the truncation is the choice of $M$ in Eq.~\eqref{eqn:multi-channel_BornOppenheimer}, which corresponds to the number of potential energy curves in Eq.~\eqref{eqn:effective_Schroendinger}. Additionally, the quality of the calculated non-adiabatic derivative couplings $A_{kl}(x_I)$ and potential renormalization $V^{\rm ren}_{k l}(x_I)$ is dictated by the choice of the primitive basis. Here we have chosen an exponential DVR with $\mathcal{M} = 1024$ points permitting derivative evaluations by the fast Fourier transform. By detailed analysis of the corresponding $A_{kl}(x_I)$ and $V^{\rm ren}_{k l}(x_I)$ matrices we have deemed that this choice is adequate for the convergence of the corresponding matrices.

Finally, we compare the results of both variational methods with the Configuration Interaction (CI) (or exact diagonalization) approach with energy pruning, that recently has attracted considerable attention in the few-fermion literature~\cite{RammelmuellerHuber2023a, RammelmuellerHuber2023b, SowinskiGarciaMarch2019}. 
Within this approach we generate the set of all number states with non-interacting energy less than a given energy cutoff, $E_{\rm cut}$ and then diagonalize the Hamiltonian, Eq.~\eqref{eqn:Hamilt_Full}, in the subspace spanned by them. In our case, the Hamiltonian in the space spanned by the non-interacting harmonic oscillator functions, $\psi_{n,\sigma}(x)$, reads
\begin{equation}
\begin{split}
    \hat{H} =&  \sum_{n = 0}^{\infty} \hbar \omega_B \bigg( n + \frac{1}{2} \bigg) \hat{c}^{\dagger}_{n, B} \hat{c}_{n, B} \\
    &+\sum_{n = 0}^{\infty} \hbar \omega_I \bigg( n + \frac{1}{2} \bigg) \hat{c}^{\dagger}_{n, I} \hat{c}_{n, I} \\
    &+\sum_{n,l,m,k = 0}^{\infty} U_{n l m k} \hat{c}^{\dagger}_{n, B} \hat{c}^{\dagger}_{l, I} \hat{c}_{m, I} \hat{c}_{k, B},
\end{split}
\label{hamiltonianCI}
\end{equation}
where $\hat{c}^{\dagger}_{n,\sigma}$ and $\hat{c}_{n,\sigma}$ are the operators that create and annihilate a species $\sigma$ particle in the $n$-th single particle eigenstate respectively and $U_{n l m k} = \int {\rm d}x~\psi^*_{n,B}(x)\psi^*_{l,I}(x)\psi_{m,I}(x)\psi_{k,B}(x)$ which can be calculated efficiently via employing the Gaussian quadrature.
As the basis of the many-body subspace we use the number states $| n_1^B, n_2^B, \dots, n^N_B; n^I \rangle$, that satisfy 
\begin{equation}
\begin{split}
    &\langle n_1^B, n_2^B, \dots, n^N_B; n^I | \hat{H} | n_1^B, n_2^B, \dots, n^N_B; n^I \rangle = \\
    &\hspace{0.2cm}\bigg( \frac{N}{2} + \sum_{j = 1}^N n_j^B \bigg) \hbar \omega_B + \bigg(n^I + \frac{1}{2} \bigg) \hbar \omega_I < E_{\rm cut}.
\end{split}
\end{equation}
Notice that this choice implies that there is a maximum value of $n_i^B$ and $n^I$ that appears in this subspace and as a consequence we have to calculate a finite number of $U_{nlmk}$ elements.
The only non-diagonal terms in this basis correspond to the interaction terms of Eq.~\eqref{hamiltonianCI}. The matrix elements of $\hat{c}^{\dagger}_{n, B} \hat{c}^{\dagger}_{l, I} \hat{c}_{m, I} \hat{c}_{k, B}$ can be evaluated by using the indexing rules of fermionic states described in Ref.~\cite{StreltsovAlon2010}.
From the above it is clear that the only approximation the energy pruned CI uses is the value of the energy cutoff $E_{\rm cut}$ that determines the size of the corresponding subspace where the Hamiltonian of Eq.~\eqref{eqn:Hamilt_Full} is diagonalized.
\begin{figure*}
    \centering
    \includegraphics[width=0.9\textwidth]{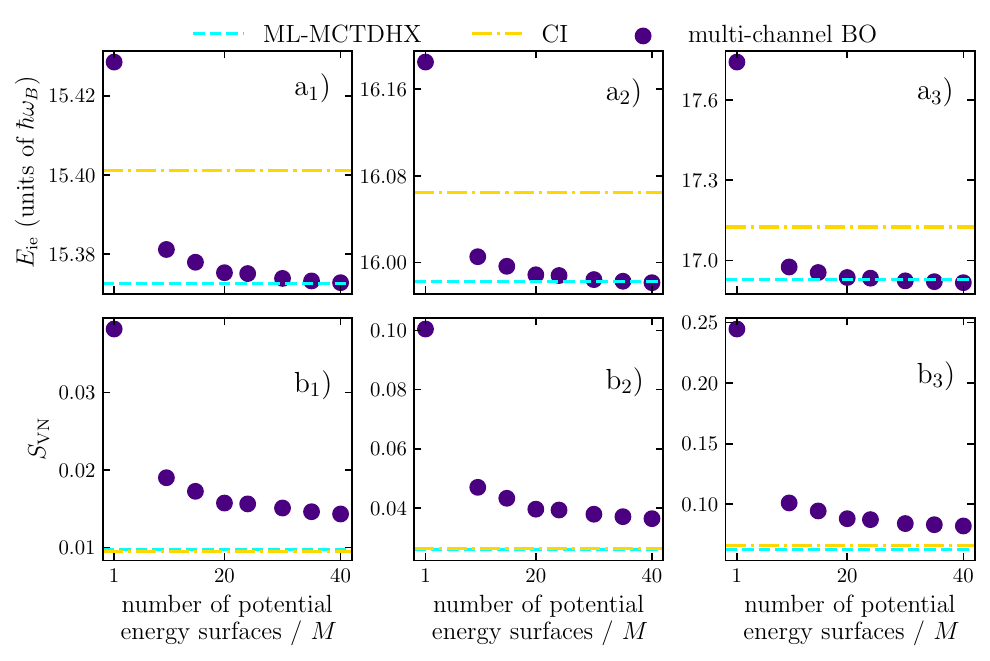}
    \caption{(a$_i$) Impurity interaction energy and (b$_i$) von Neumann entropy as a function of the number of potential energy curves, $M$, within the multi-channel BO approach for different interaction strengths $g = g_i$, with $g_1 = 1$, $g_1 = 2$ and $g_1 = 5$. The dashed lines provide the corresponding results within the ML-MCTDHX and Energy-Pruned Configuration Interaction (CI) approaches with the highest accuracy. This refers to $D =d^I = 12$ and $d^B = 18$ within ML-MCTDHX and $E_{\rm cut}=70\hbar \omega_B$ within the energy pruned configuration interaction approach.}
    \label{fig:Convergence_Check}
\end{figure*}

To estimate the convergence pattern of the multi-channel BO we compare how the energy and von Neumann entropy converge to the exact value as the number of potential energy curves increases. We observe that already for $M=2$ the multi-channel BO approach possesses a significantly lower energy than the energy pruned CI with $E_{\rm cut} = 70 \hbar \omega_B$ and on par with ML-MCTDHX with orbital configuration $C = (10; 10; 10)$, see Table~\ref{tab:convergence_check}, for all interaction strengths we have studied, see Fig.~\ref{fig:Convergence_Check} $(a_i)$ with $i=1,2,3$.
Energy convergence is observed for $M > 24$, where it shows almost the same value of energy and ML-MCTDHX with orbital configuration $C = (12, 18, 12)$.
In contrast, even the less accurate versions of the ML-MCTDHX and CI possess a von Neumann entropy much closer to their converged results than the corresponding multi-channel BO result for $M = 2$ see Table~\ref{tab:convergence_check} and Fig.~\ref{fig:Convergence_Check} $(b_i)$ with $i=1,2,3$.
In particular, we observe that multi-channel BO does not show a von Neumann entropy convergence with the value of $S_{VN}$ improving only by a factor of roughly two with respect to the other two approaches even for the largest $M = 40$ value we have studied.

Therefore, we rely on ML-MCTDHF for our exact numerical results. Nevertheless, this comparison confirms the accuracy and convergence of both used methods against the same result for increasing accuracy control parameters.

\section{Proof of Theorem~\ref{theorem}}
\label{proof_theorem}

%Here, we work out the proof of the theorem \ref{theorem} from... .
%\begin{proof}
%\textcolor{black}{[Additional points should clarify the open statements. However, this might be too detailed. It might make sense to skip some parts or put them in the appendix. We can discuss this.]}
The outline of the proof of Theorem~\ref{theorem} is as follows. First we generate a complete many-body basis for the $(N_B + 1)$--body system. Then we show that the parameter $\Delta N$ characterizing this basis is a good quantum number for the Hamiltonian of Eq.~\eqref{H0r}. Finally, by using the parity symmetry property of Eq.~\eqref{H0r} we demonstrate that all eigenstates with $\Delta N \neq 0$ are necessarily degenerate with at least one eigenstate with the opposite sign of $\Delta N$. This proves Theorem~\ref{theorem} for $N_B$ odd since in this case $\Delta N = 0$ is impossible.

We begin the proof of Theorem~\ref{theorem} by generating a complete many-body basis for the $(N_B + 1)$--body system.
For $g \to \infty$ the single-particle system defined by Eq.~\eqref{H0r} for $N_B = 1$, splits into two subsystems referring to $r > 0$ and $r < 0$. These are separated by an impenetrable wall at $r = 0$, where the wavefunctions have to vanish. Therefore, the single-particle eigenstates of the system can be expressed in terms of the eigenstates of the individual subsystems, $\psi_{j L}(r) \neq 0$ for $r < 0$ and $\psi_{j R}(r) \neq 0$ for $r > 0$ with $j = 0, 1, \dots$. Subsequently, the Hamiltonian of the many-body system can be expanded in terms of the number-states (Slater determinants) spanned by $\psi_{j L}(r)$ and $\psi_{j R}(r)$. Each of these many-body states is characterized by a definite value of the particle imbalance among the subsystems, $\Delta N = N_L - N_R$, where $N_L$ and $N_R$ are the number of particles in the left and right subsystem respectively. Of course, notice that $N_L + N_R = N_B$ holds.

Then we continue by showing that $\Delta N$ is a good quantum number for the Hamiltonian of Eq.~\eqref{H0r}. It can be proven that for any two Slater determinants, $|\Psi_k \rangle$, involving the contribution of the single-particle states $I^k_m \in (\mathbb{N}, \{L, R\})$, with $m = 1, 2, \dots, N_B$, the derivative interaction term reads
\begin{widetext}
\begin{equation}
\begin{split}
    \langle \Psi_k | 
    %-\frac{\hbar^2}{2 m_I} \sum_{j = 1}^{N_B} \sum_{{j' = 1}\atop{j' \neq j}}^{N_B} \frac{\partial^2}{\partial r_j \partial r_{j'}}
    \hat{H}_{P_{\rm CM}}
    | \Psi_{k'} \rangle =& 
    -\frac{\hbar^2}{m_I} \sum_{n = 1}^N \sum_{l = 1}^N \sum_{m = 1}^{n-1}  \sum_{r = 1}^{l-1} (-1)^{n + m + l + r}  \\
    & \times \bigg( 
    \left\langle \psi_{I^k_l} \left| \frac{\partial}{\partial r} \right| \psi_{I^{k'}_n} \right\rangle 
    \left\langle \psi_{I^k_r} \left| \frac{\partial}{\partial r} \right| \psi_{I^{k'}_m} \right\rangle  \\
    &~~~~~~ - \left\langle \psi_{I^k_r} \left| \frac{\partial}{\partial r} \right| \psi_{I^{k'}_n} \right\rangle 
        \left\langle \psi_{I^k_l} \left| \frac{\partial}{\partial r} \right| \psi_{I^{k'}_m} \right\rangle \bigg) \\
    & \times \delta_{% 
        \{I_1^k, \dots, I_{r-1}^k, I_{r+1}^k, \dots, I_{l-1}^k, I_{l+1}^k, \dots, I_{N}^k\} 
        \{I_1^{k'}, \dots, I_{m
        -1}^{k'}, I_{m+1}^{k'}, \dots, I_{n-1}^{k'}, I_{n+1}^{k'}, \dots, I_{N}^{k'} \}} \\
    &-\frac{\hbar^2}{m_I} \sum_{n = 1}^N \sum_{l = 1}^N (-1)^{n + m} 
       \left\langle \psi_{I^k_m} \left| \frac{\partial^2}{\partial r^2} \right| \psi_{I^{k'}_n} \right\rangle \\
    & \times \delta_{% 
        \{I_1^k, \dots, I_{m-1}^k, I_{m+1}^k, \dots, I_{N}^k\} 
        \{I_1^{k'}, \dots, I_{n-1}^{k'}, I_{n+1}^{k'}, \dots, I_{N}^{k'} \}},
\end{split}
\label{matrix_elements_interaction}
\end{equation}
\end{widetext}
where the Kronecker delta symbol $\delta_{S_1 S_2}$ yields zero for two different sets $S_1$ and $S_2$, and one in the case that they are equivalent. Notice that since $\psi_{j L}(r)$ and $\psi_{j R}(r)$ are non-vanishing in entirely separated spatial domains $\langle \psi_{j L} | \frac{\partial}{\partial r} |\psi_{k R} \rangle = 0$ holds for all $j$ and $k$. This fact can also be verified independently by evaluating the limits of the analytical solutions for finite $g$ in the case $g \to \infty$, see Appendix~\ref{appendix-jahn-teller}. Therefore, the quantity inside the parenthesis of Eq.~\eqref{matrix_elements_interaction} vanishes if there is a different number of $L$ and $R$ states among the $\{\psi_{I^k_l}(r), \psi_{I^k_r}(r)\}$ and  $\{\psi_{I^{k'}_n}(r), \psi_{I^{k'}_m}(r)\}$ sets. This implies that if $\Delta N$ is different for $| \Psi_k \rangle$ and $| \Psi_{k'} \rangle$ then the corresponding interaction matrix element, Eq.~\eqref{matrix_elements_interaction}, is zero. Thus a given number-state couples only with number states with the same value of $\Delta N$ and consequently $\Delta N$ is a good quantum number. This implies that eigenstates of $\hat{H}_{0r} + \hat{H}_{P_{\rm CM}}$ can be characterized in terms of $\Delta N$.

Then by considering the symmetry properties of Eq.~\eqref{H0r} Theorem~\ref{theorem} can be explicitly proven. Since, $\hat{H}_{0r}$ is parity symmetric $[\hat{H}_{0r}, \hat{\mathcal{P}}_r] = 0$ holds, where $\hat{\mathcal{P}}_r r_i = - r_i$. Let us assume an eigenstate $( \hat{H}_{0r} + \hat{H}_{P_{\rm CM}} ) | \tilde{\Psi}_k \rangle = E_k | \tilde{\Psi}_k \rangle$ with definite value of $\Delta N = \Delta N_k$. Due to the fact that $\hat{\mathcal{P}}_r \psi_{j L}(r) = \psi_{j R}(r)$ (for an appropriate choice of the overall phases of the involved single-particle states), the action of $\hat{\mathcal{P}}_r$ on $| \tilde{\Psi}_k \rangle$ results in the shift of particle imbalance $\Delta N_k \to - \Delta N_k$. This shows that for $\Delta N_k \neq 0$ the eigenstates $| \tilde{\Psi}_k \rangle$ and $\hat{\mathcal{P}}_r | \tilde{\Psi}_k \rangle$ are distinct and degenerate. Therefore, owing to the fact that $\Delta N = 0$ can only hold for even $N_B$, the ground state of $\hat{H}_{0r}$ is {\it always} degenerate for odd $N_B$ and in the $g \to \infty$ limit independently of all other system parameters, which proves the theorem. \textcolor{black}{In the case of even $N_B$ the above would hold only if the ground state possesses $\Delta N_{k} \neq 0$, but since in this situation $\Delta N_k = 0$ is possible, a degeneracy does not necessarily occur.}
%\end{proof}
    
\section{Single-particle properties of $\hat{H}_{0r}$}
\label{appendix-jahn-teller}

The symmetry properties of $\hat{H}_{0r}$, Eq.~\eqref{H0r}, were extensively analyzed in Sec.~\ref{pjte-general} and Appendix~\ref{proof_theorem} where several theoretical insights were obtained without having to consider the precise form of its eigenspectrum. The purpose of this Appendix is to review the basic properties of the analytically-obtained single-particle ($N_B=1$) eigenspectrum of $\hat{H}_{0r}$\textcolor{black}{, which will be used in Appendix~\ref{conical_appendix} to illustrate the emergence of a conical intersection in the vicinity of $1/g=x_I=0$.}

The eigenfuctions of $\hat{H}_{0r}$ for $N_B=1$ read~\cite{BuschEnglert1998, BudewigMistakidis2019}
\begin{equation}
\begin{split}
        \psi^{\rm B}_{2 n}(r;g) =& \frac{A_{n}(g) \Gamma(-\epsilon_n(g))}{2 \sqrt{\pi \ell_{\rm B}}} U\left( -\epsilon_n(g), \frac{1}{2},\frac{r^2}{\ell_{\rm B}^2} \right) e^{-\frac{r^2}{2 \ell_{\rm B}^2}}, \\
    \psi^{\rm B}_{2 n + 1}(r;g) =&\psi^{\rm B}_{2 n + 1}(r) = \\
    &\frac{(\pi l_B^2)^{-1/2}}{\sqrt{2^{2 n+1} (2 n + 1)!}} H_{2 n+1}\left( \frac{r}{\ell_B} \right) e^{-\frac{r^2}{2 \ell_{\rm B}^2}},
\end{split}
\end{equation}
where $H_n(x)$ denotes the $n$-th degree Hermite polynomial, $\Gamma(x)$ is the gamma and $U(\alpha, \beta, x)$ is the confluent hypergeometric function. The relevant length scale is $\ell_{\rm B} = \sqrt{\hbar/(m_B \omega_B)}$. The normalization factor of parity-even states reads
\begin{equation}
\begin{split}
    A_{n}(g) =& 2 \sqrt{\frac{\Gamma\left( \frac{1}{2} - \epsilon_n(g) \right)}{\Gamma(-\epsilon_n(g))}
     \frac{1}{\psi( \frac{1}{2} - \epsilon_n(g) ) - \psi(-\epsilon_n(g))}},
\end{split}
\label{eqn:A_k}
\end{equation}
where $\psi(x)$ is the digamma function.
Finally, the effective order $\epsilon_n(g)$ satisfies the consistency equation
\begin{equation}
    \frac{\Gamma\left( \frac{1}{2} - \epsilon_n(g) \right)}{\Gamma(-\epsilon_n(g))} = - \frac{g}{2 g_0},
\end{equation}
with $g_0 = \sqrt{\frac{\hbar^3 \omega_B}{m_B}}$.
This self consistency equation shows that $\frac{n}{2} \leq \epsilon_n(g) \leq \frac{n+1}{2}$ for $n \geq 1$ and $-\infty < \epsilon_0(g) \leq 1/2$. The upper bound of these inequalities gets saturated for $g \to + \infty$ while the lower saturates for $g \to - \infty$. The energy of the system is a function of the effective order reads $E_{2 n}^{\rm B}(g) = \hbar \omega_r \left(2 \epsilon_n(g) + 1/2 \right)$ for parity-even states, while $E^B_{2n+1}(g) = E^B_{2n+1}= \hbar \omega_r \left(2 n + 1 +1/2 \right)$ for parity odd states.

The position, $R_{n,m}(g) \equiv \langle \psi^B_{n}(g) | \hat{x} |\psi^B_{m}(g) \rangle$, and momentum, $P_{n,m}(g) \equiv \langle \psi^B_{n}(g) | \hat{p} |\psi^B_{m}(g) \rangle$, single-particle matrix elements are non-vanishing only in the case that states of different parity are involved. The non-zero elements of these matrices read
\begin{equation}
\begin{split}
    R_{2\lambda + 1,2\kappa}(g) =& \frac{(-1)^\lambda \ell_B A_{\kappa}(g)}{\sqrt{2} \pi^{1/2}} \frac{\sqrt{(2\lambda +1)!}}{2^{\lambda + 1} \lambda!}\\ 
    &\times \frac{1}{(\lambda - \epsilon_{\kappa}(g))(\lambda + 1 - \epsilon_{\kappa}(g))}, \\
    P_{2\lambda + 1,2\kappa}(g) =& \frac{i (-1)^\lambda \hbar A_{\kappa}(g)}{\sqrt{2} \ell_B \pi^{1/2}} \frac{\sqrt{(2\lambda +1)!}}{2^{\lambda + 1} \lambda!}\\ 
    &\times \frac{2 \lambda + 1 - 2 \epsilon_{\kappa}(g)}{(\lambda - \epsilon_{\kappa}(g))(\lambda + 1 - \epsilon_{\kappa}(g))},
\end{split}
\label{matrix_elements_normal_Busch}
\end{equation}
for $\lambda,\kappa = 0, 1, \dots$. The final relevant for us property of $| \psi_{n}^{\rm B}(g) \rangle$ are their transformation properties for a shift of $g \to g'$ unveiled in Ref.~\cite{BudewigMistakidis2019}
\begin{equation}
\begin{split}
    | \psi_{2n}^{\rm B}(g') \rangle &= \sum_{m=0}^{\infty} \frac{A_n(g') A_m(g)}{E_{2m}^{\rm B}(g)-E_{2n}^{\rm B}(g')} \left(\frac{1}{g'} - \frac{1}{g} \right) | \psi_{2m}^{\rm B}(g) \rangle, \\
    | \psi_{2n+1}^{\rm B}(g') \rangle &= | \psi_{2n+1}^{\rm B}(g) \rangle.
\end{split}
\label{transformation_budewig}
\end{equation}

\textcolor{black}{Let us briefly focus on the regime of $g \to \infty$, which is particularly interesting for discussing the pseudo Jahn-Teller effect. In this case we can obtain an analytic asymptotic expression for the effective order that reads
\begin{equation}
   \epsilon_{n + s}(g) = \frac{2 n + 1}{2} - \mathcal{E}_n \frac{g_0}{g} + \mathcal{O} \left( \frac{g_0^2}{g^2} \right), 
   \label{infinite_energies}
\end{equation}
where $\mathcal{E}_n = [2(n+1)]!/[2^{2n+1} n! (n+1)! \sqrt{\pi}]$ and $s = 0$ for $g>0$, $s = 1$ for $g<0$. This shift in the index accounts for the continuous transformation of the $\psi^{B}_{2n}(x)$ state to the $\psi^{B}_{2(n+1)}(x)$ one as the $g \to \infty$ (Tonk-Girardeau~\cite{Girardeau1960}) limit is crossed from strong repulsive to attractive interactions. Notice that we do not provide an expression for $\epsilon_0$ as $g \to -\infty$ since in this case the bound state energy diverges since $\epsilon_0 \to -\infty$.
Finally, by using Eq.~\eqref{transformation_budewig} we can show that
\begin{equation}
\begin{split}
   \psi_{2(n+\sigma)}^B(r;g) =& \psi^{B}_{2 n + 1}(|r|) \\
   &+ \frac{g_0}{g} \sum_{{m = 0} \atop {m\neq n}}^{\infty} \frac{\sqrt{\mathcal{E}_n \mathcal{E}_m}}{m - n} \psi^{B}_{2 m + 1}(|r|) \\
   &+ \mathcal{O}\left( \frac{g_0^2}{g^2} \right).
\end{split}
\label{wavefunction_expansion_infinite}
\end{equation}}

The description of the behavior of the system for strong interactions can be greatly simplified by transforming to a basis where the fermions are localized as much as possible on the left or right side of the $x=0$ barrier. This can be achieved by the following unitary transformation
\begin{equation}
\begin{split}
    \psi_{\kappa L}(r;g) &= - \frac{\psi^B_{2 \kappa + 1}(r;g) + (-1)^\kappa \psi^B_{2 \kappa}(r;g)}{\sqrt{2}}, \\
    \psi_{\kappa R}(r;g) &=   \frac{\psi^B_{2 \kappa + 1}(r;g) - (-1)^\kappa \psi^B_{2 \kappa}(r;g)}{\sqrt{2}}.
\end{split}
\end{equation}
By the use of this basis and Eq.~\eqref{matrix_elements_normal_Busch} we can verify several important properties of these maximally localized states which will be elucidated further in Appendix~\ref{conical_appendix}, see Eq.~\eqref{relevant_matrix_elements}.

\section{Synthetic conical intersection at $1/g=x_I=0$}
\label{conical_appendix}

\begin{figure}
    \centering
    \includegraphics[width=1.0\columnwidth]{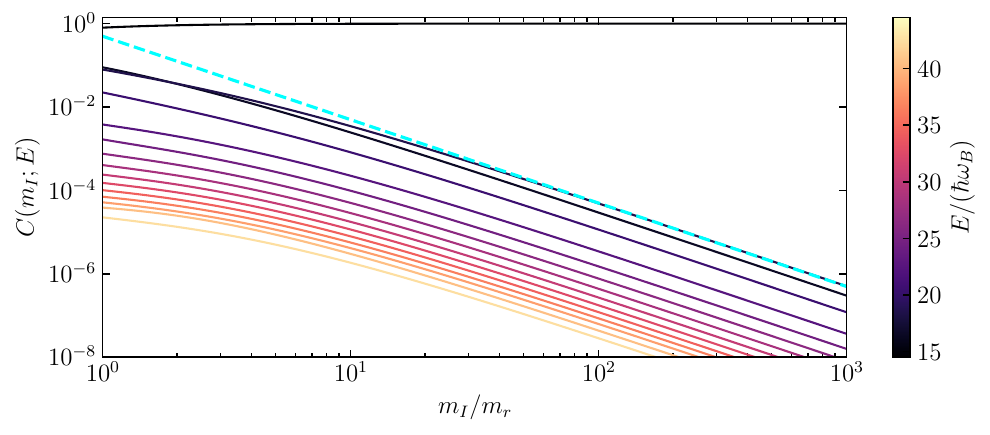}
    \caption{The contribution, $C(m_I;E)$, of the $m_I \to \infty$ eigenstates, $| \Psi_k \rangle$, of a given energetic class characterized by $E$, to the finite $m_I$ ground state $| \tilde{\Psi}_0(m_I) \rangle$ for varying $m_I/m_B$. In all cases $N_B = 5$ and $1/g = x_I =0$ is considered. Notice the logarithmic scaling of both axes, the dashed line corresponds to $\frac{ m_B^2}{2m_I^2}$ and is provided as a guide to the eye.}
    \label{fig:pop-exc-states}
\end{figure}

\textcolor{black}{
Based on the analytic properties of the $N_B = 1$ eigenstates of $\hat{H}_{0r}$ (see Appendix~\ref{appendix-jahn-teller}), characterizing the system at $x_I = 1/g =0$, and by employing perturbation theory we can demonstrate that our system maps to a $E \otimes \epsilon$ in the vicinity of $x_I = 1/g =0$ \footnote{We denote here the $E\otimes\epsilon$ case, since we are considering here both $x_I$ and $1/g$ as synthetic coordinates \cite{Bersuker2013}. For a fixed $1/g$ the system reduces to the $E\otimes b$ model as in Sec. \ref{pjte-potential-energy-surfaces}.}. Such a proof is convoluted in the case that $m_I$ is finite. However, in the infinite impurity mass case the physical situation is substantially simplified. Then by simple numerical arguments we can demonstrate that the finite $m_I$ case behaves similarly to $m_I \to \infty$ provided that $m_I > m_B$.}

\textcolor{black}{The main simplification for $m_I \to \infty$ is that $\hat{H}_{P_{\rm CM}}$ vanishes and as a consequence the eigenstates for $g \to \infty$ can be expressed as a single Slater-determinant constructed from the $\psi_{\kappa L}(r;g \to \infty)$ and $\psi_{\kappa R}(r;g \to \infty)$ states.
In particular, the degenerate ground states at $g \to \infty$, $| \Psi_L \rangle$ and $| \Psi_R \rangle$ (guaranteed to exist for odd $N_B$ owing to Theorem~\ref{theorem}), are characterized by the occupation numbers}
\begin{equation}
\begin{split}
     I^L_j =  
    \begin{cases*}
    (j, L) & for $j \leq \frac{N_B+1}{2}$, \\
    (j - \frac{N_B+3}{2}, R) & for $\frac{N_B+1}{2} < j \leq N_B$,
    \end{cases*} \\
    I^R_j =  
    \begin{cases*}
    (j, L) & for $j \leq \frac{N_B-1}{2}$, \\
    (j - \frac{N_B+1}{2}, R) & for $\frac{N_B-1}{2} < j \leq N_B$.
    \end{cases*}
\end{split}
\end{equation}
\textcolor{black}{Therefore, within first-order perturbation theory we just need to evaluate the matrix elements among these states, since contributions outside this degenerate manifold are at least second order in perturbation theory. It can be easily verified that couplings among the above states can be induced by $\hat{H}_{\rm coup}$ and $\hat{H}_{0r}$. The relevant matrix elements among the localized single-particle states up to linear order in $1/g$ read}
\begin{widetext}
\begin{equation}
\begin{split}
    \langle \psi_{n L} | \hat{H}_{0r} | \psi_{n L} \rangle &= \langle \psi_{n R} | \hat{H}_{0r} | \psi_{n R} \rangle \\
    &=\hbar \omega_B \left( 2 n + \frac{3}{2} \right) - \frac{1}{2} \hbar \omega_B \mathcal{E}_n \frac{g_0}{g} + \mathcal{O} \left( \frac{g_0^2}{g^2} \right),\\
    \langle \psi_{n L} | \hat{H}_{0r} | \psi_{n R} \rangle & \equiv t_n = 
    - \frac{1}{2} \hbar \omega_B \mathcal{E}_n \frac{g_0}{g} + \mathcal{O} \left( \frac{g_0^2}{g^2} \right),\\
    \langle \psi_{n L} | \hat{x} | \psi_{n L} \rangle &= - \langle \psi_{n R} | \hat{x} | \psi_{n R} \rangle \\ 
    &= -\ell \left( \mathcal{X}^2_n + \frac{g_0}{g} \sum_{{m =0} \atop {m \neq n}}^{\infty} \frac{\sqrt{\mathcal{E}_n \mathcal{E}_m}}{2(n-m)} \frac{\sqrt{\mathcal{X}_n \mathcal{X}_m}}{(n-m)^2 - \frac{1}{4}} \right) + \mathcal{O}\left( \frac{g_0^2}{g^2} \right)\\
    \langle \psi_{n L} | \hat{x} | \psi_{n R} \rangle &= \langle \psi_{n L} | \hat{p} | \psi_{n L} \rangle = \langle \psi_{n L} | \hat{p} | \psi_{n R} \rangle = \langle \psi_{n R} | \hat{p} | \psi_{n R} \rangle = \mathcal{O} \left( \frac{g_0^2}{g^2} \right), \\ 
\end{split}
\label{relevant_matrix_elements}
\end{equation}
\end{widetext}
\textcolor{black}{with $\mathcal{X}_n = \sqrt{(2n+1)!}/(2^{n-1}n! \sqrt{\pi})$.
Hence, the perturbative Hamiltonian for the system reads}
\begin{equation}
    \hat{H}_{\rm per} = E_0 -\delta E \frac{g_0}{g} + J \hat{\sigma}_x \frac{g_0}{g}  + \Delta \hat{\sigma}_z \hat{x}_I + \mathcal{O}\left( \frac{g_0^2}{g^2} \right),
    \label{near_infinity_effective_hamiltonian}
\end{equation}
\textcolor{black}{where $E_0 = \hbar \omega_B (N_B + N_B +1)/2$, $\delta E = \frac{2^{-(N_B+1)}}{3 \sqrt{\pi}} \frac{(2N_B +1)!(N_B-1)!}{\left( \frac{N_B-1}{2} \right)! \left( \frac{N_B+1}{2} \right)!}\hbar \omega_B$, $J = -\frac{1}{2} \hbar \omega_B \mathcal{E}_{\frac{N_B+1}{2}}$ and $\Delta= - \ell \mathcal{X}^2_{\frac{N_B+1}{2}}$. In addition, we have mapped $| \Psi_L \rangle$ to the pseudo-spin-$\uparrow$ and $| \Psi_R \rangle$ to the pseudo-spin-$\downarrow$ states, with the Pauli matrices $\hat{\sigma}_{\mu}$, $\mu \in \{x, y, z\}$ acting in the standard way in the corresponding pseudo-spin-$1/2$ space.
Notice that the $| \Psi_L \rangle$ and $| \Psi_R \rangle$ states define a well-behaved pseudo-spin-$1/2$ subspace, in particular each state of the corresponding Bloch-sphere, $| \theta, \phi \rangle$, reads}
\begin{widetext}
\begin{equation}
\begin{split}
   \langle x_1, \dots, x_{N_B} | \theta, \phi \rangle &= \frac{1}{\sqrt{N_B!}} \sum_{j = 1}^{N!} {\rm sign}\left( P_j \right) \\
       &\times \psi_{0 L}(x_{P_j(1)}) \psi_{0 R}(x_{P_j(2)}) \dots \times \psi_{\frac{N_B-1}{2} L}(x_{P_j(N_B-2)}) \psi_{\frac{N_B-1}{2} R}(x_{P_j(N_B-1)}) \\
       &\times \left[ \cos \theta~\psi_{\frac{N_B+1}{2} L}(x_{P_j(N_B)}) + e^{i \phi} \sin \theta~\psi_{\frac{N_B+1}{2} R}(x_{P_j(N_B)}) \right].
\end{split}
\label{Bloch_sphere}
\end{equation}
\end{widetext}

\textcolor{black}{The above indicates that the system exhibits an $E \otimes \epsilon$ conical intersection at $g_0/g = 0$ and $x_I = 0$. All the related symmetry requirements are satisfied, since, first, the degenerate $| \Psi_L \rangle$ and $| \Psi_R \rangle$ states give rise to a two-dimensional representation of the SU(2) symmetry, see Eq.~\eqref{Bloch_sphere}. Second, the two vibrational coordinates $x_I$ and $g_0/g$ couple to this subspace so that to favor different superpositions of the degenerate states.}

\textcolor{black}{In the case that $m_I$ is finite, the degenerate ground states of the system guaranteed for $N_B$ odd owing to Theorem~\ref{theorem} do not consist of a single Slater determinant due to the correlations induced by $\hat{H}_{P_{\rm CM}}$. Therefore, the argumentation above does not generalize straightforwardly in this case. However, as we show in Fig.~\ref{fig:pop-exc-states} the ground states of the system for finite $m_I$ are almost equivalent to the case of $m_I \to \infty$ for a wide range of masses. More specifically, Fig.~\ref{fig:pop-exc-states} shows the contribution of the $m_I \to \infty$ eigenstates, $\hat{H}_{0r}| \Psi_k \rangle = E | \Psi_k \rangle$, belonging to different energetic classes characterized by their eigenenergy $E$, to the ground state of the system for finite $m_I$, $( \hat{H}_{0r} + \hat{H}_{R_{\rm CM}} )| \tilde{\Psi}_0(m_I) \rangle = E_0(m_I) | \tilde{\Psi}_0(m_I) \rangle$. This contribution reads}
\begin{equation}
C(m_I;E) = \sum_{|\Psi_k\rangle:~ \hat{H}_{0r} |\Psi_k\rangle = E |\Psi_k\rangle} |\langle \Psi_k | \tilde{\Psi}_0(m_I) \rangle|^2.
\end{equation} 

\textcolor{black}{Notice that, in all cases, $N_B = 5$ and $1/g = x_I =0$ is considered and $| \tilde{\Psi}_0(m_I) \rangle$ is calculated via exact diagonalization (see also Appendix~\ref{sec:convergence}). 
Figure~\ref{fig:pop-exc-states} shows that states apart from $| \Psi_L \rangle$ and $| \Psi_R \rangle$ belonging to the class $E = 14.5$ contribute negligibly to the ground state of the system, since $C(m_I; E = 14.5) > 0.8$ even in the case $m_I = m_B$. In addition, the population of $E \ge 16.5$ exhibit a behaviour $\propto m_I^{-1/2}$ for $m_I > 10 m_B$. This indicates the fact that first-order perturbation theory, $\langle \Psi_k | \Psi_L \rangle \approx \frac{\langle \Psi_k | \hat{H}_{\rm coup} | \Psi_L \rangle}{E_0(m_I \to \infty) - E_k} \propto \frac{1}{m_I}$ is adequate to account for their population. Motivated by these numerical evidence we can show that the effective $E \otimes \epsilon$ Hamiltonian of Eq.~\eqref{near_infinity_effective_hamiltonian} carries over within first order perturbation theory in $1/m_I$, albeit with modified coefficients (not shown here for brevity).}

        \label{wronskian_theorem}
%\end{equation}
%Therefore, if the eigenstates, i.e. $E_1 = E_2$ are degenerate then 
%\begin{equation}
%        \frac{\mathrm{d}}{\mathrm{d}x} \left( \psi_2(x) \frac{\mathrm{d}}{\mathrm{d}x} \psi_1(x) - \psi_1(x) \frac{\mathrm{d}}{\mathrm{d}x} \psi_2(x) \right) = 0
        \label{wronskian_theorem2}
\bibliography{bibliography}

\begin{thebibliography}{114}%
\makeatletter
\providecommand \@ifxundefined [1]{%
 \@ifx{#1\undefined}
}%
\providecommand \@ifnum [1]{%
 \ifnum #1\expandafter \@firstoftwo
 \else \expandafter \@secondoftwo
 \fi
}%
\providecommand \@ifx [1]{%
 \ifx #1\expandafter \@firstoftwo
 \else \expandafter \@secondoftwo
 \fi
}%
\providecommand \natexlab [1]{#1}%
\providecommand \enquote  [1]{``#1''}%
\providecommand \bibnamefont  [1]{#1}%
\providecommand \bibfnamefont [1]{#1}%
\providecommand \citenamefont [1]{#1}%
\providecommand \href@noop [0]{\@secondoftwo}%
\providecommand \href [0]{\begingroup \@sanitize@url \@href}%
\providecommand \@href[1]{\@@startlink{#1}\@@href}%
\providecommand \@@href[1]{\endgroup#1\@@endlink}%
\providecommand \@sanitize@url [0]{\catcode `\\12\catcode `\$12\catcode
  `\&12\catcode `\#12\catcode `\^12\catcode `\_12\catcode `\%12\relax}%
\providecommand \@@startlink[1]{}%
\providecommand \@@endlink[0]{}%
\providecommand \url  [0]{\begingroup\@sanitize@url \@url }%
\providecommand \@url [1]{\endgroup\@href {#1}{\urlprefix }}%
\providecommand \urlprefix  [0]{URL }%
\providecommand \Eprint [0]{\href }%
\providecommand \doibase [0]{https://doi.org/}%
\providecommand \selectlanguage [0]{\@gobble}%
\providecommand \bibinfo  [0]{\@secondoftwo}%
\providecommand \bibfield  [0]{\@secondoftwo}%
\providecommand \translation [1]{[#1]}%
\providecommand \BibitemOpen [0]{}%
\providecommand \bibitemStop [0]{}%
\providecommand \bibitemNoStop [0]{.\EOS\space}%
\providecommand \EOS [0]{\spacefactor3000\relax}%
\providecommand \BibitemShut  [1]{\csname bibitem#1\endcsname}%
\let\auto@bib@innerbib\@empty
%</preamble>
\bibitem [{\citenamefont {Jahn}\ \emph {et~al.}(1937)\citenamefont {Jahn},
  \citenamefont {Teller},\ and\ \citenamefont {Donnan}}]{JahnTeller1937}%
  \BibitemOpen
  \bibfield  {author} {\bibinfo {author} {\bibfnamefont {H.~A.}\ \bibnamefont
  {Jahn}}, \bibinfo {author} {\bibfnamefont {E.}~\bibnamefont {Teller}},\ and\
  \bibinfo {author} {\bibfnamefont {F.~G.}\ \bibnamefont {Donnan}},\ }\bibfield
   {title} {\bibinfo {title} {Stability of polyatomic molecules in degenerate
  electronic states - {I}—orbital degeneracy},\ }\href
  {https://doi.org/10.1098/rspa.1937.0142} {\bibfield  {journal} {\bibinfo
  {journal} {Proc. Math. Phys. Sci.}\ }\textbf {\bibinfo {volume} {161}},\
  \bibinfo {pages} {220} (\bibinfo {year} {1937})}\BibitemShut {NoStop}%
\bibitem [{\citenamefont {Englman}(1972)}]{Englman1972}%
  \BibitemOpen
  \bibfield  {author} {\bibinfo {author} {\bibfnamefont {R.}~\bibnamefont
  {Englman}},\ }\href@noop {} {\emph {\bibinfo {title} {The Jahn-Teller effect
  in molecules and crystals}}},\ Vol.~\bibinfo {volume} {1}\ (\bibinfo
  {publisher} {Wiley-Interscience},\ \bibinfo {year} {1972})\BibitemShut
  {NoStop}%
\bibitem [{\citenamefont {Bersuker}(2006)}]{Bersuker2006}%
  \BibitemOpen
  \bibfield  {author} {\bibinfo {author} {\bibfnamefont {I.}~\bibnamefont
  {Bersuker}},\ }\href@noop {} {\emph {\bibinfo {title} {The Jahn-Teller
  effect}}},\ Vol.~\bibinfo {volume} {1}\ (\bibinfo  {publisher} {Cambridge
  University Press},\ \bibinfo {year} {2006})\BibitemShut {NoStop}%
\bibitem [{\citenamefont {Köppel~H.}(2009)}]{Köppel2009}%
  \BibitemOpen
  \bibfield  {author} {\bibinfo {author} {\bibfnamefont {B.~H.}\ \bibnamefont
  {Köppel~H.}, \bibfnamefont {Yarkony~D.}},\ }\href@noop {} {\emph {\bibinfo
  {title} {The Jahn-Teller Effect}}},\ Vol.~\bibinfo {volume} {1}\ (\bibinfo
  {publisher} {Springer-Verlag Berlin},\ \bibinfo {year} {2009})\BibitemShut
  {NoStop}%
\bibitem [{\citenamefont {Bersuker}(2021)}]{Bersuker2021}%
  \BibitemOpen
  \bibfield  {author} {\bibinfo {author} {\bibfnamefont {I.~B.}\ \bibnamefont
  {Bersuker}},\ }\bibfield  {title} {\bibinfo {title} {{Jahn--Teller and
  Pseudo-Jahn--Teller effects: From particular features to general tools in
  exploring molecular and solid state properties}},\ }\href
  {https://doi.org/10.1021/acs.chemrev.0c00718} {\bibfield  {journal} {\bibinfo
   {journal} {Chem. Rev.}\ }\textbf {\bibinfo {volume} {121}},\ \bibinfo
  {pages} {1463} (\bibinfo {year} {2021})}\BibitemShut {NoStop}%
\bibitem [{\citenamefont {Kahn}\ and\ \citenamefont
  {Martinez}(1998)}]{Kahn1998}%
  \BibitemOpen
  \bibfield  {author} {\bibinfo {author} {\bibfnamefont {O.}~\bibnamefont
  {Kahn}}\ and\ \bibinfo {author} {\bibfnamefont {C.~J.}\ \bibnamefont
  {Martinez}},\ }\bibfield  {title} {\bibinfo {title} {Spin-transition
  polymers: From molecular materials toward memory devices},\ }\href
  {https://doi.org/10.1126/science.279.5347.44} {\bibfield  {journal} {\bibinfo
   {journal} {Science}\ }\textbf {\bibinfo {volume} {279}},\ \bibinfo {pages}
  {44} (\bibinfo {year} {1998})}\BibitemShut {NoStop}%
\bibitem [{\citenamefont {Feringa}\ \emph {et~al.}(2000)\citenamefont
  {Feringa}, \citenamefont {van Delden}, \citenamefont {Koumura},\ and\
  \citenamefont {Geertsema}}]{Feringa2000}%
  \BibitemOpen
  \bibfield  {author} {\bibinfo {author} {\bibfnamefont {B.~L.}\ \bibnamefont
  {Feringa}}, \bibinfo {author} {\bibfnamefont {R.~A.}\ \bibnamefont {van
  Delden}}, \bibinfo {author} {\bibfnamefont {N.}~\bibnamefont {Koumura}},\
  and\ \bibinfo {author} {\bibfnamefont {E.~M.}\ \bibnamefont {Geertsema}},\
  }\bibfield  {title} {\bibinfo {title} {Chiroptical molecular switches},\
  }\href {https://doi.org/10.1021/cr9900228} {\bibfield  {journal} {\bibinfo
  {journal} {Chem. Rev.}\ }\textbf {\bibinfo {volume} {100}},\ \bibinfo {pages}
  {1789} (\bibinfo {year} {2000})}\BibitemShut {NoStop}%
\bibitem [{\citenamefont {Rocha}\ \emph {et~al.}(2005)\citenamefont {Rocha},
  \citenamefont {García-Suárez}, \citenamefont {Bailey}, \citenamefont
  {Lambert}, \citenamefont {Ferrer},\ and\ \citenamefont
  {Sanvito}}]{Rocha2005}%
  \BibitemOpen
  \bibfield  {author} {\bibinfo {author} {\bibfnamefont {A.~R.}\ \bibnamefont
  {Rocha}}, \bibinfo {author} {\bibfnamefont {V.~M.}\ \bibnamefont
  {García-Suárez}}, \bibinfo {author} {\bibfnamefont {S.~W.}\ \bibnamefont
  {Bailey}}, \bibinfo {author} {\bibfnamefont {C.~J.}\ \bibnamefont {Lambert}},
  \bibinfo {author} {\bibfnamefont {J.}~\bibnamefont {Ferrer}},\ and\ \bibinfo
  {author} {\bibfnamefont {S.}~\bibnamefont {Sanvito}},\ }\bibfield  {title}
  {\bibinfo {title} {Towards molecular spintronics},\ }\href
  {https://doi.org/https://doi.org/10.1038/nmat1349} {\bibfield  {journal}
  {\bibinfo  {journal} {Nat. Mater.}\ }\textbf {\bibinfo {volume} {4}},\
  \bibinfo {pages} {335} (\bibinfo {year} {2005})}\BibitemShut {NoStop}%
\bibitem [{\citenamefont {Magoni}\ \emph {et~al.}(2023)\citenamefont {Magoni},
  \citenamefont {Joshi},\ and\ \citenamefont {Lesanovsky}}]{magoni2023}%
  \BibitemOpen
  \bibfield  {author} {\bibinfo {author} {\bibfnamefont {M.}~\bibnamefont
  {Magoni}}, \bibinfo {author} {\bibfnamefont {R.}~\bibnamefont {Joshi}},\ and\
  \bibinfo {author} {\bibfnamefont {I.}~\bibnamefont {Lesanovsky}},\
  }\href@noop {} {\bibinfo {title} {{Rydberg tweezer molecules: Spin-phonon
  entanglement and Jahn-Teller effect}}} (\bibinfo {year} {2023}),\ \Eprint
  {https://arxiv.org/abs/2303.08861} {arXiv:2303.08861 [quant-ph]} \BibitemShut
  {NoStop}%
\bibitem [{\citenamefont {Angeli}\ \emph {et~al.}(2019)\citenamefont {Angeli},
  \citenamefont {Tosatti},\ and\ \citenamefont {Fabrizio}}]{AngeliTosatti2019}%
  \BibitemOpen
  \bibfield  {author} {\bibinfo {author} {\bibfnamefont {M.}~\bibnamefont
  {Angeli}}, \bibinfo {author} {\bibfnamefont {E.}~\bibnamefont {Tosatti}},\
  and\ \bibinfo {author} {\bibfnamefont {M.}~\bibnamefont {Fabrizio}},\
  }\bibfield  {title} {\bibinfo {title} {{Valley Jahn-Teller Effect in twisted
  bilayer graphene}},\ }\href {https://doi.org/10.1103/PhysRevX.9.041010}
  {\bibfield  {journal} {\bibinfo  {journal} {Phys. Rev. X}\ }\textbf {\bibinfo
  {volume} {9}},\ \bibinfo {pages} {041010} (\bibinfo {year}
  {2019})}\BibitemShut {NoStop}%
\bibitem [{\citenamefont {Freitag}\ and\ \citenamefont
  {Conradie}(2013)}]{Freitag2013}%
  \BibitemOpen
  \bibfield  {author} {\bibinfo {author} {\bibfnamefont {R.}~\bibnamefont
  {Freitag}}\ and\ \bibinfo {author} {\bibfnamefont {J.}~\bibnamefont
  {Conradie}},\ }\bibfield  {title} {\bibinfo {title} {{Understanding the
  Jahn--Teller effect in octahedral transition-metal complexes: a molecular
  orbital view of the Mn($\beta$-diketonato)3 Complex}},\ }\href
  {https://doi.org/10.1021/ed400370p} {\bibfield  {journal} {\bibinfo
  {journal} {J. Chem. Educ.}\ }\textbf {\bibinfo {volume} {90}},\ \bibinfo
  {pages} {1692} (\bibinfo {year} {2013})}\BibitemShut {NoStop}%
\bibitem [{\citenamefont {Bacci}(1980)}]{Bacci1980}%
  \BibitemOpen
  \bibfield  {author} {\bibinfo {author} {\bibfnamefont {M.}~\bibnamefont
  {Bacci}},\ }\bibfield  {title} {\bibinfo {title} {{J}ahn-{T}eller effect in
  biomolecules},\ }\href
  {https://doi.org/https://doi.org/10.1016/0301-4622(80)85006-X} {\bibfield
  {journal} {\bibinfo  {journal} {Biophys. Chem.}\ }\textbf {\bibinfo {volume}
  {11}},\ \bibinfo {pages} {39} (\bibinfo {year} {1980})}\BibitemShut {NoStop}%
\bibitem [{\citenamefont {Öpik}\ and\ \citenamefont
  {Pryce}(1957)}]{OepikPryce1957}%
  \BibitemOpen
  \bibfield  {author} {\bibinfo {author} {\bibfnamefont {U.}~\bibnamefont
  {Öpik}}\ and\ \bibinfo {author} {\bibfnamefont {M.~H.~L.}\ \bibnamefont
  {Pryce}},\ }\bibfield  {title} {\bibinfo {title} {{Studies of the Jahn-Teller
  effect. I. A survey of the static problem}},\ }\href
  {https://doi.org/10.1098/rspa.1957.0010} {\bibfield  {journal} {\bibinfo
  {journal} {Proc. Math. Phys. Sci.}\ }\textbf {\bibinfo {volume} {238}},\
  \bibinfo {pages} {425} (\bibinfo {year} {1957})}\BibitemShut {NoStop}%
\bibitem [{\citenamefont {Bersuker}\ and\ \citenamefont
  {Polinger}(1989)}]{BersukerPolinger1989}%
  \BibitemOpen
  \bibfield  {author} {\bibinfo {author} {\bibfnamefont {I.~B.}\ \bibnamefont
  {Bersuker}}\ and\ \bibinfo {author} {\bibfnamefont {V.~Z.}\ \bibnamefont
  {Polinger}},\ }\href@noop {} {\emph {\bibinfo {title} {Vibronic Interactions
  in Molecules and Crystals}}},\ Springer Series in Chemical Physics\ (\bibinfo
   {publisher} {Springer},\ \bibinfo {address} {Berlin, Heidelberg},\ \bibinfo
  {year} {1989})\BibitemShut {NoStop}%
\bibitem [{\citenamefont {Bersuker}(2013)}]{Bersuker2013}%
  \BibitemOpen
  \bibfield  {author} {\bibinfo {author} {\bibfnamefont {I.~B.}\ \bibnamefont
  {Bersuker}},\ }\bibfield  {title} {\bibinfo {title} {{Pseudo-Jahn-teller}
  effect--a two-state paradigm in formation, deformation, and transformation of
  molecular systems and solids},\ }\href@noop {} {\bibfield  {journal}
  {\bibinfo  {journal} {Chem. Rev.}\ }\textbf {\bibinfo {volume} {113}},\
  \bibinfo {pages} {1351} (\bibinfo {year} {2013})}\BibitemShut {NoStop}%
\bibitem [{\citenamefont {Bersuker}(2016)}]{Bersuker2016}%
  \BibitemOpen
  \bibfield  {author} {\bibinfo {author} {\bibfnamefont {I.~B.}\ \bibnamefont
  {Bersuker}},\ }\bibinfo {title} {Spontaneous symmetry breaking in matter
  induced by degeneracies and pseudodegeneracies},\ in\ \href
  {https://doi.org/https://doi.org/10.1002/9781119165156.ch3} {\emph {\bibinfo
  {booktitle} {Advances in Chemical Physics}}}\ (\bibinfo  {publisher} {John
  Wiley \& Sons, Ltd},\ \bibinfo {year} {2016})\ pp.\ \bibinfo {pages}
  {159--208}\BibitemShut {NoStop}%
\bibitem [{\citenamefont {Pethick}\ and\ \citenamefont
  {Smith}(2001)}]{PethickSmith2001}%
  \BibitemOpen
  \bibfield  {author} {\bibinfo {author} {\bibfnamefont {C.~J.}\ \bibnamefont
  {Pethick}}\ and\ \bibinfo {author} {\bibfnamefont {H.}~\bibnamefont
  {Smith}},\ }\href {https://doi.org/10.1017/CBO9780511755583} {\emph {\bibinfo
  {title} {Bose–Einstein Condensation in Dilute Gases}}}\ (\bibinfo
  {publisher} {Cambridge University Press},\ \bibinfo {year}
  {2001})\BibitemShut {NoStop}%
\bibitem [{\citenamefont {Pitaevskii L.~P.}(2016)}]{PitaevskiiStringari}%
  \BibitemOpen
  \bibfield  {author} {\bibinfo {author} {\bibfnamefont {S.~S.}\ \bibnamefont
  {Pitaevskii L.~P.}},\ }\href@noop {} {\emph {\bibinfo {title} {Bose-Einstein
  Condensation and Superfluidity}}}\ (\bibinfo  {publisher} {Oxford University
  Press},\ \bibinfo {year} {2016})\BibitemShut {NoStop}%
\bibitem [{\citenamefont {Bloch}\ \emph {et~al.}(2008)\citenamefont {Bloch},
  \citenamefont {Dalibard},\ and\ \citenamefont {Zwerger}}]{Bloch_2008}%
  \BibitemOpen
  \bibfield  {author} {\bibinfo {author} {\bibfnamefont {I.}~\bibnamefont
  {Bloch}}, \bibinfo {author} {\bibfnamefont {J.}~\bibnamefont {Dalibard}},\
  and\ \bibinfo {author} {\bibfnamefont {W.}~\bibnamefont {Zwerger}},\
  }\bibfield  {title} {\bibinfo {title} {Many-body physics with ultracold
  gases},\ }\href {https://doi.org/10.1103/revmodphys.80.885} {\bibfield
  {journal} {\bibinfo  {journal} {Rev. Mod. Phys.}\ }\textbf {\bibinfo {volume}
  {80}},\ \bibinfo {pages} {885} (\bibinfo {year} {2008})}\BibitemShut
  {NoStop}%
\bibitem [{\citenamefont {Bloch}\ \emph {et~al.}(2012)\citenamefont {Bloch},
  \citenamefont {Dalibard},\ and\ \citenamefont {Nascimbène}}]{Bloch2012}%
  \BibitemOpen
  \bibfield  {author} {\bibinfo {author} {\bibfnamefont {I.}~\bibnamefont
  {Bloch}}, \bibinfo {author} {\bibfnamefont {J.}~\bibnamefont {Dalibard}},\
  and\ \bibinfo {author} {\bibfnamefont {S.}~\bibnamefont {Nascimbène}},\
  }\bibfield  {title} {\bibinfo {title} {Quantum simulations with ultracold
  quantum gases},\ }\href {https://doi.org/https://doi.org/10.1038/nphys2259}
  {\bibfield  {journal} {\bibinfo  {journal} {Nat. Phys.}\ }\textbf {\bibinfo
  {volume} {8}},\ \bibinfo {pages} {267} (\bibinfo {year} {2012})}\BibitemShut
  {NoStop}%
\bibitem [{\citenamefont {Gross}\ and\ \citenamefont
  {Bloch}(2017)}]{Gross2017}%
  \BibitemOpen
  \bibfield  {author} {\bibinfo {author} {\bibfnamefont {C.}~\bibnamefont
  {Gross}}\ and\ \bibinfo {author} {\bibfnamefont {I.}~\bibnamefont {Bloch}},\
  }\bibfield  {title} {\bibinfo {title} {Quantum simulations with ultracold
  atoms in optical lattices},\ }\href {https://doi.org/10.1126/science.aal3837}
  {\bibfield  {journal} {\bibinfo  {journal} {Science}\ }\textbf {\bibinfo
  {volume} {357}},\ \bibinfo {pages} {995} (\bibinfo {year}
  {2017})}\BibitemShut {NoStop}%
\bibitem [{\citenamefont {Sowiński}\ and\ \citenamefont {Ángel
  García-March}(2019)}]{SowinskiGarciaMarch2019}%
  \BibitemOpen
  \bibfield  {author} {\bibinfo {author} {\bibfnamefont {T.}~\bibnamefont
  {Sowiński}}\ and\ \bibinfo {author} {\bibfnamefont {M.}~\bibnamefont {Ángel
  García-March}},\ }\bibfield  {title} {\bibinfo {title} {One-dimensional
  mixtures of several ultracold atoms: a review},\ }\href
  {https://doi.org/10.1088/1361-6633/ab3a80} {\bibfield  {journal} {\bibinfo
  {journal} {Rep. Prog. Phys.}\ }\textbf {\bibinfo {volume} {82}},\ \bibinfo
  {pages} {104401} (\bibinfo {year} {2019})}\BibitemShut {NoStop}%
\bibitem [{\citenamefont {Schäfer}\ \emph {et~al.}(2020)\citenamefont
  {Schäfer}, \citenamefont {Fukuhara},\ and\ \citenamefont
  {Sugawa}}]{Schaefer2020}%
  \BibitemOpen
  \bibfield  {author} {\bibinfo {author} {\bibfnamefont {F.}~\bibnamefont
  {Schäfer}}, \bibinfo {author} {\bibfnamefont {T.}~\bibnamefont {Fukuhara}},\
  and\ \bibinfo {author} {\bibfnamefont {S.}~\bibnamefont {Sugawa}},\
  }\bibfield  {title} {\bibinfo {title} {Tools for quantum simulation with
  ultracold atoms in optical lattices},\ }\href
  {https://doi.org/https://doi.org/10.1038/s42254-020-0195-3} {\bibfield
  {journal} {\bibinfo  {journal} {Nat. Rev. Phys.}\ }\textbf {\bibinfo {volume}
  {2}},\ \bibinfo {pages} {411} (\bibinfo {year} {2020})}\BibitemShut {NoStop}%
\bibitem [{\citenamefont {Mistakidis}\ \emph {et~al.}(2022)\citenamefont
  {Mistakidis}, \citenamefont {Volosniev}, \citenamefont {Barfknecht},
  \citenamefont {Fogarty}, \citenamefont {Busch}, \citenamefont {Foerster},
  \citenamefont {Schmelcher},\ and\ \citenamefont
  {Zinner}}]{mistakidis_volosniev2022}%
  \BibitemOpen
  \bibfield  {author} {\bibinfo {author} {\bibfnamefont {S.~I.}\ \bibnamefont
  {Mistakidis}}, \bibinfo {author} {\bibfnamefont {A.~G.}\ \bibnamefont
  {Volosniev}}, \bibinfo {author} {\bibfnamefont {R.~E.}\ \bibnamefont
  {Barfknecht}}, \bibinfo {author} {\bibfnamefont {T.}~\bibnamefont {Fogarty}},
  \bibinfo {author} {\bibfnamefont {T.}~\bibnamefont {Busch}}, \bibinfo
  {author} {\bibfnamefont {A.}~\bibnamefont {Foerster}}, \bibinfo {author}
  {\bibfnamefont {P.}~\bibnamefont {Schmelcher}},\ and\ \bibinfo {author}
  {\bibfnamefont {N.~T.}\ \bibnamefont {Zinner}},\ }\href@noop {} {\bibinfo
  {title} {Cold atoms in low dimensions -- a laboratory for quantum dynamics}}
  (\bibinfo {year} {2022}),\ \Eprint {https://arxiv.org/abs/2202.11071}
  {arXiv:2202.11071 [cond-mat.quant-gas]} \BibitemShut {NoStop}%
\bibitem [{\citenamefont {Tempere}\ \emph {et~al.}(2009)\citenamefont
  {Tempere}, \citenamefont {Casteels}, \citenamefont {Oberthaler},
  \citenamefont {Knoop}, \citenamefont {Timmermans},\ and\ \citenamefont
  {Devreese}}]{Tempere2009}%
  \BibitemOpen
  \bibfield  {author} {\bibinfo {author} {\bibfnamefont {J.}~\bibnamefont
  {Tempere}}, \bibinfo {author} {\bibfnamefont {W.}~\bibnamefont {Casteels}},
  \bibinfo {author} {\bibfnamefont {M.~K.}\ \bibnamefont {Oberthaler}},
  \bibinfo {author} {\bibfnamefont {S.}~\bibnamefont {Knoop}}, \bibinfo
  {author} {\bibfnamefont {E.}~\bibnamefont {Timmermans}},\ and\ \bibinfo
  {author} {\bibfnamefont {J.~T.}\ \bibnamefont {Devreese}},\ }\bibfield
  {title} {\bibinfo {title} {{Feynman path-integral treatment of the
  BEC-impurity polaron}},\ }\href {https://doi.org/10.1103/PhysRevB.80.184504}
  {\bibfield  {journal} {\bibinfo  {journal} {Phys. Rev. B}\ }\textbf {\bibinfo
  {volume} {80}},\ \bibinfo {pages} {184504} (\bibinfo {year}
  {2009})}\BibitemShut {NoStop}%
\bibitem [{\citenamefont {Scelle}\ \emph {et~al.}(2013)\citenamefont {Scelle},
  \citenamefont {Rentrop}, \citenamefont {Trautmann}, \citenamefont
  {Schuster},\ and\ \citenamefont {Oberthaler}}]{Scelle2013}%
  \BibitemOpen
  \bibfield  {author} {\bibinfo {author} {\bibfnamefont {R.}~\bibnamefont
  {Scelle}}, \bibinfo {author} {\bibfnamefont {T.}~\bibnamefont {Rentrop}},
  \bibinfo {author} {\bibfnamefont {A.}~\bibnamefont {Trautmann}}, \bibinfo
  {author} {\bibfnamefont {T.}~\bibnamefont {Schuster}},\ and\ \bibinfo
  {author} {\bibfnamefont {M.~K.}\ \bibnamefont {Oberthaler}},\ }\bibfield
  {title} {\bibinfo {title} {{Motional coherence of Fermions immersed in a Bose
  gas}},\ }\href {https://doi.org/10.1103/PhysRevLett.111.070401} {\bibfield
  {journal} {\bibinfo  {journal} {Phys. Rev. Lett.}\ }\textbf {\bibinfo
  {volume} {111}},\ \bibinfo {pages} {070401} (\bibinfo {year}
  {2013})}\BibitemShut {NoStop}%
\bibitem [{\citenamefont {Lemeshko}\ and\ \citenamefont
  {Schmidt}(2017)}]{Lemeshko2016}%
  \BibitemOpen
  \bibfield  {author} {\bibinfo {author} {\bibfnamefont {M.}~\bibnamefont
  {Lemeshko}}\ and\ \bibinfo {author} {\bibfnamefont {R.}~\bibnamefont
  {Schmidt}},\ }\href {https://doi.org/10.1039/9781782626800-00444} {\emph
  {\bibinfo {title} {{Cold Chemistry: Molecular Scattering and Reactivity Near
  Absolute Zero}}}}\ (\bibinfo  {publisher} {The Royal Society of Chemistry},\
  \bibinfo {year} {2017})\BibitemShut {NoStop}%
\bibitem [{\citenamefont {Erdmann}\ \emph {et~al.}(2018)\citenamefont
  {Erdmann}, \citenamefont {Mistakidis},\ and\ \citenamefont
  {Schmelcher}}]{Erdmann2018}%
  \BibitemOpen
  \bibfield  {author} {\bibinfo {author} {\bibfnamefont {J.}~\bibnamefont
  {Erdmann}}, \bibinfo {author} {\bibfnamefont {S.~I.}\ \bibnamefont
  {Mistakidis}},\ and\ \bibinfo {author} {\bibfnamefont {P.}~\bibnamefont
  {Schmelcher}},\ }\bibfield  {title} {\bibinfo {title} {Correlated tunneling
  dynamics of an ultracold fermi-fermi mixture confined in a double well},\
  }\href {https://doi.org/10.1103/PhysRevA.98.053614} {\bibfield  {journal}
  {\bibinfo  {journal} {Phys. Rev. A}\ }\textbf {\bibinfo {volume} {98}},\
  \bibinfo {pages} {053614} (\bibinfo {year} {2018})}\BibitemShut {NoStop}%
\bibitem [{\citenamefont {Erdmann}\ \emph {et~al.}(2019)\citenamefont
  {Erdmann}, \citenamefont {Mistakidis},\ and\ \citenamefont
  {Schmelcher}}]{ErdmannMistakidis2019}%
  \BibitemOpen
  \bibfield  {author} {\bibinfo {author} {\bibfnamefont {J.}~\bibnamefont
  {Erdmann}}, \bibinfo {author} {\bibfnamefont {S.~I.}\ \bibnamefont
  {Mistakidis}},\ and\ \bibinfo {author} {\bibfnamefont {P.}~\bibnamefont
  {Schmelcher}},\ }\bibfield  {title} {\bibinfo {title} {Phase-separation
  dynamics induced by an interaction quench of a correlated fermi-fermi mixture
  in a double well},\ }\href {https://doi.org/10.1103/PhysRevA.99.013605}
  {\bibfield  {journal} {\bibinfo  {journal} {Phys. Rev. A}\ }\textbf {\bibinfo
  {volume} {99}},\ \bibinfo {pages} {013605} (\bibinfo {year}
  {2019})}\BibitemShut {NoStop}%
\bibitem [{\citenamefont {Rammelmüller}\ \emph
  {et~al.}(2023{\natexlab{a}})\citenamefont {Rammelmüller}, \citenamefont
  {Huber}, \citenamefont {Čufar}, \citenamefont {Brand}, \citenamefont
  {Hammer},\ and\ \citenamefont {Volosniev}}]{RammelmuellerHuber2023a}%
  \BibitemOpen
  \bibfield  {author} {\bibinfo {author} {\bibfnamefont {L.}~\bibnamefont
  {Rammelmüller}}, \bibinfo {author} {\bibfnamefont {D.}~\bibnamefont
  {Huber}}, \bibinfo {author} {\bibfnamefont {M.}~\bibnamefont {Čufar}},
  \bibinfo {author} {\bibfnamefont {J.}~\bibnamefont {Brand}}, \bibinfo
  {author} {\bibfnamefont {H.-W.}\ \bibnamefont {Hammer}},\ and\ \bibinfo
  {author} {\bibfnamefont {A.~G.}\ \bibnamefont {Volosniev}},\ }\bibfield
  {title} {\bibinfo {title} {{Magnetic impurity in a one-dimensional
  few-fermion system}},\ }\href {https://doi.org/10.21468/SciPostPhys.14.1.006}
  {\bibfield  {journal} {\bibinfo  {journal} {SciPost Phys.}\ }\textbf
  {\bibinfo {volume} {14}},\ \bibinfo {pages} {6} (\bibinfo {year}
  {2023}{\natexlab{a}})}\BibitemShut {NoStop}%
\bibitem [{\citenamefont {Rammelmüller}\ \emph
  {et~al.}(2023{\natexlab{b}})\citenamefont {Rammelmüller}, \citenamefont
  {Huber},\ and\ \citenamefont {Volosniev}}]{RammelmuellerHuber2023b}%
  \BibitemOpen
  \bibfield  {author} {\bibinfo {author} {\bibfnamefont {L.}~\bibnamefont
  {Rammelmüller}}, \bibinfo {author} {\bibfnamefont {D.}~\bibnamefont
  {Huber}},\ and\ \bibinfo {author} {\bibfnamefont {A.~G.}\ \bibnamefont
  {Volosniev}},\ }\bibfield  {title} {\bibinfo {title} {{A modular
  implementation of an effective interaction approach for harmonically trapped
  fermions in 1D}},\ }\href {https://doi.org/10.21468/SciPostPhysCodeb.12}
  {\bibfield  {journal} {\bibinfo  {journal} {SciPost Phys. Codebases}\ ,\
  \bibinfo {pages} {12}} (\bibinfo {year} {2023}{\natexlab{b}})}\BibitemShut
  {NoStop}%
\bibitem [{\citenamefont {Chevy}\ and\ \citenamefont {Mora}(2010)}]{Chevy2010}%
  \BibitemOpen
  \bibfield  {author} {\bibinfo {author} {\bibfnamefont {F.}~\bibnamefont
  {Chevy}}\ and\ \bibinfo {author} {\bibfnamefont {C.}~\bibnamefont {Mora}},\
  }\bibfield  {title} {\bibinfo {title} {{Ultra-cold polarized Fermi gases}},\
  }\href {https://doi.org/10.1088/0034-4885/73/11/112401} {\bibfield  {journal}
  {\bibinfo  {journal} {Rep. Prog. Phys.}\ }\textbf {\bibinfo {volume} {73}},\
  \bibinfo {pages} {112401} (\bibinfo {year} {2010})}\BibitemShut {NoStop}%
\bibitem [{\citenamefont {Spethmann}\ \emph {et~al.}(2012)\citenamefont
  {Spethmann}, \citenamefont {Kindermann}, \citenamefont {John}, \citenamefont
  {Weber}, \citenamefont {Meschede},\ and\ \citenamefont
  {Widera}}]{Spethmann2012}%
  \BibitemOpen
  \bibfield  {author} {\bibinfo {author} {\bibfnamefont {N.}~\bibnamefont
  {Spethmann}}, \bibinfo {author} {\bibfnamefont {F.}~\bibnamefont
  {Kindermann}}, \bibinfo {author} {\bibfnamefont {S.}~\bibnamefont {John}},
  \bibinfo {author} {\bibfnamefont {C.}~\bibnamefont {Weber}}, \bibinfo
  {author} {\bibfnamefont {D.}~\bibnamefont {Meschede}},\ and\ \bibinfo
  {author} {\bibfnamefont {A.}~\bibnamefont {Widera}},\ }\bibfield  {title}
  {\bibinfo {title} {Dynamics of single neutral impurity atoms immersed in an
  ultracold gas},\ }\href {https://doi.org/10.1103/PhysRevLett.109.235301}
  {\bibfield  {journal} {\bibinfo  {journal} {Phys. Rev. Lett.}\ }\textbf
  {\bibinfo {volume} {109}},\ \bibinfo {pages} {235301} (\bibinfo {year}
  {2012})}\BibitemShut {NoStop}%
\bibitem [{\citenamefont {Massignan}\ \emph {et~al.}(2013)\citenamefont
  {Massignan}, \citenamefont {Yu},\ and\ \citenamefont
  {Bruun}}]{Massignan2013}%
  \BibitemOpen
  \bibfield  {author} {\bibinfo {author} {\bibfnamefont {P.}~\bibnamefont
  {Massignan}}, \bibinfo {author} {\bibfnamefont {Z.}~\bibnamefont {Yu}},\ and\
  \bibinfo {author} {\bibfnamefont {G.~M.}\ \bibnamefont {Bruun}},\ }\bibfield
  {title} {\bibinfo {title} {Itinerant ferromagnetism in a polarized
  two-component fermi gas},\ }\href
  {https://doi.org/10.1103/PhysRevLett.110.230401} {\bibfield  {journal}
  {\bibinfo  {journal} {Phys. Rev. Lett.}\ }\textbf {\bibinfo {volume} {110}},\
  \bibinfo {pages} {230401} (\bibinfo {year} {2013})}\BibitemShut {NoStop}%
\bibitem [{\citenamefont {Massignan}\ \emph {et~al.}(2014)\citenamefont
  {Massignan}, \citenamefont {Zaccanti},\ and\ \citenamefont
  {Bruun}}]{Massignan2014}%
  \BibitemOpen
  \bibfield  {author} {\bibinfo {author} {\bibfnamefont {P.}~\bibnamefont
  {Massignan}}, \bibinfo {author} {\bibfnamefont {M.}~\bibnamefont
  {Zaccanti}},\ and\ \bibinfo {author} {\bibfnamefont {G.~M.}\ \bibnamefont
  {Bruun}},\ }\bibfield  {title} {\bibinfo {title} {{Polarons, dressed
  molecules and itinerant ferromagnetism in ultracold Fermi gases}},\ }\href
  {https://doi.org/10.1088/0034-4885/77/3/034401} {\bibfield  {journal}
  {\bibinfo  {journal} {Rep. Prog. Phys.}\ }\textbf {\bibinfo {volume} {77}},\
  \bibinfo {pages} {034401} (\bibinfo {year} {2014})}\BibitemShut {NoStop}%
\bibitem [{\citenamefont {Fröhlich}(1954)}]{Froehlich1954}%
  \BibitemOpen
  \bibfield  {author} {\bibinfo {author} {\bibfnamefont {H.}~\bibnamefont
  {Fröhlich}},\ }\bibfield  {title} {\bibinfo {title} {Electrons in lattice
  fields},\ }\href {https://doi.org/10.1080/00018735400101213} {\bibfield
  {journal} {\bibinfo  {journal} {Adv. Phys.}\ }\textbf {\bibinfo {volume}
  {3}},\ \bibinfo {pages} {325} (\bibinfo {year} {1954})}\BibitemShut {NoStop}%
\bibitem [{\citenamefont {Feynman}(1955)}]{Feynman1955}%
  \BibitemOpen
  \bibfield  {author} {\bibinfo {author} {\bibfnamefont {R.~P.}\ \bibnamefont
  {Feynman}},\ }\bibfield  {title} {\bibinfo {title} {Slow electrons in a polar
  crystal},\ }\href {https://doi.org/10.1103/PhysRev.97.660} {\bibfield
  {journal} {\bibinfo  {journal} {Phys. Rev.}\ }\textbf {\bibinfo {volume}
  {97}},\ \bibinfo {pages} {660} (\bibinfo {year} {1955})}\BibitemShut
  {NoStop}%
\bibitem [{\citenamefont {Alexandrov}\ and\ \citenamefont
  {Devreese}(2010)}]{AlexandrovDevreese2010}%
  \BibitemOpen
  \bibfield  {author} {\bibinfo {author} {\bibfnamefont {A.~S.}\ \bibnamefont
  {Alexandrov}}\ and\ \bibinfo {author} {\bibfnamefont {J.~T.}\ \bibnamefont
  {Devreese}},\ }\href {https://doi.org/10.1007/978-3-642-01896-1_1} {\emph
  {\bibinfo {title} {Advances in Polaron Physics}}}\ (\bibinfo  {publisher}
  {Springer},\ \bibinfo {address} {Berlin},\ \bibinfo {year}
  {2010})\BibitemShut {NoStop}%
\bibitem [{\citenamefont {Sidler}\ \emph {et~al.}(2016)\citenamefont {Sidler},
  \citenamefont {Back}, \citenamefont {Cotlet}, \citenamefont {Srivastava},
  \citenamefont {Fink}, \citenamefont {Kroner}, \citenamefont {Demler},\ and\
  \citenamefont {Imamoglu}}]{Sidler2016}%
  \BibitemOpen
  \bibfield  {author} {\bibinfo {author} {\bibfnamefont {M.}~\bibnamefont
  {Sidler}}, \bibinfo {author} {\bibfnamefont {P.}~\bibnamefont {Back}},
  \bibinfo {author} {\bibfnamefont {O.}~\bibnamefont {Cotlet}}, \bibinfo
  {author} {\bibfnamefont {A.}~\bibnamefont {Srivastava}}, \bibinfo {author}
  {\bibfnamefont {T.}~\bibnamefont {Fink}}, \bibinfo {author} {\bibfnamefont
  {M.}~\bibnamefont {Kroner}}, \bibinfo {author} {\bibfnamefont
  {E.}~\bibnamefont {Demler}},\ and\ \bibinfo {author} {\bibfnamefont
  {A.}~\bibnamefont {Imamoglu}},\ }\bibfield  {title} {\bibinfo {title} {Fermi
  polaron-polaritons in charge-tunable atomically thin semiconductors},\ }\href
  {https://doi.org/10.1038/nphys3949} {\bibfield  {journal} {\bibinfo
  {journal} {Nat. Phys.}\ }\textbf {\bibinfo {volume} {13}},\ \bibinfo {pages}
  {255} (\bibinfo {year} {2016})}\BibitemShut {NoStop}%
\bibitem [{\citenamefont {Schirotzek}\ \emph {et~al.}(2009)\citenamefont
  {Schirotzek}, \citenamefont {Wu}, \citenamefont {Sommer},\ and\ \citenamefont
  {Zwierlein}}]{Schirotek2009}%
  \BibitemOpen
  \bibfield  {author} {\bibinfo {author} {\bibfnamefont {A.}~\bibnamefont
  {Schirotzek}}, \bibinfo {author} {\bibfnamefont {C.-H.}\ \bibnamefont {Wu}},
  \bibinfo {author} {\bibfnamefont {A.}~\bibnamefont {Sommer}},\ and\ \bibinfo
  {author} {\bibfnamefont {M.~W.}\ \bibnamefont {Zwierlein}},\ }\bibfield
  {title} {\bibinfo {title} {{Observation of Fermi polarons in a tunable Fermi
  liquid of ultracold atoms}},\ }\href
  {https://doi.org/10.1103/PhysRevLett.102.230402} {\bibfield  {journal}
  {\bibinfo  {journal} {Phys. Rev. Lett.}\ }\textbf {\bibinfo {volume} {102}},\
  \bibinfo {pages} {230402} (\bibinfo {year} {2009})}\BibitemShut {NoStop}%
\bibitem [{\citenamefont {Nascimb\`ene}\ \emph {et~al.}(2009)\citenamefont
  {Nascimb\`ene}, \citenamefont {Navon}, \citenamefont {Jiang}, \citenamefont
  {Tarruell}, \citenamefont {Teichmann}, \citenamefont {McKeever},
  \citenamefont {Chevy},\ and\ \citenamefont {Salomon}}]{Nascimbene2009}%
  \BibitemOpen
  \bibfield  {author} {\bibinfo {author} {\bibfnamefont {S.}~\bibnamefont
  {Nascimb\`ene}}, \bibinfo {author} {\bibfnamefont {N.}~\bibnamefont {Navon}},
  \bibinfo {author} {\bibfnamefont {K.~J.}\ \bibnamefont {Jiang}}, \bibinfo
  {author} {\bibfnamefont {L.}~\bibnamefont {Tarruell}}, \bibinfo {author}
  {\bibfnamefont {M.}~\bibnamefont {Teichmann}}, \bibinfo {author}
  {\bibfnamefont {J.}~\bibnamefont {McKeever}}, \bibinfo {author}
  {\bibfnamefont {F.}~\bibnamefont {Chevy}},\ and\ \bibinfo {author}
  {\bibfnamefont {C.}~\bibnamefont {Salomon}},\ }\bibfield  {title} {\bibinfo
  {title} {{Collective oscillations of an imbalanced Fermi gas: axial
  compression modes and polaron effective mass}},\ }\href
  {https://doi.org/10.1103/PhysRevLett.103.170402} {\bibfield  {journal}
  {\bibinfo  {journal} {Phys. Rev. Lett.}\ }\textbf {\bibinfo {volume} {103}},\
  \bibinfo {pages} {170402} (\bibinfo {year} {2009})}\BibitemShut {NoStop}%
\bibitem [{\citenamefont {Palzer}\ \emph {et~al.}(2009)\citenamefont {Palzer},
  \citenamefont {Zipkes}, \citenamefont {Sias},\ and\ \citenamefont
  {K\"ohl}}]{Palzer2009}%
  \BibitemOpen
  \bibfield  {author} {\bibinfo {author} {\bibfnamefont {S.}~\bibnamefont
  {Palzer}}, \bibinfo {author} {\bibfnamefont {C.}~\bibnamefont {Zipkes}},
  \bibinfo {author} {\bibfnamefont {C.}~\bibnamefont {Sias}},\ and\ \bibinfo
  {author} {\bibfnamefont {M.}~\bibnamefont {K\"ohl}},\ }\bibfield  {title}
  {\bibinfo {title} {{Quantum transport through a Tonks-Girardeau gas}},\
  }\href {https://doi.org/10.1103/PhysRevLett.103.150601} {\bibfield  {journal}
  {\bibinfo  {journal} {Phys. Rev. Lett.}\ }\textbf {\bibinfo {volume} {103}},\
  \bibinfo {pages} {150601} (\bibinfo {year} {2009})}\BibitemShut {NoStop}%
\bibitem [{\citenamefont {Punk}\ \emph {et~al.}(2009)\citenamefont {Punk},
  \citenamefont {Dumitrescu},\ and\ \citenamefont {Zwerger}}]{Punk2009}%
  \BibitemOpen
  \bibfield  {author} {\bibinfo {author} {\bibfnamefont {M.}~\bibnamefont
  {Punk}}, \bibinfo {author} {\bibfnamefont {P.~T.}\ \bibnamefont
  {Dumitrescu}},\ and\ \bibinfo {author} {\bibfnamefont {W.}~\bibnamefont
  {Zwerger}},\ }\bibfield  {title} {\bibinfo {title} {{Polaron-to-molecule
  transition in a strongly imbalanced Fermi gas}},\ }\href
  {https://doi.org/10.1103/PhysRevA.80.053605} {\bibfield  {journal} {\bibinfo
  {journal} {Phys. Rev. A}\ }\textbf {\bibinfo {volume} {80}},\ \bibinfo
  {pages} {053605} (\bibinfo {year} {2009})}\BibitemShut {NoStop}%
\bibitem [{\citenamefont {Massignan}\ and\ \citenamefont
  {Bruun}(2011)}]{Massignan2011}%
  \BibitemOpen
  \bibfield  {author} {\bibinfo {author} {\bibfnamefont {P.}~\bibnamefont
  {Massignan}}\ and\ \bibinfo {author} {\bibfnamefont {G.~M.}\ \bibnamefont
  {Bruun}},\ }\bibfield  {title} {\bibinfo {title} {{Repulsive polarons and
  itinerant ferromagnetism in strongly polarized Fermi gases}},\ }\href
  {https://doi.org/10.1140/epjd/e2011-20084-5} {\bibfield  {journal} {\bibinfo
  {journal} {Eur. Phys. J. D}\ }\textbf {\bibinfo {volume} {65}},\ \bibinfo
  {pages} {83} (\bibinfo {year} {2011})}\BibitemShut {NoStop}%
\bibitem [{\citenamefont {Schmidt}\ and\ \citenamefont
  {Enss}(2011)}]{Schmidt2011}%
  \BibitemOpen
  \bibfield  {author} {\bibinfo {author} {\bibfnamefont {R.}~\bibnamefont
  {Schmidt}}\ and\ \bibinfo {author} {\bibfnamefont {T.}~\bibnamefont {Enss}},\
  }\bibfield  {title} {\bibinfo {title} {Excitation spectra and rf response
  near the polaron-to-molecule transition from the functional renormalization
  group},\ }\href {https://doi.org/10.1103/PhysRevA.83.063620} {\bibfield
  {journal} {\bibinfo  {journal} {Phys. Rev. A}\ }\textbf {\bibinfo {volume}
  {83}},\ \bibinfo {pages} {063620} (\bibinfo {year} {2011})}\BibitemShut
  {NoStop}%
\bibitem [{\citenamefont {Schmidt}\ \emph {et~al.}(2012)\citenamefont
  {Schmidt}, \citenamefont {Enss}, \citenamefont {Pietil\"a},\ and\
  \citenamefont {Demler}}]{Schmidt2012}%
  \BibitemOpen
  \bibfield  {author} {\bibinfo {author} {\bibfnamefont {R.}~\bibnamefont
  {Schmidt}}, \bibinfo {author} {\bibfnamefont {T.}~\bibnamefont {Enss}},
  \bibinfo {author} {\bibfnamefont {V.}~\bibnamefont {Pietil\"a}},\ and\
  \bibinfo {author} {\bibfnamefont {E.}~\bibnamefont {Demler}},\ }\bibfield
  {title} {\bibinfo {title} {Fermi polarons in two dimensions},\ }\href
  {https://doi.org/10.1103/PhysRevA.85.021602} {\bibfield  {journal} {\bibinfo
  {journal} {Phys. Rev. A}\ }\textbf {\bibinfo {volume} {85}},\ \bibinfo
  {pages} {021602} (\bibinfo {year} {2012})}\BibitemShut {NoStop}%
\bibitem [{\citenamefont {Kohstall}\ \emph {et~al.}(2012)\citenamefont
  {Kohstall}, \citenamefont {Zaccanti}, \citenamefont {Jag}, \citenamefont
  {Trenkwalder}, \citenamefont {Massignan}, \citenamefont {Bruun},
  \citenamefont {Schreck},\ and\ \citenamefont {Grimm}}]{Kohstall2012}%
  \BibitemOpen
  \bibfield  {author} {\bibinfo {author} {\bibfnamefont {C.}~\bibnamefont
  {Kohstall}}, \bibinfo {author} {\bibfnamefont {M.}~\bibnamefont {Zaccanti}},
  \bibinfo {author} {\bibfnamefont {M.}~\bibnamefont {Jag}}, \bibinfo {author}
  {\bibfnamefont {A.}~\bibnamefont {Trenkwalder}}, \bibinfo {author}
  {\bibfnamefont {P.}~\bibnamefont {Massignan}}, \bibinfo {author}
  {\bibfnamefont {G.~M.}\ \bibnamefont {Bruun}}, \bibinfo {author}
  {\bibfnamefont {F.}~\bibnamefont {Schreck}},\ and\ \bibinfo {author}
  {\bibfnamefont {R.}~\bibnamefont {Grimm}},\ }\bibfield  {title} {\bibinfo
  {title} {{Metastability and coherence of repulsive polarons in a strongly
  interacting Fermi mixture}},\ }\href {https://doi.org/10.1038/Nat.11065}
  {\bibfield  {journal} {\bibinfo  {journal} {Nature}\ }\textbf {\bibinfo
  {volume} {485}},\ \bibinfo {pages} {615} (\bibinfo {year}
  {2012})}\BibitemShut {NoStop}%
\bibitem [{\citenamefont {Ngampruetikorn}\ \emph {et~al.}(2012)\citenamefont
  {Ngampruetikorn}, \citenamefont {Levinsen},\ and\ \citenamefont
  {Parish}}]{Ngampruetikorn2012}%
  \BibitemOpen
  \bibfield  {author} {\bibinfo {author} {\bibfnamefont {V.}~\bibnamefont
  {Ngampruetikorn}}, \bibinfo {author} {\bibfnamefont {J.}~\bibnamefont
  {Levinsen}},\ and\ \bibinfo {author} {\bibfnamefont {M.~M.}\ \bibnamefont
  {Parish}},\ }\bibfield  {title} {\bibinfo {title} {{Repulsive polarons in
  two-dimensional Fermi gases}},\ }\href
  {https://doi.org/10.1209/0295-5075/98/30005} {\bibfield  {journal} {\bibinfo
  {journal} {Eur. Phys. Let.}\ }\textbf {\bibinfo {volume} {98}},\ \bibinfo
  {pages} {30005} (\bibinfo {year} {2012})}\BibitemShut {NoStop}%
\bibitem [{\citenamefont {Zhang}\ \emph {et~al.}(2012)\citenamefont {Zhang},
  \citenamefont {Ong}, \citenamefont {Arakelyan},\ and\ \citenamefont
  {Thomas}}]{Zhang2012}%
  \BibitemOpen
  \bibfield  {author} {\bibinfo {author} {\bibfnamefont {Y.}~\bibnamefont
  {Zhang}}, \bibinfo {author} {\bibfnamefont {W.}~\bibnamefont {Ong}}, \bibinfo
  {author} {\bibfnamefont {I.}~\bibnamefont {Arakelyan}},\ and\ \bibinfo
  {author} {\bibfnamefont {J.~E.}\ \bibnamefont {Thomas}},\ }\bibfield  {title}
  {\bibinfo {title} {{Polaron-to-Polaron transitions in the radio-frequency
  spectrum of a quasi-two-dimensional Fermi gas}},\ }\href
  {https://doi.org/10.1103/PhysRevLett.108.235302} {\bibfield  {journal}
  {\bibinfo  {journal} {Phys. Rev. Lett.}\ }\textbf {\bibinfo {volume} {108}},\
  \bibinfo {pages} {235302} (\bibinfo {year} {2012})}\BibitemShut {NoStop}%
\bibitem [{\citenamefont {Catani}\ \emph {et~al.}(2012)\citenamefont {Catani},
  \citenamefont {Lamporesi}, \citenamefont {Naik}, \citenamefont {Gring},
  \citenamefont {Inguscio}, \citenamefont {Minardi}, \citenamefont {Kantian},\
  and\ \citenamefont {Giamarchi}}]{Catani2012}%
  \BibitemOpen
  \bibfield  {author} {\bibinfo {author} {\bibfnamefont {J.}~\bibnamefont
  {Catani}}, \bibinfo {author} {\bibfnamefont {G.}~\bibnamefont {Lamporesi}},
  \bibinfo {author} {\bibfnamefont {D.}~\bibnamefont {Naik}}, \bibinfo {author}
  {\bibfnamefont {M.}~\bibnamefont {Gring}}, \bibinfo {author} {\bibfnamefont
  {M.}~\bibnamefont {Inguscio}}, \bibinfo {author} {\bibfnamefont
  {F.}~\bibnamefont {Minardi}}, \bibinfo {author} {\bibfnamefont
  {A.}~\bibnamefont {Kantian}},\ and\ \bibinfo {author} {\bibfnamefont
  {T.}~\bibnamefont {Giamarchi}},\ }\bibfield  {title} {\bibinfo {title}
  {Quantum dynamics of impurities in a one-dimensional bose gas},\ }\href
  {https://doi.org/10.1103/PhysRevA.85.023623} {\bibfield  {journal} {\bibinfo
  {journal} {Phys. Rev. A}\ }\textbf {\bibinfo {volume} {85}},\ \bibinfo
  {pages} {023623} (\bibinfo {year} {2012})}\BibitemShut {NoStop}%
\bibitem [{\citenamefont {Fukuhara}\ \emph {et~al.}(2013)\citenamefont
  {Fukuhara}, \citenamefont {Kantian}, \citenamefont {Endres}, \citenamefont
  {Cheneau}, \citenamefont {Schau{\ss}}, \citenamefont {Hild}, \citenamefont
  {Bellem}, \citenamefont {Schollwöck}, \citenamefont {Giamarchi},
  \citenamefont {Gross}, \citenamefont {Bloch},\ and\ \citenamefont
  {Kuhr}}]{Fukuhara2013}%
  \BibitemOpen
  \bibfield  {author} {\bibinfo {author} {\bibfnamefont {T.}~\bibnamefont
  {Fukuhara}}, \bibinfo {author} {\bibfnamefont {A.}~\bibnamefont {Kantian}},
  \bibinfo {author} {\bibfnamefont {M.}~\bibnamefont {Endres}}, \bibinfo
  {author} {\bibfnamefont {M.}~\bibnamefont {Cheneau}}, \bibinfo {author}
  {\bibfnamefont {P.}~\bibnamefont {Schau{\ss}}}, \bibinfo {author}
  {\bibfnamefont {S.}~\bibnamefont {Hild}}, \bibinfo {author} {\bibfnamefont
  {D.}~\bibnamefont {Bellem}}, \bibinfo {author} {\bibfnamefont
  {U.}~\bibnamefont {Schollwöck}}, \bibinfo {author} {\bibfnamefont
  {T.}~\bibnamefont {Giamarchi}}, \bibinfo {author} {\bibfnamefont
  {C.}~\bibnamefont {Gross}}, \bibinfo {author} {\bibfnamefont
  {I.}~\bibnamefont {Bloch}},\ and\ \bibinfo {author} {\bibfnamefont
  {S.}~\bibnamefont {Kuhr}},\ }\bibfield  {title} {\bibinfo {title} {Quantum
  dynamics of a mobile spin impurity},\ }\href
  {https://doi.org/10.1038/nphys2561} {\bibfield  {journal} {\bibinfo
  {journal} {Nat. Phys.}\ }\textbf {\bibinfo {volume} {9}},\ \bibinfo {pages}
  {235} (\bibinfo {year} {2013})}\BibitemShut {NoStop}%
\bibitem [{\citenamefont {Burovski}\ \emph {et~al.}(2014)\citenamefont
  {Burovski}, \citenamefont {Cheianov}, \citenamefont {Gamayun},\ and\
  \citenamefont {Lychkovskiy}}]{Burovski2014}%
  \BibitemOpen
  \bibfield  {author} {\bibinfo {author} {\bibfnamefont {E.}~\bibnamefont
  {Burovski}}, \bibinfo {author} {\bibfnamefont {V.}~\bibnamefont {Cheianov}},
  \bibinfo {author} {\bibfnamefont {O.}~\bibnamefont {Gamayun}},\ and\ \bibinfo
  {author} {\bibfnamefont {O.}~\bibnamefont {Lychkovskiy}},\ }\bibfield
  {title} {\bibinfo {title} {Momentum relaxation of a mobile impurity in a
  one-dimensional quantum gas},\ }\href
  {https://doi.org/10.1103/PhysRevA.89.041601} {\bibfield  {journal} {\bibinfo
  {journal} {Phys. Rev. A}\ }\textbf {\bibinfo {volume} {89}},\ \bibinfo
  {pages} {041601} (\bibinfo {year} {2014})}\BibitemShut {NoStop}%
\bibitem [{\citenamefont {Ardila}\ and\ \citenamefont
  {Giorgini}(2016)}]{Ardila2016}%
  \BibitemOpen
  \bibfield  {author} {\bibinfo {author} {\bibfnamefont {L.~A. P.~n.}\
  \bibnamefont {Ardila}}\ and\ \bibinfo {author} {\bibfnamefont
  {S.}~\bibnamefont {Giorgini}},\ }\bibfield  {title} {\bibinfo {title} {Bose
  polaron problem: Effect of mass imbalance on binding energy},\ }\href
  {https://doi.org/10.1103/PhysRevA.94.063640} {\bibfield  {journal} {\bibinfo
  {journal} {Phys. Rev. A}\ }\textbf {\bibinfo {volume} {94}},\ \bibinfo
  {pages} {063640} (\bibinfo {year} {2016})}\BibitemShut {NoStop}%
\bibitem [{\citenamefont {Schmidt}\ \emph {et~al.}(2016)\citenamefont
  {Schmidt}, \citenamefont {Sadeghpour},\ and\ \citenamefont
  {Demler}}]{Schmidt2016B}%
  \BibitemOpen
  \bibfield  {author} {\bibinfo {author} {\bibfnamefont {R.}~\bibnamefont
  {Schmidt}}, \bibinfo {author} {\bibfnamefont {H.~R.}\ \bibnamefont
  {Sadeghpour}},\ and\ \bibinfo {author} {\bibfnamefont {E.}~\bibnamefont
  {Demler}},\ }\bibfield  {title} {\bibinfo {title} {{Mesoscopic Rydberg
  impurity in an atomic quantum gas}},\ }\href
  {https://doi.org/10.1103/PhysRevLett.116.105302} {\bibfield  {journal}
  {\bibinfo  {journal} {Phys. Rev. Lett.}\ }\textbf {\bibinfo {volume} {116}},\
  \bibinfo {pages} {105302} (\bibinfo {year} {2016})}\BibitemShut {NoStop}%
\bibitem [{\citenamefont {Grusdt}\ \emph {et~al.}(2017)\citenamefont {Grusdt},
  \citenamefont {Schmidt}, \citenamefont {Shchadilova},\ and\ \citenamefont
  {Demler}}]{Grusdt2017}%
  \BibitemOpen
  \bibfield  {author} {\bibinfo {author} {\bibfnamefont {F.}~\bibnamefont
  {Grusdt}}, \bibinfo {author} {\bibfnamefont {R.}~\bibnamefont {Schmidt}},
  \bibinfo {author} {\bibfnamefont {Y.~E.}\ \bibnamefont {Shchadilova}},\ and\
  \bibinfo {author} {\bibfnamefont {E.}~\bibnamefont {Demler}},\ }\bibfield
  {title} {\bibinfo {title} {{Strong-coupling Bose polarons in a Bose-Einstein
  condensate}},\ }\href {https://doi.org/10.1103/PhysRevA.96.013607} {\bibfield
   {journal} {\bibinfo  {journal} {Phys. Rev. A}\ }\textbf {\bibinfo {volume}
  {96}},\ \bibinfo {pages} {013607} (\bibinfo {year} {2017})}\BibitemShut
  {NoStop}%
\bibitem [{\citenamefont {Volosniev}\ and\ \citenamefont
  {Hammer}(2017)}]{Volosniev2017}%
  \BibitemOpen
  \bibfield  {author} {\bibinfo {author} {\bibfnamefont {A.~G.}\ \bibnamefont
  {Volosniev}}\ and\ \bibinfo {author} {\bibfnamefont {H.-W.}\ \bibnamefont
  {Hammer}},\ }\bibfield  {title} {\bibinfo {title} {Analytical approach to the
  bose-polaron problem in one dimension},\ }\href
  {https://doi.org/10.1103/PhysRevA.96.031601} {\bibfield  {journal} {\bibinfo
  {journal} {Phys. Rev. A}\ }\textbf {\bibinfo {volume} {96}},\ \bibinfo
  {pages} {031601} (\bibinfo {year} {2017})}\BibitemShut {NoStop}%
\bibitem [{\citenamefont {Scazza}\ \emph {et~al.}(2017)\citenamefont {Scazza},
  \citenamefont {Valtolina}, \citenamefont {Massignan}, \citenamefont {Recati},
  \citenamefont {Amico}, \citenamefont {Burchianti}, \citenamefont {Fort},
  \citenamefont {Inguscio}, \citenamefont {Zaccanti},\ and\ \citenamefont
  {Roati}}]{ScazzaValtolina2017}%
  \BibitemOpen
  \bibfield  {author} {\bibinfo {author} {\bibfnamefont {F.}~\bibnamefont
  {Scazza}}, \bibinfo {author} {\bibfnamefont {G.}~\bibnamefont {Valtolina}},
  \bibinfo {author} {\bibfnamefont {P.}~\bibnamefont {Massignan}}, \bibinfo
  {author} {\bibfnamefont {A.}~\bibnamefont {Recati}}, \bibinfo {author}
  {\bibfnamefont {A.}~\bibnamefont {Amico}}, \bibinfo {author} {\bibfnamefont
  {A.}~\bibnamefont {Burchianti}}, \bibinfo {author} {\bibfnamefont
  {C.}~\bibnamefont {Fort}}, \bibinfo {author} {\bibfnamefont {M.}~\bibnamefont
  {Inguscio}}, \bibinfo {author} {\bibfnamefont {M.}~\bibnamefont {Zaccanti}},\
  and\ \bibinfo {author} {\bibfnamefont {G.}~\bibnamefont {Roati}},\ }\bibfield
   {title} {\bibinfo {title} {{Repulsive Fermi polarons in a resonant mixture
  of ultracold $^{6}\mathrm{Li}$ atoms}},\ }\href
  {https://doi.org/10.1103/PhysRevLett.118.083602} {\bibfield  {journal}
  {\bibinfo  {journal} {Phys. Rev. Lett.}\ }\textbf {\bibinfo {volume} {118}},\
  \bibinfo {pages} {083602} (\bibinfo {year} {2017})}\BibitemShut {NoStop}%
\bibitem [{\citenamefont {Gamayun}\ \emph {et~al.}(2018)\citenamefont
  {Gamayun}, \citenamefont {Lychkovskiy}, \citenamefont {Burovski},
  \citenamefont {Malcomson}, \citenamefont {Cheianov},\ and\ \citenamefont
  {Zvonarev}}]{Gamayun2018}%
  \BibitemOpen
  \bibfield  {author} {\bibinfo {author} {\bibfnamefont {O.}~\bibnamefont
  {Gamayun}}, \bibinfo {author} {\bibfnamefont {O.}~\bibnamefont
  {Lychkovskiy}}, \bibinfo {author} {\bibfnamefont {E.}~\bibnamefont
  {Burovski}}, \bibinfo {author} {\bibfnamefont {M.}~\bibnamefont {Malcomson}},
  \bibinfo {author} {\bibfnamefont {V.~V.}\ \bibnamefont {Cheianov}},\ and\
  \bibinfo {author} {\bibfnamefont {M.~B.}\ \bibnamefont {Zvonarev}},\
  }\bibfield  {title} {\bibinfo {title} {Impact of the injection protocol on an
  impurity's stationary state},\ }\href
  {https://doi.org/10.1103/PhysRevLett.120.220605} {\bibfield  {journal}
  {\bibinfo  {journal} {Phys. Rev. Lett.}\ }\textbf {\bibinfo {volume} {120}},\
  \bibinfo {pages} {220605} (\bibinfo {year} {2018})}\BibitemShut {NoStop}%
\bibitem [{\citenamefont {Guenther}\ \emph {et~al.}(2018)\citenamefont
  {Guenther}, \citenamefont {Massignan}, \citenamefont {Lewenstein},\ and\
  \citenamefont {Bruun}}]{Guenther2018}%
  \BibitemOpen
  \bibfield  {author} {\bibinfo {author} {\bibfnamefont {N.-E.}\ \bibnamefont
  {Guenther}}, \bibinfo {author} {\bibfnamefont {P.}~\bibnamefont {Massignan}},
  \bibinfo {author} {\bibfnamefont {M.}~\bibnamefont {Lewenstein}},\ and\
  \bibinfo {author} {\bibfnamefont {G.~M.}\ \bibnamefont {Bruun}},\ }\bibfield
  {title} {\bibinfo {title} {Bose polarons at finite temperature and strong
  coupling},\ }\href {https://doi.org/10.1103/PhysRevLett.120.050405}
  {\bibfield  {journal} {\bibinfo  {journal} {Phys. Rev. Lett.}\ }\textbf
  {\bibinfo {volume} {120}},\ \bibinfo {pages} {050405} (\bibinfo {year}
  {2018})}\BibitemShut {NoStop}%
\bibitem [{\citenamefont {Schmidt}\ \emph {et~al.}(2018)\citenamefont
  {Schmidt}, \citenamefont {Knap}, \citenamefont {Ivanov}, \citenamefont {You},
  \citenamefont {Cetina},\ and\ \citenamefont {Demler}}]{Schmidt_2018}%
  \BibitemOpen
  \bibfield  {author} {\bibinfo {author} {\bibfnamefont {R.}~\bibnamefont
  {Schmidt}}, \bibinfo {author} {\bibfnamefont {M.}~\bibnamefont {Knap}},
  \bibinfo {author} {\bibfnamefont {D.~A.}\ \bibnamefont {Ivanov}}, \bibinfo
  {author} {\bibfnamefont {J.-S.}\ \bibnamefont {You}}, \bibinfo {author}
  {\bibfnamefont {M.}~\bibnamefont {Cetina}},\ and\ \bibinfo {author}
  {\bibfnamefont {E.}~\bibnamefont {Demler}},\ }\bibfield  {title} {\bibinfo
  {title} {{Universal many-body response of heavy impurities coupled to a Fermi
  sea: a review of recent progress}},\ }\href
  {https://doi.org/10.1088/1361-6633/aa9593} {\bibfield  {journal} {\bibinfo
  {journal} {Rep. Prog. Phys.}\ }\textbf {\bibinfo {volume} {81}},\ \bibinfo
  {pages} {024401} (\bibinfo {year} {2018})}\BibitemShut {NoStop}%
\bibitem [{\citenamefont {Mistakidis}\ \emph
  {et~al.}(2019{\natexlab{a}})\citenamefont {Mistakidis}, \citenamefont
  {Katsimiga}, \citenamefont {Koutentakis},\ and\ \citenamefont
  {Schmelcher}}]{MistakidisKatsimigaKoutentakis2019}%
  \BibitemOpen
  \bibfield  {author} {\bibinfo {author} {\bibfnamefont {S.~I.}\ \bibnamefont
  {Mistakidis}}, \bibinfo {author} {\bibfnamefont {G.~C.}\ \bibnamefont
  {Katsimiga}}, \bibinfo {author} {\bibfnamefont {G.~M.}\ \bibnamefont
  {Koutentakis}},\ and\ \bibinfo {author} {\bibfnamefont {P.}~\bibnamefont
  {Schmelcher}},\ }\bibfield  {title} {\bibinfo {title} {{Repulsive Fermi
  polarons and their induced interactions in binary mixtures of ultracold
  atoms}},\ }\href {https://doi.org/10.1088/1367-2630/ab1045} {\bibfield
  {journal} {\bibinfo  {journal} {New J. Phys.}\ }\textbf {\bibinfo {volume}
  {21}},\ \bibinfo {pages} {043032} (\bibinfo {year}
  {2019}{\natexlab{a}})}\BibitemShut {NoStop}%
\bibitem [{\citenamefont {Mistakidis}\ \emph
  {et~al.}(2019{\natexlab{b}})\citenamefont {Mistakidis}, \citenamefont
  {Katsimiga}, \citenamefont {Koutentakis}, \citenamefont {Busch},\ and\
  \citenamefont {Schmelcher}}]{MistakidisKatsimiga2019}%
  \BibitemOpen
  \bibfield  {author} {\bibinfo {author} {\bibfnamefont {S.~I.}\ \bibnamefont
  {Mistakidis}}, \bibinfo {author} {\bibfnamefont {G.~C.}\ \bibnamefont
  {Katsimiga}}, \bibinfo {author} {\bibfnamefont {G.~M.}\ \bibnamefont
  {Koutentakis}}, \bibinfo {author} {\bibfnamefont {T.}~\bibnamefont {Busch}},\
  and\ \bibinfo {author} {\bibfnamefont {P.}~\bibnamefont {Schmelcher}},\
  }\bibfield  {title} {\bibinfo {title} {{Quench dynamics and orthogonality
  catastrophe of Bose polarons}},\ }\href
  {https://doi.org/10.1103/PhysRevLett.122.183001} {\bibfield  {journal}
  {\bibinfo  {journal} {Phys. Rev. Lett.}\ }\textbf {\bibinfo {volume} {122}},\
  \bibinfo {pages} {183001} (\bibinfo {year} {2019}{\natexlab{b}})}\BibitemShut
  {NoStop}%
\bibitem [{\citenamefont {Mistakidis}\ \emph {et~al.}(2021)\citenamefont
  {Mistakidis}, \citenamefont {Koutentakis}, \citenamefont {Grusdt},
  \citenamefont {Sadeghpour},\ and\ \citenamefont
  {Schmelcher}}]{MistakidisKoutentakis2021}%
  \BibitemOpen
  \bibfield  {author} {\bibinfo {author} {\bibfnamefont {S.~I.}\ \bibnamefont
  {Mistakidis}}, \bibinfo {author} {\bibfnamefont {G.~M.}\ \bibnamefont
  {Koutentakis}}, \bibinfo {author} {\bibfnamefont {F.}~\bibnamefont {Grusdt}},
  \bibinfo {author} {\bibfnamefont {H.~R.}\ \bibnamefont {Sadeghpour}},\ and\
  \bibinfo {author} {\bibfnamefont {P.}~\bibnamefont {Schmelcher}},\ }\bibfield
   {title} {\bibinfo {title} {{Radiofrequency spectroscopy of one-dimensional
  trapped Bose polarons: crossover from the adiabatic to the diabatic
  regime}},\ }\href {https://doi.org/10.1088/1367-2630/abe9d5} {\bibfield
  {journal} {\bibinfo  {journal} {New J. Phys.}\ }\textbf {\bibinfo {volume}
  {23}},\ \bibinfo {pages} {043051} (\bibinfo {year} {2021})}\BibitemShut
  {NoStop}%
\bibitem [{\citenamefont {Taglieber}\ \emph {et~al.}(2008)\citenamefont
  {Taglieber}, \citenamefont {Voigt}, \citenamefont {Aoki}, \citenamefont
  {H\"ansch},\ and\ \citenamefont {Dieckmann}}]{Taglieber2008}%
  \BibitemOpen
  \bibfield  {author} {\bibinfo {author} {\bibfnamefont {M.}~\bibnamefont
  {Taglieber}}, \bibinfo {author} {\bibfnamefont {A.-C.}\ \bibnamefont
  {Voigt}}, \bibinfo {author} {\bibfnamefont {T.}~\bibnamefont {Aoki}},
  \bibinfo {author} {\bibfnamefont {T.~W.}\ \bibnamefont {H\"ansch}},\ and\
  \bibinfo {author} {\bibfnamefont {K.}~\bibnamefont {Dieckmann}},\ }\bibfield
  {title} {\bibinfo {title} {{Quantum degenerate two-species Fermi-Fermi
  mixture coexisting with a Bose-Einstein condensate}},\ }\href
  {https://doi.org/10.1103/PhysRevLett.100.010401} {\bibfield  {journal}
  {\bibinfo  {journal} {Phys. Rev. Lett.}\ }\textbf {\bibinfo {volume} {100}},\
  \bibinfo {pages} {010401} (\bibinfo {year} {2008})}\BibitemShut {NoStop}%
\bibitem [{\citenamefont {Tiecke}\ \emph {et~al.}(2010)\citenamefont {Tiecke},
  \citenamefont {Goosen}, \citenamefont {Ludewig}, \citenamefont {Gensemer},
  \citenamefont {Kraft}, \citenamefont {Kokkelmans},\ and\ \citenamefont
  {Walraven}}]{Tiecke2010}%
  \BibitemOpen
  \bibfield  {author} {\bibinfo {author} {\bibfnamefont {T.~G.}\ \bibnamefont
  {Tiecke}}, \bibinfo {author} {\bibfnamefont {M.~R.}\ \bibnamefont {Goosen}},
  \bibinfo {author} {\bibfnamefont {A.}~\bibnamefont {Ludewig}}, \bibinfo
  {author} {\bibfnamefont {S.~D.}\ \bibnamefont {Gensemer}}, \bibinfo {author}
  {\bibfnamefont {S.}~\bibnamefont {Kraft}}, \bibinfo {author} {\bibfnamefont
  {S.}~\bibnamefont {Kokkelmans}},\ and\ \bibinfo {author} {\bibfnamefont
  {J.~T.~M.}\ \bibnamefont {Walraven}},\ }\bibfield  {title} {\bibinfo {title}
  {Broad feshbach resonance in the
  $^{6}\mathrm{Li}\mathrm{\text{\ensuremath{-}}}^{40}\mathrm{K}$ mixture},\
  }\href {https://doi.org/10.1103/PhysRevLett.104.053202} {\bibfield  {journal}
  {\bibinfo  {journal} {Phys. Rev. Lett.}\ }\textbf {\bibinfo {volume} {104}},\
  \bibinfo {pages} {053202} (\bibinfo {year} {2010})}\BibitemShut {NoStop}%
\bibitem [{\citenamefont {Naik}\ \emph {et~al.}(2011)\citenamefont {Naik},
  \citenamefont {Trenkwalder}, \citenamefont {Kohstall}, \citenamefont
  {Spiegelhalder}, \citenamefont {Zaccanti}, \citenamefont {Hendl},
  \citenamefont {Schreck}, \citenamefont {Grimm}, \citenamefont {Hanna},\ and\
  \citenamefont {Julienne}}]{Naik_2011}%
  \BibitemOpen
  \bibfield  {author} {\bibinfo {author} {\bibfnamefont {D.}~\bibnamefont
  {Naik}}, \bibinfo {author} {\bibfnamefont {A.}~\bibnamefont {Trenkwalder}},
  \bibinfo {author} {\bibfnamefont {C.}~\bibnamefont {Kohstall}}, \bibinfo
  {author} {\bibfnamefont {F.~M.}\ \bibnamefont {Spiegelhalder}}, \bibinfo
  {author} {\bibfnamefont {M.}~\bibnamefont {Zaccanti}}, \bibinfo {author}
  {\bibfnamefont {G.}~\bibnamefont {Hendl}}, \bibinfo {author} {\bibfnamefont
  {F.}~\bibnamefont {Schreck}}, \bibinfo {author} {\bibfnamefont
  {R.}~\bibnamefont {Grimm}}, \bibinfo {author} {\bibfnamefont {T.~M.}\
  \bibnamefont {Hanna}},\ and\ \bibinfo {author} {\bibfnamefont {P.~S.}\
  \bibnamefont {Julienne}},\ }\bibfield  {title} {\bibinfo {title} {{Feshbach
  resonances in the $^6$Li-$^{40}$K Fermi-Fermi mixture: elastic versus
  inelastic interactions}},\ }\href
  {https://doi.org/10.1140/epjd/e2010-10591-2} {\bibfield  {journal} {\bibinfo
  {journal} {Europhys. J. D}\ }\textbf {\bibinfo {volume} {65}},\ \bibinfo
  {pages} {55} (\bibinfo {year} {2011})}\BibitemShut {NoStop}%
\bibitem [{\citenamefont {Cetina}\ \emph {et~al.}(2016)\citenamefont {Cetina},
  \citenamefont {Jag}, \citenamefont {Lous}, \citenamefont {Fritsche},
  \citenamefont {Walraven}, \citenamefont {Grimm}, \citenamefont {Levinsen},
  \citenamefont {Parish}, \citenamefont {Schmidt}, \citenamefont {Knap},\ and\
  \citenamefont {Demler}}]{Cetina2016}%
  \BibitemOpen
  \bibfield  {author} {\bibinfo {author} {\bibfnamefont {M.}~\bibnamefont
  {Cetina}}, \bibinfo {author} {\bibfnamefont {M.}~\bibnamefont {Jag}},
  \bibinfo {author} {\bibfnamefont {R.~S.}\ \bibnamefont {Lous}}, \bibinfo
  {author} {\bibfnamefont {I.}~\bibnamefont {Fritsche}}, \bibinfo {author}
  {\bibfnamefont {J.~T.~M.}\ \bibnamefont {Walraven}}, \bibinfo {author}
  {\bibfnamefont {R.}~\bibnamefont {Grimm}}, \bibinfo {author} {\bibfnamefont
  {J.}~\bibnamefont {Levinsen}}, \bibinfo {author} {\bibfnamefont {M.~M.}\
  \bibnamefont {Parish}}, \bibinfo {author} {\bibfnamefont {R.}~\bibnamefont
  {Schmidt}}, \bibinfo {author} {\bibfnamefont {M.}~\bibnamefont {Knap}},\ and\
  \bibinfo {author} {\bibfnamefont {E.}~\bibnamefont {Demler}},\ }\bibfield
  {title} {\bibinfo {title} {{Ultrafast many-body interferometry of impurities
  coupled to a Fermi sea}},\ }\href {https://doi.org/10.1126/science.aaf5134}
  {\bibfield  {journal} {\bibinfo  {journal} {Science}\ }\textbf {\bibinfo
  {volume} {354}},\ \bibinfo {pages} {96} (\bibinfo {year} {2016})}\BibitemShut
  {NoStop}%
\bibitem [{\citenamefont {Voigt}\ \emph {et~al.}(2009)\citenamefont {Voigt},
  \citenamefont {Taglieber}, \citenamefont {Costa}, \citenamefont {Aoki},
  \citenamefont {Wieser}, \citenamefont {H\"ansch},\ and\ \citenamefont
  {Dieckmann}}]{Voigt2009}%
  \BibitemOpen
  \bibfield  {author} {\bibinfo {author} {\bibfnamefont {A.-C.}\ \bibnamefont
  {Voigt}}, \bibinfo {author} {\bibfnamefont {M.}~\bibnamefont {Taglieber}},
  \bibinfo {author} {\bibfnamefont {L.}~\bibnamefont {Costa}}, \bibinfo
  {author} {\bibfnamefont {T.}~\bibnamefont {Aoki}}, \bibinfo {author}
  {\bibfnamefont {W.}~\bibnamefont {Wieser}}, \bibinfo {author} {\bibfnamefont
  {T.~W.}\ \bibnamefont {H\"ansch}},\ and\ \bibinfo {author} {\bibfnamefont
  {K.}~\bibnamefont {Dieckmann}},\ }\bibfield  {title} {\bibinfo {title}
  {{Ultracold heteronuclear Fermi-Fermi molecules}},\ }\href
  {https://doi.org/10.1103/PhysRevLett.102.020405} {\bibfield  {journal}
  {\bibinfo  {journal} {Phys. Rev. Lett.}\ }\textbf {\bibinfo {volume} {102}},\
  \bibinfo {pages} {020405} (\bibinfo {year} {2009})}\BibitemShut {NoStop}%
\bibitem [{\citenamefont {Nelson}\ \emph {et~al.}(2007)\citenamefont {Nelson},
  \citenamefont {Li},\ and\ \citenamefont {Weiss}}]{Nelson2007}%
  \BibitemOpen
  \bibfield  {author} {\bibinfo {author} {\bibfnamefont {K.~D.}\ \bibnamefont
  {Nelson}}, \bibinfo {author} {\bibfnamefont {X.}~\bibnamefont {Li}},\ and\
  \bibinfo {author} {\bibfnamefont {D.~S.}\ \bibnamefont {Weiss}},\ }\bibfield
  {title} {\bibinfo {title} {Imaging single atoms in a three-dimensional
  array},\ }\href {https://doi.org/10.1038/nphys645} {\bibfield  {journal}
  {\bibinfo  {journal} {Nat. Phys.}\ }\textbf {\bibinfo {volume} {3}},\
  \bibinfo {pages} {556} (\bibinfo {year} {2007})}\BibitemShut {NoStop}%
\bibitem [{\citenamefont {Sherson}\ \emph {et~al.}(2010)\citenamefont
  {Sherson}, \citenamefont {Weitenberg}, \citenamefont {Endres}, \citenamefont
  {Cheneau}, \citenamefont {Bloch},\ and\ \citenamefont {Kuhr}}]{Sherson2010}%
  \BibitemOpen
  \bibfield  {author} {\bibinfo {author} {\bibfnamefont {J.~F.}\ \bibnamefont
  {Sherson}}, \bibinfo {author} {\bibfnamefont {C.}~\bibnamefont {Weitenberg}},
  \bibinfo {author} {\bibfnamefont {M.}~\bibnamefont {Endres}}, \bibinfo
  {author} {\bibfnamefont {M.}~\bibnamefont {Cheneau}}, \bibinfo {author}
  {\bibfnamefont {I.}~\bibnamefont {Bloch}},\ and\ \bibinfo {author}
  {\bibfnamefont {S.}~\bibnamefont {Kuhr}},\ }\bibfield  {title} {\bibinfo
  {title} {{Single-atom-resolved fluorescence imaging of an atomic Mott
  insulator}},\ }\href {https://doi.org/10.1038/nature09378} {\bibfield
  {journal} {\bibinfo  {journal} {Nature}\ }\textbf {\bibinfo {volume} {467}},\
  \bibinfo {pages} {68} (\bibinfo {year} {2010})}\BibitemShut {NoStop}%
\bibitem [{\citenamefont {Bakr}\ \emph {et~al.}(2010)\citenamefont {Bakr},
  \citenamefont {Peng}, \citenamefont {Tai}, \citenamefont {Ma}, \citenamefont
  {Simon}, \citenamefont {Gillen}, \citenamefont {Fölling}, \citenamefont
  {Pollet},\ and\ \citenamefont {Greiner}}]{Bakr2011}%
  \BibitemOpen
  \bibfield  {author} {\bibinfo {author} {\bibfnamefont {W.~S.}\ \bibnamefont
  {Bakr}}, \bibinfo {author} {\bibfnamefont {A.}~\bibnamefont {Peng}}, \bibinfo
  {author} {\bibfnamefont {M.~E.}\ \bibnamefont {Tai}}, \bibinfo {author}
  {\bibfnamefont {R.}~\bibnamefont {Ma}}, \bibinfo {author} {\bibfnamefont
  {J.}~\bibnamefont {Simon}}, \bibinfo {author} {\bibfnamefont {J.~I.}\
  \bibnamefont {Gillen}}, \bibinfo {author} {\bibfnamefont {S.}~\bibnamefont
  {Fölling}}, \bibinfo {author} {\bibfnamefont {L.}~\bibnamefont {Pollet}},\
  and\ \bibinfo {author} {\bibfnamefont {M.}~\bibnamefont {Greiner}},\
  }\bibfield  {title} {\bibinfo {title} {{Probing the superfluid–to–Mott
  insulator transition at the single-atom level}},\ }\href
  {https://doi.org/10.1126/science.1192368} {\bibfield  {journal} {\bibinfo
  {journal} {Science}\ }\textbf {\bibinfo {volume} {329}},\ \bibinfo {pages}
  {547} (\bibinfo {year} {2010})}\BibitemShut {NoStop}%
\bibitem [{\citenamefont {Serwane}\ \emph {et~al.}(2011)\citenamefont
  {Serwane}, \citenamefont {Zürn}, \citenamefont {Lompe}, \citenamefont
  {Ottenstein}, \citenamefont {Wenz},\ and\ \citenamefont
  {Jochim}}]{Serwane2011}%
  \BibitemOpen
  \bibfield  {author} {\bibinfo {author} {\bibfnamefont {F.}~\bibnamefont
  {Serwane}}, \bibinfo {author} {\bibfnamefont {G.}~\bibnamefont {Zürn}},
  \bibinfo {author} {\bibfnamefont {T.}~\bibnamefont {Lompe}}, \bibinfo
  {author} {\bibfnamefont {T.~B.}\ \bibnamefont {Ottenstein}}, \bibinfo
  {author} {\bibfnamefont {A.~N.}\ \bibnamefont {Wenz}},\ and\ \bibinfo
  {author} {\bibfnamefont {S.}~\bibnamefont {Jochim}},\ }\bibfield  {title}
  {\bibinfo {title} {Deterministic preparation of a tunable few-fermion
  system},\ }\href {https://doi.org/10.1126/science.1201351} {\bibfield
  {journal} {\bibinfo  {journal} {Science}\ }\textbf {\bibinfo {volume}
  {332}},\ \bibinfo {pages} {336} (\bibinfo {year} {2011})}\BibitemShut
  {NoStop}%
\bibitem [{\citenamefont {Holten}\ \emph {et~al.}(2022)\citenamefont {Holten},
  \citenamefont {Bayha}, \citenamefont {Subramanian}, \citenamefont
  {Brandstetter}, \citenamefont {Heintze}, \citenamefont {Lunt}, \citenamefont
  {Preiss},\ and\ \citenamefont {Jochim}}]{Holten2022}%
  \BibitemOpen
  \bibfield  {author} {\bibinfo {author} {\bibfnamefont {M.}~\bibnamefont
  {Holten}}, \bibinfo {author} {\bibfnamefont {L.}~\bibnamefont {Bayha}},
  \bibinfo {author} {\bibfnamefont {K.}~\bibnamefont {Subramanian}}, \bibinfo
  {author} {\bibfnamefont {S.}~\bibnamefont {Brandstetter}}, \bibinfo {author}
  {\bibfnamefont {C.}~\bibnamefont {Heintze}}, \bibinfo {author} {\bibfnamefont
  {P.}~\bibnamefont {Lunt}}, \bibinfo {author} {\bibfnamefont {P.~M.}\
  \bibnamefont {Preiss}},\ and\ \bibinfo {author} {\bibfnamefont
  {S.}~\bibnamefont {Jochim}},\ }\bibfield  {title} {\bibinfo {title}
  {Observation of cooper pairs in a mesoscopic two-dimensional fermi gas},\
  }\href {https://doi.org/10.1038/s41586-022-04678-1} {\bibfield  {journal}
  {\bibinfo  {journal} {Nature}\ }\textbf {\bibinfo {volume} {606}},\ \bibinfo
  {pages} {287} (\bibinfo {year} {2022})}\BibitemShut {NoStop}%
\bibitem [{\citenamefont {Z\"urn}\ \emph {et~al.}(2013)\citenamefont {Z\"urn},
  \citenamefont {Wenz}, \citenamefont {Murmann}, \citenamefont {Bergschneider},
  \citenamefont {Lompe},\ and\ \citenamefont {Jochim}}]{ZuernWenzMurmann2013}%
  \BibitemOpen
  \bibfield  {author} {\bibinfo {author} {\bibfnamefont {G.}~\bibnamefont
  {Z\"urn}}, \bibinfo {author} {\bibfnamefont {A.~N.}\ \bibnamefont {Wenz}},
  \bibinfo {author} {\bibfnamefont {S.}~\bibnamefont {Murmann}}, \bibinfo
  {author} {\bibfnamefont {A.}~\bibnamefont {Bergschneider}}, \bibinfo {author}
  {\bibfnamefont {T.}~\bibnamefont {Lompe}},\ and\ \bibinfo {author}
  {\bibfnamefont {S.}~\bibnamefont {Jochim}},\ }\bibfield  {title} {\bibinfo
  {title} {Pairing in few-fermion systems with attractive interactions},\
  }\href {https://doi.org/10.1103/PhysRevLett.111.175302} {\bibfield  {journal}
  {\bibinfo  {journal} {Phys. Rev. Lett.}\ }\textbf {\bibinfo {volume} {111}},\
  \bibinfo {pages} {175302} (\bibinfo {year} {2013})}\BibitemShut {NoStop}%
\bibitem [{\citenamefont {Wenz}\ \emph {et~al.}(2013)\citenamefont {Wenz},
  \citenamefont {Zürn}, \citenamefont {Murmann}, \citenamefont {Brouzos},
  \citenamefont {Lompe},\ and\ \citenamefont {Jochim}}]{Wenz_2013}%
  \BibitemOpen
  \bibfield  {author} {\bibinfo {author} {\bibfnamefont {A.~N.}\ \bibnamefont
  {Wenz}}, \bibinfo {author} {\bibfnamefont {G.}~\bibnamefont {Zürn}},
  \bibinfo {author} {\bibfnamefont {S.}~\bibnamefont {Murmann}}, \bibinfo
  {author} {\bibfnamefont {I.}~\bibnamefont {Brouzos}}, \bibinfo {author}
  {\bibfnamefont {T.}~\bibnamefont {Lompe}},\ and\ \bibinfo {author}
  {\bibfnamefont {S.}~\bibnamefont {Jochim}},\ }\bibfield  {title} {\bibinfo
  {title} {From few to many: Observing the formation of a fermi sea one atom at
  a time},\ }\href {https://doi.org/10.1126/science.1240516} {\bibfield
  {journal} {\bibinfo  {journal} {Science}\ }\textbf {\bibinfo {volume}
  {342}},\ \bibinfo {pages} {457} (\bibinfo {year} {2013})}\BibitemShut
  {NoStop}%
\bibitem [{\citenamefont {Murmann}\ \emph {et~al.}(2015)\citenamefont
  {Murmann}, \citenamefont {Deuretzbacher}, \citenamefont {Z\"urn},
  \citenamefont {Bjerlin}, \citenamefont {Reimann}, \citenamefont {Santos},
  \citenamefont {Lompe},\ and\ \citenamefont {Jochim}}]{Murmann2015}%
  \BibitemOpen
  \bibfield  {author} {\bibinfo {author} {\bibfnamefont {S.}~\bibnamefont
  {Murmann}}, \bibinfo {author} {\bibfnamefont {F.}~\bibnamefont
  {Deuretzbacher}}, \bibinfo {author} {\bibfnamefont {G.}~\bibnamefont
  {Z\"urn}}, \bibinfo {author} {\bibfnamefont {J.}~\bibnamefont {Bjerlin}},
  \bibinfo {author} {\bibfnamefont {S.~M.}\ \bibnamefont {Reimann}}, \bibinfo
  {author} {\bibfnamefont {L.}~\bibnamefont {Santos}}, \bibinfo {author}
  {\bibfnamefont {T.}~\bibnamefont {Lompe}},\ and\ \bibinfo {author}
  {\bibfnamefont {S.}~\bibnamefont {Jochim}},\ }\bibfield  {title} {\bibinfo
  {title} {Antiferromagnetic heisenberg spin chain of a few cold atoms in a
  one-dimensional trap},\ }\href
  {https://doi.org/10.1103/PhysRevLett.115.215301} {\bibfield  {journal}
  {\bibinfo  {journal} {Phys. Rev. Lett.}\ }\textbf {\bibinfo {volume} {115}},\
  \bibinfo {pages} {215301} (\bibinfo {year} {2015})}\BibitemShut {NoStop}%
\bibitem [{\citenamefont {Bayha}\ \emph {et~al.}(2020)\citenamefont {Bayha},
  \citenamefont {Holten}, \citenamefont {Klemt}, \citenamefont {Subramanian},
  \citenamefont {Bjerlin}, \citenamefont {Reimann}, \citenamefont {Bruun},
  \citenamefont {Preiss},\ and\ \citenamefont {Jochim}}]{Bayha2020}%
  \BibitemOpen
  \bibfield  {author} {\bibinfo {author} {\bibfnamefont {L.}~\bibnamefont
  {Bayha}}, \bibinfo {author} {\bibfnamefont {M.}~\bibnamefont {Holten}},
  \bibinfo {author} {\bibfnamefont {R.}~\bibnamefont {Klemt}}, \bibinfo
  {author} {\bibfnamefont {K.}~\bibnamefont {Subramanian}}, \bibinfo {author}
  {\bibfnamefont {J.}~\bibnamefont {Bjerlin}}, \bibinfo {author} {\bibfnamefont
  {S.~M.}\ \bibnamefont {Reimann}}, \bibinfo {author} {\bibfnamefont {G.~M.}\
  \bibnamefont {Bruun}}, \bibinfo {author} {\bibfnamefont {P.~M.}\ \bibnamefont
  {Preiss}},\ and\ \bibinfo {author} {\bibfnamefont {S.}~\bibnamefont
  {Jochim}},\ }\bibfield  {title} {\bibinfo {title} {Observing the emergence of
  a quantum phase transition shell by shell},\ }\href
  {https://doi.org/10.1038/s41586-020-2936-y} {\bibfield  {journal} {\bibinfo
  {journal} {Nature}\ }\textbf {\bibinfo {volume} {587}},\ \bibinfo {pages}
  {583} (\bibinfo {year} {2020})}\BibitemShut {NoStop}%
\bibitem [{\citenamefont {Holten}\ \emph {et~al.}(2021)\citenamefont {Holten},
  \citenamefont {Bayha}, \citenamefont {Subramanian}, \citenamefont {Heintze},
  \citenamefont {Preiss},\ and\ \citenamefont {Jochim}}]{Holten2021}%
  \BibitemOpen
  \bibfield  {author} {\bibinfo {author} {\bibfnamefont {M.}~\bibnamefont
  {Holten}}, \bibinfo {author} {\bibfnamefont {L.}~\bibnamefont {Bayha}},
  \bibinfo {author} {\bibfnamefont {K.}~\bibnamefont {Subramanian}}, \bibinfo
  {author} {\bibfnamefont {C.}~\bibnamefont {Heintze}}, \bibinfo {author}
  {\bibfnamefont {P.~M.}\ \bibnamefont {Preiss}},\ and\ \bibinfo {author}
  {\bibfnamefont {S.}~\bibnamefont {Jochim}},\ }\bibfield  {title} {\bibinfo
  {title} {Observation of pauli crystals},\ }\href
  {https://doi.org/10.1103/PhysRevLett.126.020401} {\bibfield  {journal}
  {\bibinfo  {journal} {Phys. Rev. Lett.}\ }\textbf {\bibinfo {volume} {126}},\
  \bibinfo {pages} {020401} (\bibinfo {year} {2021})}\BibitemShut {NoStop}%
\bibitem [{\citenamefont {Krönke}\ \emph {et~al.}(2013)\citenamefont
  {Krönke}, \citenamefont {Cao}, \citenamefont {Vendrell},\ and\ \citenamefont
  {Schmelcher}}]{KrönkeCao2013}%
  \BibitemOpen
  \bibfield  {author} {\bibinfo {author} {\bibfnamefont {S.}~\bibnamefont
  {Krönke}}, \bibinfo {author} {\bibfnamefont {L.}~\bibnamefont {Cao}},
  \bibinfo {author} {\bibfnamefont {O.}~\bibnamefont {Vendrell}},\ and\
  \bibinfo {author} {\bibfnamefont {P.}~\bibnamefont {Schmelcher}},\ }\bibfield
   {title} {\bibinfo {title} {{Non-equilibrium quantum dynamics of ultra-cold
  atomic mixtures: the multi-layer multi-configuration time-dependent Hartree
  method for bosons}},\ }\href {https://doi.org/10.1088/1367-2630/15/6/063018}
  {\bibfield  {journal} {\bibinfo  {journal} {New J. Phys.}\ }\textbf {\bibinfo
  {volume} {15}},\ \bibinfo {pages} {063018} (\bibinfo {year}
  {2013})}\BibitemShut {NoStop}%
\bibitem [{\citenamefont {Cao}\ \emph {et~al.}(2013)\citenamefont {Cao},
  \citenamefont {Krönke}, \citenamefont {Vendrell},\ and\ \citenamefont
  {Schmelcher}}]{CaoLushuai2013}%
  \BibitemOpen
  \bibfield  {author} {\bibinfo {author} {\bibfnamefont {L.}~\bibnamefont
  {Cao}}, \bibinfo {author} {\bibfnamefont {S.}~\bibnamefont {Krönke}},
  \bibinfo {author} {\bibfnamefont {O.}~\bibnamefont {Vendrell}},\ and\
  \bibinfo {author} {\bibfnamefont {P.}~\bibnamefont {Schmelcher}},\ }\bibfield
   {title} {\bibinfo {title} {{The multi-layer multi-configuration
  time-dependent Hartree method for bosons: Theory, implementation, and
  applications}},\ }\href {https://doi.org/10.1063/1.4821350} {\bibfield
  {journal} {\bibinfo  {journal} {Chem. Phys.}\ }\textbf {\bibinfo {volume}
  {139}},\ \bibinfo {pages} {134103} (\bibinfo {year} {2013})}\BibitemShut
  {NoStop}%
\bibitem [{\citenamefont {Cao}\ \emph {et~al.}(2017)\citenamefont {Cao},
  \citenamefont {Bolsinger}, \citenamefont {Mistakidis}, \citenamefont
  {Koutentakis}, \citenamefont {Krönke}, \citenamefont {Schurer},\ and\
  \citenamefont {Schmelcher}}]{CaoBolsinger2017}%
  \BibitemOpen
  \bibfield  {author} {\bibinfo {author} {\bibfnamefont {L.}~\bibnamefont
  {Cao}}, \bibinfo {author} {\bibfnamefont {V.}~\bibnamefont {Bolsinger}},
  \bibinfo {author} {\bibfnamefont {S.~I.}\ \bibnamefont {Mistakidis}},
  \bibinfo {author} {\bibfnamefont {G.~M.}\ \bibnamefont {Koutentakis}},
  \bibinfo {author} {\bibfnamefont {S.}~\bibnamefont {Krönke}}, \bibinfo
  {author} {\bibfnamefont {J.~M.}\ \bibnamefont {Schurer}},\ and\ \bibinfo
  {author} {\bibfnamefont {P.}~\bibnamefont {Schmelcher}},\ }\bibfield  {title}
  {\bibinfo {title} {A unified ab initio approach to the correlated quantum
  dynamics of ultracold fermionic and bosonic mixtures},\ }\href
  {https://doi.org/10.1063/1.4993512} {\bibfield  {journal} {\bibinfo
  {journal} {J. Chem. Phys.}\ }\textbf {\bibinfo {volume} {147}},\ \bibinfo
  {pages} {044106} (\bibinfo {year} {2017})}\BibitemShut {NoStop}%
\bibitem [{\citenamefont {Hummel}\ \emph {et~al.}(2021)\citenamefont {Hummel},
  \citenamefont {Eiles},\ and\ \citenamefont {Schmelcher}}]{Hummel2021}%
  \BibitemOpen
  \bibfield  {author} {\bibinfo {author} {\bibfnamefont {F.}~\bibnamefont
  {Hummel}}, \bibinfo {author} {\bibfnamefont {M.~T.}\ \bibnamefont {Eiles}},\
  and\ \bibinfo {author} {\bibfnamefont {P.}~\bibnamefont {Schmelcher}},\
  }\bibfield  {title} {\bibinfo {title} {{Synthetic dimension-induced conical
  intersections in Rydberg molecules}},\ }\href
  {https://doi.org/10.1103/PhysRevLett.127.023003} {\bibfield  {journal}
  {\bibinfo  {journal} {Phys. Rev. Lett.}\ }\textbf {\bibinfo {volume} {127}},\
  \bibinfo {pages} {023003} (\bibinfo {year} {2021})}\BibitemShut {NoStop}%
\bibitem [{\citenamefont {Levi}\ \emph {et~al.}(2011)\citenamefont {Levi},
  \citenamefont {Rechtsman}, \citenamefont {Freedman}, \citenamefont
  {Schwartz}, \citenamefont {Manela},\ and\ \citenamefont {Segev}}]{Levi2011}%
  \BibitemOpen
  \bibfield  {author} {\bibinfo {author} {\bibfnamefont {L.}~\bibnamefont
  {Levi}}, \bibinfo {author} {\bibfnamefont {M.}~\bibnamefont {Rechtsman}},
  \bibinfo {author} {\bibfnamefont {B.}~\bibnamefont {Freedman}}, \bibinfo
  {author} {\bibfnamefont {T.}~\bibnamefont {Schwartz}}, \bibinfo {author}
  {\bibfnamefont {O.}~\bibnamefont {Manela}},\ and\ \bibinfo {author}
  {\bibfnamefont {M.}~\bibnamefont {Segev}},\ }\bibfield  {title} {\bibinfo
  {title} {{Disorder-enhanced transport in photonic quasicrystals}},\ }\href
  {https://doi.org/10.1126/science.1202977} {\bibfield  {journal} {\bibinfo
  {journal} {Science}\ }\textbf {\bibinfo {volume} {332}},\ \bibinfo {pages}
  {1541} (\bibinfo {year} {2011})}\BibitemShut {NoStop}%
\bibitem [{\citenamefont {Kolkowitz}\ \emph {et~al.}(2016)\citenamefont
  {Kolkowitz}, \citenamefont {Bromley}, \citenamefont {Bothwell}, \citenamefont
  {Wall}, \citenamefont {Marti}, \citenamefont {Koller}, \citenamefont {Zhang},
  \citenamefont {Rey},\ and\ \citenamefont {Ye}}]{Kolkowitz2016}%
  \BibitemOpen
  \bibfield  {author} {\bibinfo {author} {\bibfnamefont {S.}~\bibnamefont
  {Kolkowitz}}, \bibinfo {author} {\bibfnamefont {S.~L.}\ \bibnamefont
  {Bromley}}, \bibinfo {author} {\bibfnamefont {T.}~\bibnamefont {Bothwell}},
  \bibinfo {author} {\bibfnamefont {M.~L.}\ \bibnamefont {Wall}}, \bibinfo
  {author} {\bibfnamefont {G.~E.}\ \bibnamefont {Marti}}, \bibinfo {author}
  {\bibfnamefont {A.~P.}\ \bibnamefont {Koller}}, \bibinfo {author}
  {\bibfnamefont {X.}~\bibnamefont {Zhang}}, \bibinfo {author} {\bibfnamefont
  {A.~M.}\ \bibnamefont {Rey}},\ and\ \bibinfo {author} {\bibfnamefont
  {J.}~\bibnamefont {Ye}},\ }\bibfield  {title} {\bibinfo {title}
  {Spin{\textendash}orbit-coupled fermions in an optical lattice clock},\
  }\href {https://doi.org/10.1038/nature20811} {\bibfield  {journal} {\bibinfo
  {journal} {Nature}\ }\textbf {\bibinfo {volume} {542}},\ \bibinfo {pages}
  {66} (\bibinfo {year} {2016})}\BibitemShut {NoStop}%
\bibitem [{\citenamefont {Ozawa}\ \emph {et~al.}(2019)\citenamefont {Ozawa},
  \citenamefont {Price}, \citenamefont {Amo}, \citenamefont {Goldman},
  \citenamefont {Hafezi}, \citenamefont {Lu}, \citenamefont {Rechtsman},
  \citenamefont {Schuster}, \citenamefont {Simon}, \citenamefont {Zilberberg},\
  and\ \citenamefont {Carusotto}}]{Ozawa2019}%
  \BibitemOpen
  \bibfield  {author} {\bibinfo {author} {\bibfnamefont {T.}~\bibnamefont
  {Ozawa}}, \bibinfo {author} {\bibfnamefont {H.~M.}\ \bibnamefont {Price}},
  \bibinfo {author} {\bibfnamefont {A.}~\bibnamefont {Amo}}, \bibinfo {author}
  {\bibfnamefont {N.}~\bibnamefont {Goldman}}, \bibinfo {author} {\bibfnamefont
  {M.}~\bibnamefont {Hafezi}}, \bibinfo {author} {\bibfnamefont
  {L.}~\bibnamefont {Lu}}, \bibinfo {author} {\bibfnamefont {M.~C.}\
  \bibnamefont {Rechtsman}}, \bibinfo {author} {\bibfnamefont {D.}~\bibnamefont
  {Schuster}}, \bibinfo {author} {\bibfnamefont {J.}~\bibnamefont {Simon}},
  \bibinfo {author} {\bibfnamefont {O.}~\bibnamefont {Zilberberg}},\ and\
  \bibinfo {author} {\bibfnamefont {I.}~\bibnamefont {Carusotto}},\ }\bibfield
  {title} {\bibinfo {title} {Topological photonics},\ }\href
  {https://doi.org/10.1103/RevModPhys.91.015006} {\bibfield  {journal}
  {\bibinfo  {journal} {Rev. Mod. Phys.}\ }\textbf {\bibinfo {volume} {91}},\
  \bibinfo {pages} {015006} (\bibinfo {year} {2019})}\BibitemShut {NoStop}%
\bibitem [{\citenamefont {Celi}\ \emph {et~al.}(2014)\citenamefont {Celi},
  \citenamefont {Massignan}, \citenamefont {Ruseckas}, \citenamefont {Goldman},
  \citenamefont {Spielman}, \citenamefont {Juzeli\ifmmode~\bar{u}\else
  \={u}\fi{}nas},\ and\ \citenamefont {Lewenstein}}]{Celi2014}%
  \BibitemOpen
  \bibfield  {author} {\bibinfo {author} {\bibfnamefont {A.}~\bibnamefont
  {Celi}}, \bibinfo {author} {\bibfnamefont {P.}~\bibnamefont {Massignan}},
  \bibinfo {author} {\bibfnamefont {J.}~\bibnamefont {Ruseckas}}, \bibinfo
  {author} {\bibfnamefont {N.}~\bibnamefont {Goldman}}, \bibinfo {author}
  {\bibfnamefont {I.~B.}\ \bibnamefont {Spielman}}, \bibinfo {author}
  {\bibfnamefont {G.}~\bibnamefont {Juzeli\ifmmode~\bar{u}\else
  \={u}\fi{}nas}},\ and\ \bibinfo {author} {\bibfnamefont {M.}~\bibnamefont
  {Lewenstein}},\ }\bibfield  {title} {\bibinfo {title} {Synthetic gauge fields
  in synthetic dimensions},\ }\href
  {https://doi.org/10.1103/PhysRevLett.112.043001} {\bibfield  {journal}
  {\bibinfo  {journal} {Phys. Rev. Lett.}\ }\textbf {\bibinfo {volume} {112}},\
  \bibinfo {pages} {043001} (\bibinfo {year} {2014})}\BibitemShut {NoStop}%
\bibitem [{\citenamefont {Boada}\ \emph {et~al.}(2012)\citenamefont {Boada},
  \citenamefont {Celi}, \citenamefont {Latorre},\ and\ \citenamefont
  {Lewenstein}}]{Boada2012}%
  \BibitemOpen
  \bibfield  {author} {\bibinfo {author} {\bibfnamefont {O.}~\bibnamefont
  {Boada}}, \bibinfo {author} {\bibfnamefont {A.}~\bibnamefont {Celi}},
  \bibinfo {author} {\bibfnamefont {J.~I.}\ \bibnamefont {Latorre}},\ and\
  \bibinfo {author} {\bibfnamefont {M.}~\bibnamefont {Lewenstein}},\ }\bibfield
   {title} {\bibinfo {title} {Quantum simulation of an extra dimension},\
  }\href {https://doi.org/10.1103/PhysRevLett.108.133001} {\bibfield  {journal}
  {\bibinfo  {journal} {Phys. Rev. Lett.}\ }\textbf {\bibinfo {volume} {108}},\
  \bibinfo {pages} {133001} (\bibinfo {year} {2012})}\BibitemShut {NoStop}%
\bibitem [{\citenamefont {Olshanii}(1998)}]{Olshanii1998}%
  \BibitemOpen
  \bibfield  {author} {\bibinfo {author} {\bibfnamefont {M.}~\bibnamefont
  {Olshanii}},\ }\bibfield  {title} {\bibinfo {title} {Atomic scattering in the
  presence of an external confinement and a gas of impenetrable bosons},\
  }\href {https://doi.org/10.1103/PhysRevLett.81.938} {\bibfield  {journal}
  {\bibinfo  {journal} {Phys. Rev. Lett.}\ }\textbf {\bibinfo {volume} {81}},\
  \bibinfo {pages} {938} (\bibinfo {year} {1998})}\BibitemShut {NoStop}%
\bibitem [{\citenamefont {Chin}\ \emph {et~al.}(2010)\citenamefont {Chin},
  \citenamefont {Grimm}, \citenamefont {Julienne},\ and\ \citenamefont
  {Tiesinga}}]{ChinChengGrimm2010}%
  \BibitemOpen
  \bibfield  {author} {\bibinfo {author} {\bibfnamefont {C.}~\bibnamefont
  {Chin}}, \bibinfo {author} {\bibfnamefont {R.}~\bibnamefont {Grimm}},
  \bibinfo {author} {\bibfnamefont {P.}~\bibnamefont {Julienne}},\ and\
  \bibinfo {author} {\bibfnamefont {E.}~\bibnamefont {Tiesinga}},\ }\bibfield
  {title} {\bibinfo {title} {Feshbach resonances in ultracold gases},\ }\href
  {https://doi.org/10.1103/RevModPhys.82.1225} {\bibfield  {journal} {\bibinfo
  {journal} {Rev. Mod. Phys.}\ }\textbf {\bibinfo {volume} {82}},\ \bibinfo
  {pages} {1225} (\bibinfo {year} {2010})}\BibitemShut {NoStop}%
\bibitem [{\citenamefont {Born}\ and\ \citenamefont
  {Oppenheimer}(1927)}]{BornOppenheimer1927}%
  \BibitemOpen
  \bibfield  {author} {\bibinfo {author} {\bibfnamefont {M.}~\bibnamefont
  {Born}}\ and\ \bibinfo {author} {\bibfnamefont {R.}~\bibnamefont
  {Oppenheimer}},\ }\bibfield  {title} {\bibinfo {title} {{Zur Quantentheorie
  der Molek\"ule}},\ }\href
  {https://doi.org/https://doi.org/10.1002/andp.19273892002} {\bibfield
  {journal} {\bibinfo  {journal} {Ann. Phys.}\ }\textbf {\bibinfo {volume}
  {389}},\ \bibinfo {pages} {457} (\bibinfo {year} {1927})}\BibitemShut
  {NoStop}%
\bibitem [{\citenamefont {Schuster}\ \emph {et~al.}(2012)\citenamefont
  {Schuster}, \citenamefont {Scelle}, \citenamefont {Trautmann}, \citenamefont
  {Knoop}, \citenamefont {Oberthaler}, \citenamefont {Haverhals}, \citenamefont
  {Goosen}, \citenamefont {Kokkelmans},\ and\ \citenamefont
  {Tiemann}}]{Schuster2012}%
  \BibitemOpen
  \bibfield  {author} {\bibinfo {author} {\bibfnamefont {T.}~\bibnamefont
  {Schuster}}, \bibinfo {author} {\bibfnamefont {R.}~\bibnamefont {Scelle}},
  \bibinfo {author} {\bibfnamefont {A.}~\bibnamefont {Trautmann}}, \bibinfo
  {author} {\bibfnamefont {S.}~\bibnamefont {Knoop}}, \bibinfo {author}
  {\bibfnamefont {M.~K.}\ \bibnamefont {Oberthaler}}, \bibinfo {author}
  {\bibfnamefont {M.~M.}\ \bibnamefont {Haverhals}}, \bibinfo {author}
  {\bibfnamefont {M.~R.}\ \bibnamefont {Goosen}}, \bibinfo {author}
  {\bibfnamefont {S.}~\bibnamefont {Kokkelmans}},\ and\ \bibinfo {author}
  {\bibfnamefont {E.}~\bibnamefont {Tiemann}},\ }\bibfield  {title} {\bibinfo
  {title} {{Feshbach spectroscopy and scattering properties of ultracold Li +
  Na mixtures}},\ }\href {https://doi.org/10.1103/PhysRevA.85.042721}
  {\bibfield  {journal} {\bibinfo  {journal} {Phys. Rev. A}\ }\textbf {\bibinfo
  {volume} {85}},\ \bibinfo {pages} {042721} (\bibinfo {year}
  {2012})}\BibitemShut {NoStop}%
\bibitem [{\citenamefont {Dirac}(1930)}]{Dirac1930annihilation}%
  \BibitemOpen
  \bibfield  {author} {\bibinfo {author} {\bibfnamefont {P.~A.~M.}\
  \bibnamefont {Dirac}},\ }\bibfield  {title} {\bibinfo {title} {On the
  annihilation of electrons and protons},\ }\href
  {https://doi.org/10.1017/S0305004100016091} {\bibfield  {journal} {\bibinfo
  {journal} {Proc. Cambridge Phil. Soc.}\ }\textbf {\bibinfo {volume} {26}},\
  \bibinfo {pages} {361–375} (\bibinfo {year} {1930})}\BibitemShut {NoStop}%
\bibitem [{\citenamefont {Frenkel}(1934)}]{frenkel1934wave}%
  \BibitemOpen
  \bibfield  {author} {\bibinfo {author} {\bibfnamefont {J.}~\bibnamefont
  {Frenkel}},\ }\href@noop {} {\emph {\bibinfo {title} {Wave mechanics,
  advanced general theory}}},\ Vol.~\bibinfo {volume} {1}\ (\bibinfo
  {publisher} {Oxford University Press},\ \bibinfo {year} {1934})\BibitemShut
  {NoStop}%
\bibitem [{\citenamefont {Köppel}\ \emph {et~al.}(1984)\citenamefont
  {Köppel}, \citenamefont {Domcke},\ and\ \citenamefont
  {Cederbaum}}]{Köppel1984}%
  \BibitemOpen
  \bibfield  {author} {\bibinfo {author} {\bibfnamefont {H.}~\bibnamefont
  {Köppel}}, \bibinfo {author} {\bibfnamefont {W.}~\bibnamefont {Domcke}},\
  and\ \bibinfo {author} {\bibfnamefont {L.~S.}\ \bibnamefont {Cederbaum}},\
  }\bibinfo {title} {{Multimode molecular dynamics beyond the Born-Oppenheimer
  approximation}},\ in\ \href
  {https://doi.org/https://doi.org/10.1002/9780470142813.ch2} {\emph {\bibinfo
  {booktitle} {Advances in Chemical Physics}}}\ (\bibinfo  {publisher} {John
  Wiley \& Sons, Ltd},\ \bibinfo {year} {1984})\ pp.\ \bibinfo {pages}
  {59--246}\BibitemShut {NoStop}%
\bibitem [{\citenamefont {Pacher}\ \emph {et~al.}(1989)\citenamefont {Pacher},
  \citenamefont {Mead}, \citenamefont {Cederbaum},\ and\ \citenamefont
  {Köppel}}]{Pacher1989}%
  \BibitemOpen
  \bibfield  {author} {\bibinfo {author} {\bibfnamefont {T.}~\bibnamefont
  {Pacher}}, \bibinfo {author} {\bibfnamefont {C.~A.}\ \bibnamefont {Mead}},
  \bibinfo {author} {\bibfnamefont {L.~S.}\ \bibnamefont {Cederbaum}},\ and\
  \bibinfo {author} {\bibfnamefont {H.}~\bibnamefont {Köppel}},\ }\bibfield
  {title} {\bibinfo {title} {{Gauge theory and quasidiabatic states in
  molecular physics}},\ }\href {https://doi.org/10.1063/1.457323} {\bibfield
  {journal} {\bibinfo  {journal} {Chem. Phys.}\ }\textbf {\bibinfo {volume}
  {91}},\ \bibinfo {pages} {7057} (\bibinfo {year} {1989})}\BibitemShut
  {NoStop}%
\bibitem [{\citenamefont {Epstein}(1966)}]{Epstein1966}%
  \BibitemOpen
  \bibfield  {author} {\bibinfo {author} {\bibfnamefont {S.~T.}\ \bibnamefont
  {Epstein}},\ }\bibfield  {title} {\bibinfo {title} {{Ground‐state energy of
  a molecule in the adiabatic approximation}},\ }\href
  {https://doi.org/10.1063/1.1726771} {\bibfield  {journal} {\bibinfo
  {journal} {Chem. Phys.}\ }\textbf {\bibinfo {volume} {44}},\ \bibinfo {pages}
  {836} (\bibinfo {year} {1966})}\BibitemShut {NoStop}%
\bibitem [{\citenamefont {Baer}(2006)}]{Baer2006}%
  \BibitemOpen
  \bibfield  {author} {\bibinfo {author} {\bibfnamefont {M.}~\bibnamefont
  {Baer}},\ }\href@noop {} {\emph {\bibinfo {title} {Beyond
  Born-Oppenheimer}}},\ \bibinfo {edition} {1st}\ ed.\ (\bibinfo  {publisher}
  {John Wiley \& Sons},\ \bibinfo {address} {Hoboken, New Jersey},\ \bibinfo
  {year} {2006})\BibitemShut {NoStop}%
\bibitem [{\citenamefont {Gurari}(1953)}]{Gurari1953}%
  \BibitemOpen
  \bibfield  {author} {\bibinfo {author} {\bibfnamefont {M.}~\bibnamefont
  {Gurari}},\ }\bibfield  {title} {\bibinfo {title} {{XXXVI. Self energy of
  slow electrons in polar materials}},\ }\href
  {https://doi.org/10.1080/14786440308520313} {\bibfield  {journal} {\bibinfo
  {journal} {London, Edinburgh, Dublin Philos. Mag. J. Sci.}\ }\textbf
  {\bibinfo {volume} {44}},\ \bibinfo {pages} {329} (\bibinfo {year}
  {1953})}\BibitemShut {NoStop}%
\bibitem [{\citenamefont {Lee}\ and\ \citenamefont
  {Pines}(1952)}]{LeePines1952}%
  \BibitemOpen
  \bibfield  {author} {\bibinfo {author} {\bibfnamefont {T.-D.}\ \bibnamefont
  {Lee}}\ and\ \bibinfo {author} {\bibfnamefont {D.}~\bibnamefont {Pines}},\
  }\bibfield  {title} {\bibinfo {title} {The motion of slow electrons in polar
  crystals},\ }\href {https://doi.org/10.1103/physrev.88.960} {\bibfield
  {journal} {\bibinfo  {journal} {Phys. Rev.}\ }\textbf {\bibinfo {volume}
  {88}},\ \bibinfo {pages} {960} (\bibinfo {year} {1952})}\BibitemShut
  {NoStop}%
\bibitem [{\citenamefont {Lee}\ \emph {et~al.}(1953)\citenamefont {Lee},
  \citenamefont {Low},\ and\ \citenamefont {Pines}}]{LeeLow1953}%
  \BibitemOpen
  \bibfield  {author} {\bibinfo {author} {\bibfnamefont {T.~D.}\ \bibnamefont
  {Lee}}, \bibinfo {author} {\bibfnamefont {F.~E.}\ \bibnamefont {Low}},\ and\
  \bibinfo {author} {\bibfnamefont {D.}~\bibnamefont {Pines}},\ }\bibfield
  {title} {\bibinfo {title} {The motion of slow electrons in a polar crystal},\
  }\href {https://doi.org/10.1103/physrev.90.297} {\bibfield  {journal}
  {\bibinfo  {journal} {Phys. Rev.}\ }\textbf {\bibinfo {volume} {90}},\
  \bibinfo {pages} {297} (\bibinfo {year} {1953})}\BibitemShut {NoStop}%
\bibitem [{\citenamefont {Busch}\ \emph {et~al.}(1998)\citenamefont {Busch},
  \citenamefont {Englert}, \citenamefont {Rza{\.{z}}ewski},\ and\ \citenamefont
  {Wilkens}}]{BuschEnglert1998}%
  \BibitemOpen
  \bibfield  {author} {\bibinfo {author} {\bibfnamefont {T.}~\bibnamefont
  {Busch}}, \bibinfo {author} {\bibfnamefont {B.-G.}\ \bibnamefont {Englert}},
  \bibinfo {author} {\bibfnamefont {K.}~\bibnamefont {Rza{\.{z}}ewski}},\ and\
  \bibinfo {author} {\bibfnamefont {M.}~\bibnamefont {Wilkens}},\ }\bibfield
  {title} {\bibinfo {title} {Two cold atoms in a harmonic trap},\ }\href
  {https://doi.org/10.1023/A:1018705520999} {\bibfield  {journal} {\bibinfo
  {journal} {Found. Phys.}\ }\textbf {\bibinfo {volume} {28}},\ \bibinfo
  {pages} {549} (\bibinfo {year} {1998})}\BibitemShut {NoStop}%
\bibitem [{\citenamefont {Farrell}\ and\ \citenamefont {van
  Zyl}(2009)}]{FarrellVanZyl2010}%
  \BibitemOpen
  \bibfield  {author} {\bibinfo {author} {\bibfnamefont {A.}~\bibnamefont
  {Farrell}}\ and\ \bibinfo {author} {\bibfnamefont {B.~P.}\ \bibnamefont {van
  Zyl}},\ }\bibfield  {title} {\bibinfo {title} {Universality of the energy
  spectrum for two interacting harmonically trapped ultra-cold atoms in one and
  two dimensions},\ }\href {https://doi.org/10.1088/1751-8113/43/1/015302}
  {\bibfield  {journal} {\bibinfo  {journal} {J. Phys. A: Math. Theor.}\
  }\textbf {\bibinfo {volume} {43}},\ \bibinfo {pages} {015302} (\bibinfo
  {year} {2009})}\BibitemShut {NoStop}%
\bibitem [{\citenamefont {Budewig}\ \emph {et~al.}(2019)\citenamefont
  {Budewig}, \citenamefont {Mistakidis},\ and\ \citenamefont
  {Schmelcher}}]{BudewigMistakidis2019}%
  \BibitemOpen
  \bibfield  {author} {\bibinfo {author} {\bibfnamefont {L.}~\bibnamefont
  {Budewig}}, \bibinfo {author} {\bibfnamefont {S.~I.}\ \bibnamefont
  {Mistakidis}},\ and\ \bibinfo {author} {\bibfnamefont {P.}~\bibnamefont
  {Schmelcher}},\ }\bibfield  {title} {\bibinfo {title} {Quench dynamics of two
  one-dimensional harmonically trapped bosons bridging attraction and
  repulsion},\ }\href {https://doi.org/10.1080/00268976.2019.1575995}
  {\bibfield  {journal} {\bibinfo  {journal} {Mol. Phys.}\ }\textbf {\bibinfo
  {volume} {117}},\ \bibinfo {pages} {2043} (\bibinfo {year}
  {2019})}\BibitemShut {NoStop}%
\bibitem [{\citenamefont {Domcke}\ \emph {et~al.}(2004)\citenamefont {Domcke},
  \citenamefont {Yarkony},\ and\ \citenamefont {K\"oppel}}]{Domcke2004}%
  \BibitemOpen
  \bibfield  {author} {\bibinfo {author} {\bibfnamefont {W.}~\bibnamefont
  {Domcke}}, \bibinfo {author} {\bibfnamefont {D.}~\bibnamefont {Yarkony}},\
  and\ \bibinfo {author} {\bibfnamefont {H.}~\bibnamefont {K\"oppel}},\
  }\href@noop {} {\emph {\bibinfo {title} {Conical Intersection: Electronic
  Structure, Dynamics \& Spectroscopy}}}\ (\bibinfo  {publisher} {World
  Scientific},\ \bibinfo {year} {2004})\BibitemShut {NoStop}%
\bibitem [{\citenamefont {Kemper}\ \emph {et~al.}(1977)\citenamefont {Kemper},
  \citenamefont {{Van Dijk}},\ and\ \citenamefont {Buck}}]{Kemper1977}%
  \BibitemOpen
  \bibfield  {author} {\bibinfo {author} {\bibfnamefont {M.}~\bibnamefont
  {Kemper}}, \bibinfo {author} {\bibfnamefont {J.}~\bibnamefont {{Van Dijk}}},\
  and\ \bibinfo {author} {\bibfnamefont {H.}~\bibnamefont {Buck}},\ }\bibfield
  {title} {\bibinfo {title} {{On the comparison between crude and adiabatic
  Born-Oppenheimer coupling elements}},\ }\href
  {https://doi.org/https://doi.org/10.1016/0009-2614(77)85100-2} {\bibfield
  {journal} {\bibinfo  {journal} {Chem. Phys. Lett.}\ }\textbf {\bibinfo
  {volume} {48}},\ \bibinfo {pages} {590} (\bibinfo {year} {1977})}\BibitemShut
  {NoStop}%
\bibitem [{\citenamefont {Bersuker}(2022)}]{Bersuker2023}%
  \BibitemOpen
  \bibfield  {author} {\bibinfo {author} {\bibfnamefont {I.}~\bibnamefont
  {Bersuker}},\ }\bibfield  {title} {\bibinfo {title} {{Four modifications of
  the Jahn-Teller effects. The problem of observables: spin-orbital
  interaction, tunneling splitting, orientational polarization of solids.}},\
  }\href {https://doi.org/10.1039/D2CP02895F} {\bibfield  {journal} {\bibinfo
  {journal} {Phys. Chem. Chem. Phys.}\ }\textbf {\bibinfo {volume} {25}},\
  \bibinfo {pages} {1556} (\bibinfo {year} {2022})}\BibitemShut {NoStop}%
\bibitem [{\citenamefont {Anderson}(1967{\natexlab{a}})}]{Anderson1967}%
  \BibitemOpen
  \bibfield  {author} {\bibinfo {author} {\bibfnamefont {P.~W.}\ \bibnamefont
  {Anderson}},\ }\bibfield  {title} {\bibinfo {title} {{Infrared catastrophe in
  Fermi gases with local scattering potentials}},\ }\href
  {https://doi.org/10.1103/PhysRevLett.18.1049} {\bibfield  {journal} {\bibinfo
   {journal} {Phys. Rev. Lett.}\ }\textbf {\bibinfo {volume} {18}},\ \bibinfo
  {pages} {1049} (\bibinfo {year} {1967}{\natexlab{a}})}\BibitemShut {NoStop}%
\bibitem [{\citenamefont {Anderson}(1967{\natexlab{b}})}]{Anderson1967_2}%
  \BibitemOpen
  \bibfield  {author} {\bibinfo {author} {\bibfnamefont {P.~W.}\ \bibnamefont
  {Anderson}},\ }\bibfield  {title} {\bibinfo {title} {Ground state of a
  magnetic impurity in a metal},\ }\href
  {https://doi.org/10.1103/PhysRev.164.352} {\bibfield  {journal} {\bibinfo
  {journal} {Phys. Rev.}\ }\textbf {\bibinfo {volume} {164}},\ \bibinfo {pages}
  {352} (\bibinfo {year} {1967}{\natexlab{b}})}\BibitemShut {NoStop}%
\bibitem [{\citenamefont {Thouless}(1961)}]{Thouless1961}%
  \BibitemOpen
  \bibfield  {author} {\bibinfo {author} {\bibfnamefont {D.~J.}\ \bibnamefont
  {Thouless}},\ }\href
  {https://doi.org/https://doi.org/10.1016/B978-1-4832-3066-5.50013-5} {\emph
  {\bibinfo {title} {The Quantum Mechanics of Many-Body Systems}}},\ \bibinfo
  {series} {Pure and Applied Physics}, Vol.~\bibinfo {volume} {11}\ (\bibinfo
  {publisher} {Elsevier},\ \bibinfo {address} {London},\ \bibinfo {year}
  {1961})\ pp.\ \bibinfo {pages} {153 -- 162}\BibitemShut {NoStop}%
\bibitem [{\citenamefont {Messiah}(1981)}]{Messiah1981}%
  \BibitemOpen
  \bibfield  {author} {\bibinfo {author} {\bibfnamefont {A.}~\bibnamefont
  {Messiah}},\ }\href@noop {} {\emph {\bibinfo {title} {Quantum Mechanics}}}\
  (\bibinfo  {publisher} {Elsevier Science},\ \bibinfo {address} {Amsterdam},\
  \bibinfo {year} {1981})\BibitemShut {NoStop}%
\bibitem [{\citenamefont {Bouvrie}\ \emph {et~al.}(2014)\citenamefont
  {Bouvrie}, \citenamefont {Majtey}, \citenamefont {Tichy}, \citenamefont
  {Dehesa},\ and\ \citenamefont {Plastino}}]{BouvrieMajtey2014}%
  \BibitemOpen
  \bibfield  {author} {\bibinfo {author} {\bibfnamefont {P.~A.}\ \bibnamefont
  {Bouvrie}}, \bibinfo {author} {\bibfnamefont {A.~P.}\ \bibnamefont {Majtey}},
  \bibinfo {author} {\bibfnamefont {M.~C.}\ \bibnamefont {Tichy}}, \bibinfo
  {author} {\bibfnamefont {J.~S.}\ \bibnamefont {Dehesa}},\ and\ \bibinfo
  {author} {\bibfnamefont {A.~R.}\ \bibnamefont {Plastino}},\ }\bibfield
  {title} {\bibinfo {title} {{Entanglement and the Born-Oppenheimer
  approximation in an exactly solvable quantum many-body system}},\ }\href
  {https://doi.org/10.1140/epjd/e2014-50349-2} {\bibfield  {journal} {\bibinfo
  {journal} {Eur. Phys. J. D}\ }\textbf {\bibinfo {volume} {68}},\ \bibinfo
  {pages} {346} (\bibinfo {year} {2014})}\BibitemShut {NoStop}%
\bibitem [{\citenamefont {Izmaylov}\ and\ \citenamefont
  {Franco}(2017)}]{IzmaylovFranco2017}%
  \BibitemOpen
  \bibfield  {author} {\bibinfo {author} {\bibfnamefont {A.~F.}\ \bibnamefont
  {Izmaylov}}\ and\ \bibinfo {author} {\bibfnamefont {I.}~\bibnamefont
  {Franco}},\ }\bibfield  {title} {\bibinfo {title} {{Entanglement in the
  Born--Oppenheimer approximation}},\ }\href
  {https://doi.org/10.1021/acs.jctc.6b00959} {\bibfield  {journal} {\bibinfo
  {journal} {J. Chem. Theory Comput.}\ }\textbf {\bibinfo {volume} {13}},\
  \bibinfo {pages} {20} (\bibinfo {year} {2017})}\BibitemShut {NoStop}%
\bibitem [{\citenamefont {Streltsov}\ \emph {et~al.}(2010)\citenamefont
  {Streltsov}, \citenamefont {Alon},\ and\ \citenamefont
  {Cederbaum}}]{StreltsovAlon2010}%
  \BibitemOpen
  \bibfield  {author} {\bibinfo {author} {\bibfnamefont {A.~I.}\ \bibnamefont
  {Streltsov}}, \bibinfo {author} {\bibfnamefont {O.~E.}\ \bibnamefont
  {Alon}},\ and\ \bibinfo {author} {\bibfnamefont {L.~S.}\ \bibnamefont
  {Cederbaum}},\ }\bibfield  {title} {\bibinfo {title} {{General mapping for
  bosonic and fermionic operators in Fock space}},\ }\href
  {https://doi.org/10.1103/PhysRevA.81.022124} {\bibfield  {journal} {\bibinfo
  {journal} {Phys. Rev. A}\ }\textbf {\bibinfo {volume} {81}},\ \bibinfo
  {pages} {022124} (\bibinfo {year} {2010})}\BibitemShut {NoStop}%
\bibitem [{\citenamefont {Girardeau}(1960)}]{Girardeau1960}%
  \BibitemOpen
  \bibfield  {author} {\bibinfo {author} {\bibfnamefont {M.}~\bibnamefont
  {Girardeau}},\ }\bibfield  {title} {\bibinfo {title} {{Relationship between
  systems of impenetrable bosons and fermions in one dimension}},\ }\href
  {https://doi.org/10.1063/1.1703687} {\bibfield  {journal} {\bibinfo
  {journal} {J. Math. Phys.}\ }\textbf {\bibinfo {volume} {1}},\ \bibinfo
  {pages} {516} (\bibinfo {year} {1960})}\BibitemShut {NoStop}%
\end{thebibliography}%
\end{document}